\documentclass[twocolumn,superscriptaddress,pra]{revtex4-2}

\usepackage{times}
\usepackage[T1]{fontenc}
\usepackage{microtype}
\usepackage[utf8]{inputenc}
\usepackage{amsmath,amsthm,amsfonts,amsbsy}
\usepackage{newtxmath,mathrsfs,bbold,bbm,dsfont}
\usepackage{enumitem}
\usepackage{graphicx,float}
\usepackage[caption=false]{subfig}
\usepackage{array,longtable,booktabs,makecell}
\usepackage{tikz,xcolor}
\usepackage{verbatim,algpseudocode,listings}
\usepackage{csquotes}
\usepackage{hyperref}
\usepackage{lipsum}

\newtheorem{definition}{Definition}

\newtheorem{proposition}{Proposition}

\newtheorem{theorem}{Theorem}

\newcommand{\1}{\mathbb{1}}

\newcommand{\cone}{\mathrm{cone}}

\newcommand{\id}{\mathrm{id}}

\newcommand{\rk}{\mathrm{rank}}

\newcommand{\supp}{\mathrm{supp}}
\newcommand{\tr}{\mathrm{Tr}}
\newcommand{\bra}[1]{\langle#1|}
\newcommand{\ket}[1]{|#1\rangle}
\newcommand{\ip}[2]{\langle#1|#2\rangle}
\newcommand{\op}[2]{\ket{#1}\bra{#2}}

\newcommand{\abb}[1]{{\textnormal{#1}}} 
\newcommand{\cc}[1]{{\overline{#1}}} 
\newcommand{\f}[1]{{\mathsf{#1}}} 
\newcommand{\m}[1]{{\mathcal{#1}}} 
\newcommand{\ens}[1]{{#1}} 
\newcommand{\idx}[1]{{\mathtt{#1}}} 
\newcommand{\ins}[1]{{\mathscr{#1}}} 
\newcommand{\opt}[1]{{\widehat{#1}}} 
\newcommand{\pd}[1]{{\mathbb{#1}}} 
\newcommand{\povm}[1]{{\mathscr{#1}}} 
\newcommand{\rpl}[1]{{\widetilde{#1}}} 
\newcommand{\s}[1]{{\mathfrak{#1}}} 
\newcommand{\spa}[1]{{\mathds{#1}}} 

\definecolor{darkblue}{rgb}{0,0,0.5}
\definecolor{darkgreen}{rgb}{0,0.5,0}
\definecolor{darkred}{rgb}{0.5,0,0}
\hypersetup{
	colorlinks = true,
	citecolor = darkgreen, 
	linkcolor = darkblue, 
	urlcolor  = blue, 
	filecolor = darkred 
}
\definecolor{cool_green}{rgb}{0.0,0.5,0.0}
\definecolor{cool_purple}{rgb}{0.5,0.0,0.5}

\usetikzlibrary{quantikz}
\allowdisplaybreaks

\begin{document}


\title{Incompatibility as a Resource for Programmable Quantum Instruments}

\author{Kaiyuan Ji}
\email{kj264@cornell.edu}
\affiliation{Center for Quantum Information, Institute for Interdisciplinary Information Sciences, Tsinghua University, Beijing 100084, China}
\affiliation{School of Electrical and Computer Engineering, Cornell University, Ithaca, New York 14850, USA}

\author{Eric Chitambar}
\email{echitamb@illinois.edu}
\affiliation{Department of Electrical and Computer Engineering, Coordinated Science Laboratory, University of Illinois at Urbana-Champaign, Urbana, Illinois 61801, USA}

\date{\today}


\begin{abstract}
Quantum instruments represent the most general type of quantum measurement, as they incorporate processes with both classical and quantum outputs.  In many scenarios, it may be desirable to have some ``on-demand'' device that is capable of implementing one of many possible instruments whenever the experimenter desires.  We refer to such objects as programmable instrument devices (PIDs), and this paper studies PIDs from a resource-theoretic perspective.  A physically important class of PIDs are those that do not require quantum memories to implement, and these are naturally ``free'' in this resource theory.  Additionally, these free objects correspond precisely to the class of unsteerable channel assemblages in the study of channel steering.  The traditional notion of measurement incompatibility emerges as a resource in this theory since any PID controlling an incompatible family of instruments requires a quantum memory to build.  We identify an incompatibility preorder between PIDs based on whether one can be transformed into another using processes that do not require additional quantum memories.  Necessary and sufficient conditions are derived for when such transformations are possible based on how well certain guessing games can be played using a given PID\@.  Ultimately our results provide an operational characterization of incompatibility, and they offer semi-device-independent tests for incompatibility in the most general types of quantum instruments.
\end{abstract}





\maketitle

\tableofcontents

\section{Introduction}
\label{sec:introduction}

Incompatibility is a quintessential feature of quantum mechanics.  Unlike classical systems in which conjugate variables have definite values at each moment in time, quantum systems are dictated by celebrated uncertainty relations, which place sharp restrictions on how well the measurement outcomes of two (or more) noncommuting observables can be predicted~\cite{Coles-2017a}.  The incompatibility of noncommuting observables has wide-ranging applications in quantum information science from quantum cryptography~\cite{Koashi-2006a, Tomamichel-2011a} to entanglement detection~\cite{Horodecki-2009a} to quantum error correction~\cite{Renes-2014a}.  For more general types of measurements beyond textbook observables, commutation relations are no longer sufficient to characterize measurement incompatibility.  One instead considers the property of joint measurability, which means that a joint probability distribution can be defined for the given collection of measurement devices, each being described by a positive operator-valued measure (POVM)~\cite{Lahti-2003a, Heinosaari-2008a, Heinosaari-2016a}.  Incompatibility between POVMs in this sense means that such joint measurability is not possible.

Whereas POVMs characterize the classical output of a quantum measurement, a more general description of the measurement process also includes the quantum output.  Here, one typically invokes the theory of quantum instruments~\cite{Davies-1970a}, with an instrument formally being defined as a family of completely positive  (CP) maps $\{\Lambda_{x_1}\}_{x_1}$ such that $\sum_{x_1}\Lambda_{x_1}$ is trace preserving (TP).  When performing an instrument on a quantum state $\rho$, a classical outcome $x_1$ is observed with probability $p_{x_1}=\tr[\Lambda_{x_1}[\rho]]$, and the postmeasurement state is then given by $\Lambda_{x_1}[\rho]/p_{x_1}$.  Note that POVMs are a special type of instrument for which $\Lambda_{x_1}[\rho]=\tr[M_{x_1}\rho]$ for some collection of positive semidefinite operators $\{M_{x_1}\}_{x_1}$ with $\sum_{x_1}M_{x_1}=\1$.  Likewise, a quantum channel (i.e., a CPTP map) is also a type of quantum instrument, having just a single classical outcome.  The notion of incompatibility can also be extended into the domain of channels and instruments~\cite{Heinosaari-2013a, Heinosaari-2016a, Heinosaari-2017a, D'Ariano-2022a}.  Similar to the case of POVMs, a family of instruments $\{\Lambda_{x_1|x_0}\}_{x_0,x_1}$ is compatible if all the constituent instruments can be simulated using a single instrument combined with classical postprocessing; incompatible instruments lack this property.

Extensive work has recently been conducted to capture incompatibility as a physical resource in quantum information processing~\cite{Pusey-2015a, Heinosaari-2015a, Heinosaari-2015b, Guerini-2017a, Oszmaniec-2019a, Uola-2019a, Skrzypczyk-2019a, Buscemi-2020a, Buscemi-2023a}.  This can be accomplished using the formal structure of a resource theory~\cite{Horodecki-2013a, Coeke-2016a, Chitambar-2019a, Takagi-2019b, Liu-2019a}, in which objects are characterized as being either free or resourceful.  Additionally, only a restricted set of physical operations can be performed by the experimenter, and these are unable to create resourceful objects from free ones.  In the case of quantum incompatibility, the free objects are compatible families of POVMs or instruments, and the incompatible ones are resourceful.

By adopting a resource-theoretic perspective, one can establish operationally meaningful measures of incompatibility such as its robustness to noise~\cite{Piani-2015a, Heinosaari-2015a, Haapasalo-2015a, Guerini-2019a, Uola-2019a, Designolle-2019a}.  The incompatibility in one family of POVMs or instruments can then be quantitatively compared to another.  Resource theories also provide tools for detecting or ``witnessing'' the incompatibility present in general measurement devices~\cite{Guerini-2019a, Carmeli-2019a, Carmeli-2019b, Heinosaari-2021a, Regula-2021a}.  This certification can also be done in a semi-device-independent way~\cite{Pusey-2015a, Carmeli-2019c, Skrzypczyk-2019a, Uola-2019a, Buscemi-2020a, Takagi-2019b, Uola-2020a, Mori-2020a}.  In other words, by attaining a certain score on some type of quantum measurement game, the experimenter can rest assured that he or she is controlling some family of incompatible POVMs or instruments without having full trust in the inner workings of the device.  Crucially, the largest achievable score using some device cannot be increased using the allowed operations of the resource theory, and the scores therefore represent resource monotones.  In many cases, these games define a complete set of monotones whose values provide necessary and sufficient conditions for convertibility of one object to another by the allowed operations~\cite{Buscemi-2012a, Buscemi-2012b, Buscemi-2016a, Rosset-2018a, Skrzypczyk-2019b, Buscemi-2020a, Regula-2021a}.  We show in Sec.~\ref{sec:game} that the same holds true for the guessing games considered in this paper, and furthermore, the advantage of using an incompatible device in these games can be quantitatively characterized by the aforementioned robustness measure.  However, the general idea of relating convertibility to guessing games can be traced back to the original work of Blackwell on statistical comparisons~\cite{Blackwell-1953a} (see Ref.~\cite{Buscemi-2012a} for more discussions).

\subsection{From programmability to nonsignaling}
\label{sec:programmability}

\begin{figure}[t]
\includegraphics[scale=0.2]{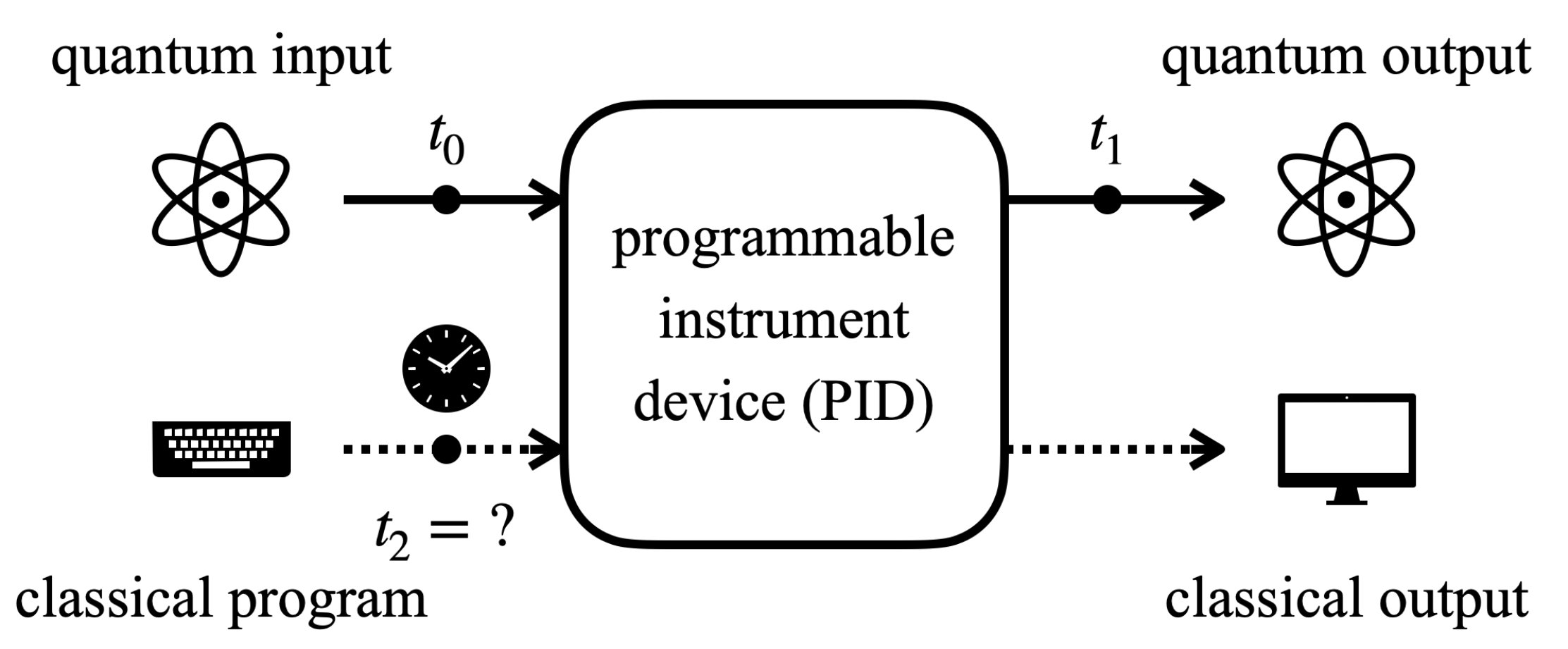}
\caption{A general controllable quantum device applies an instrument $\{\Lambda_{x_1|x_0}\}_{x_1}$ to the quantum input whenever a particular program $x_0$ is chosen.  The characteristic time $\Delta t_\abb{D}\coloneq t_1-t_0$ is known as the \emph{quantum delay time} of the device, and it measures how quickly the device functions as a q-to-q channel.  The device is said to be fully programmable if the program is free to arrive at any time $t_2$, even outside the interval $[t_0,t_1)$, and we refer to such a device as a \emph{programmable instrument device}.  Given that classical memories are freely available, the full programmability of a PID is essentially its ability to withstand an arbitrarily late arrival of the program, termed the \emph{late-program assumption}.  In the framework of Ref.~\cite{Chitambar-2021a}, a PID represents a so-called ``multiprocess.''}
\label{fig:temporal}
\end{figure}

Our analysis of quantum incompatibility is motivated by the idea of ``programmable'' quantum instruments.  Consider a generic controllable measurement device as depicted in Fig.~\ref{fig:temporal}, which is capable of implementing some family of instruments $\{\Lambda_{x_1|x_0}\}_{x_0,x_1}$.  The classical program is the input value $x_0$, which dictates that the instrument $\{\Lambda_{x_1|x_0}\}_{x_1}$ be performed on the quantum input.  We consider these devices to be modules in nature so that the classical or quantum output from one device can be connected to a classical or quantum input of another.  This introduces a critical consideration of time: for the devices to function together properly, the outputs of one device must arrive at a time when the next device is ready to receive them.  In practice, every physical device will have a characteristic \emph{quantum delay time}~\cite{Chitambar-2021a}, which measures how fast the device generates a quantum output when given a quantum input, and it corresponds to $\Delta t_\abb{D}\coloneq t_1-t_0$ in Fig.~\ref{fig:temporal}.  How about the timing of the classical program?  As typically demanded by devices with multiple inputs, one would expect that the program be synchronized with the quantum input, or at least be within the finite window $[t_0,t_1)$.  However, if the quantum delay time $\Delta t_\abb{D}$ appears to be short, such a hard constraint on timing can be unrealistic in practice, and therefore we have a particular interest in devices that give the experimenter full temporal freedom over when he or she can submit the program, a capability called \emph{programmability} in Ref.~\cite{Buscemi-2020a}.  As a consequence of programmability, the timing of the classical and quantum inputs need not be synchronized, and the classical program can arrive significantly before $t_0$ or after $t_1$.  Without loss of generality, we need only to assume that the program arrives after the quantum output time $t_1$, which we refer to as the \emph{late-program assumption}, and this is because an early arriving program can always be buffered in a classical memory before the quantum input arrives.  Clearly, not every controllable quantum device can work through the late-program assumption, and a device can do so if and only if the quantum output is independent of the classical input after coarse-graining the classical output (see Sec.~\ref{sec:PID}).  Formally, this constraint is known as \emph{nonsignaling}, which requires that
\begin{align}
\label{eq:nonsignaling}
	\sum_{x_1}\Lambda_{x_1|x_0}&=\sum_{x_1}\Lambda_{x_1|x'_0}\eqcolon\Lambda\quad\forall x_0,x'_0.
\end{align}
In other words, all the instruments in the family $\{\Lambda_{x_1|x_0}\}_{x_0,x_1}$ sum up to the same channel $\Lambda$.  Since the nonsignaling constraint in Eq.~\eqref{eq:nonsignaling} is necessary and sufficient for programmability, we naturally refer to devices satisfying this constraint as \emph{programmable instrument devices (PIDs)}.

While the quantum output at $t_1$ for a PID is independent of the classical program, the classical output will generally depend on the quantum input at $t_0$ (and certainly also on the classical input at $t_2$).  Hence under the late-program assumption, the internal quantum memory of the PID might need to store quantum information for an indefinite amount of time until the experimenter chooses to issue a program.  However, there is a special class of PIDs for which the quantum memory can be perfectly substituted with a classical memory.  These are called \emph{simple PIDs}, and they represent the ``free'' objects in the resource theory.  Remarkably, simple PIDs are precisely those in which the family of instruments being implemented is compatible.  In contrast, \emph{nonsimple PIDs}, i.e., PIDs that are not simple, require quantum memories with an indefinite lifetime to support full programmability, and thus they are resources, demonstrating incompatibility.  Of course, an indefinite lifetime of a quantum memory is an idealization and hence so is full programmability.  In practice, every realizable PID will have a quantum memory with some finite lifetime $\Delta t_\abb{QM}<\infty$.  To pinpoint the differing demands on quantum memories for programmability, throughout this paper, we may as well assume that every PID satisfies $\Delta t_\abb{D}\approx0$ and that the internal quantum memory of any nonsimple PID satisfies $\Delta t_\abb{QM}\gg\Delta t_\abb{D}$.  These assumptions are in line with our identification of only simple PIDs as being free.

\begin{figure}[t]
\includegraphics[scale=0.2]{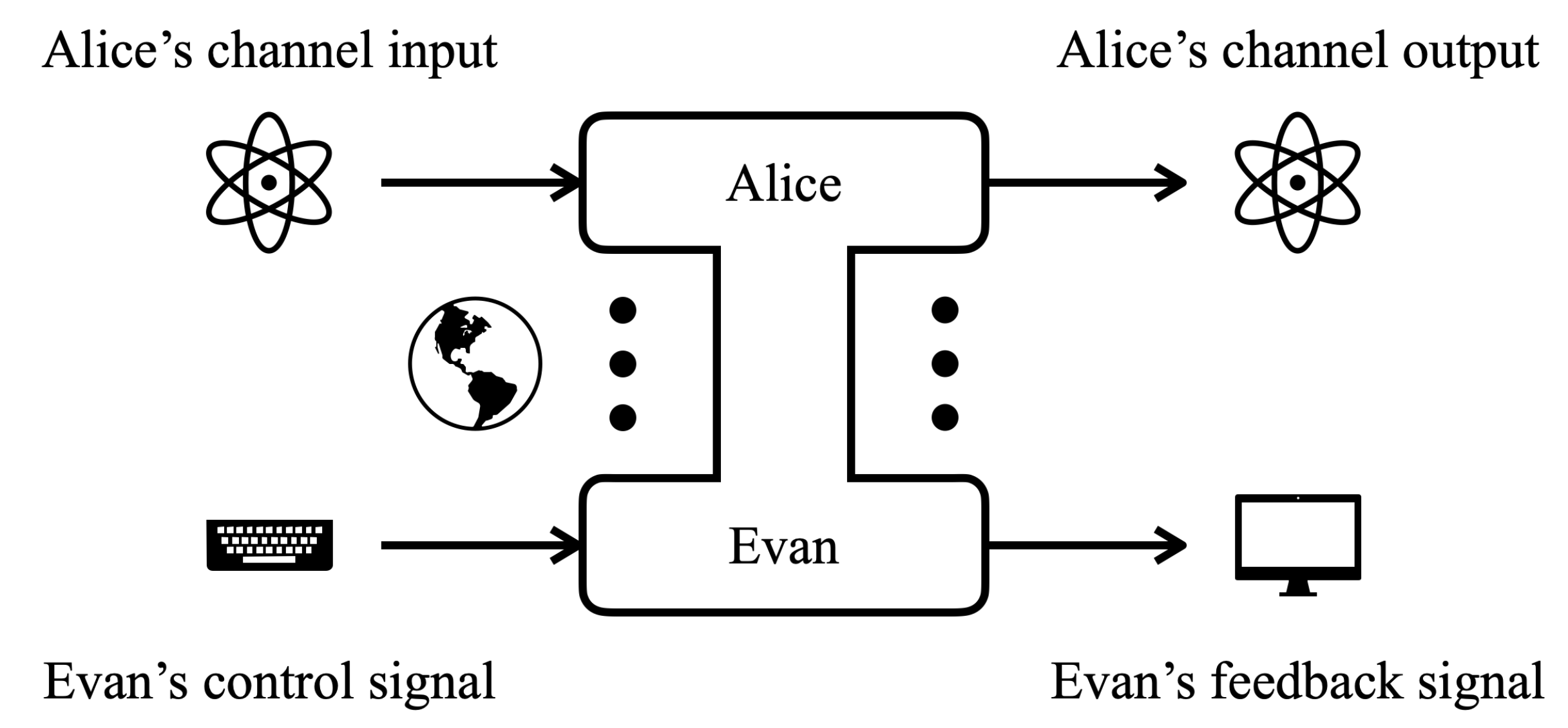}
\caption{The control device could also be split between spatially separated parties Alice and Evan.  The nonsignaling constraint from Evan's control signal to Alice's channel output naturally arises if the spatial separation is so large that Evan's signal propagation time exceeds the anticipated quantum delay time between Alice's local input and output.}
\label{fig:spatial}
\end{figure}

There is an alternative justification for imposing the nonsignaling constraint in Eq.~\eqref{eq:nonsignaling} not directly related to programmability.  One could imagine that the device in Fig.~\ref{fig:temporal} is a bipartite channel shared between spatially separated parties, with Evan controlling the classical input and output and Alice controlling the quantum input and output, as depicted in Fig.~\ref{fig:spatial}.  Alice may be unaware of the existence of the eavesdropping party Evan, and she thinks of her quantum input and output as being connected by a local channel.  She would then expect that the quantum delay time of her channel should be extremely short, limited only by the local inner workings of her device, which implies that the information from Evan's control signal should not have enough time to propagate across spacetime and influence her channel output.  Thus Alice's output would be spacelike separated from the choice of Evan's control signal, and the nonsignaling condition (from Evan to Alice) would hold, as in Eq.~\eqref{eq:nonsignaling}.  Note that this reflects a scenario known as \emph{channel steering}~\cite{Piani-2015b}, as Evan is remotely manipulating Alice's channel with his control signal without letting her detect him due to the nonsignaling constraint.  Apart from being useful for better understanding quantum incompatibility, the scenario of channel steering is also relevant for cryptographic applications and one-sided device-independent testing of coherent channel extensions~\cite{Piani-2015b}.  In fact, steerability and incompatibility are equivalent concepts when defined on nonsignaling families of instruments (see Sec.~\ref{sec:steering}), and so the resource theory of PID nonsimplicity that we develop in this paper is equivalently a resource theory of channel steering.

\subsection{Organization of the paper}
\label{sec:organization}

With this background and motivation in hand, we now conduct our resource-theoretic analysis of instrument incompatibility in terms of nonsimplicity of PIDs in more detail.  The rest of the paper is organized as follows.

In Secs.~\ref{sec:device} and \ref{sec:simulation}, we establish the basic pieces of our resource theory.  We first review the traditional concept of measurement incompatibility and its resource theory in Secs.~\ref{sec:PMD} and \ref{sec:simulation-PMD}.  Then we extend the present theory to incorporate the more general concept of incompatibility between quantum instruments.  Specifically, we formally introduce simple (i.e., free) versus nonsimple (i.e., resourceful) PIDs and the physical distinction between them in Sec.~\ref{sec:PID}.  We then propose a class of free transformations between PIDs, which is incompatibility-nonincreasing and has a clear operational meaning, in Sec.~\ref{sec:simulation-PID}.

In Sec.~\ref{sec:relationship}, we focus on discussing and discovering the relationships between PID nonsimplicity, steering, and traditional measurement incompatibility.  We explicate the spatiotemporal correspondence between channel steering and PIDs in Sec.~\ref{sec:steering}.  Then we generalize the concept of steering-equivalent observables to the scenario of channel steering in Sec.~\ref{sec:SEM}, and in doing so we derive a monotonicity theorem that signifies a fundamental connection between the resource theory of PID nonsimplicity and that of measurement incompatibility.

Finally, in Sec.~\ref{sec:game}, we provide a semi-device-independent characterization for PID nonsimplicity by designing a class of so-called ``nontransient'' guessing games as a temporal analog to the well-studied nonlocal games.  In particular, nontransient guessing games can be used to characterize the incompatibility preorder between PIDs by providing necessary and sufficient conditions for convertibility from one PID to another under the aforementioned free transformations, as shown in Sec.~\ref{sec:convertibility}.  We further show in Sec.~\ref{sec:robustness} that the operational advantage of a given PID over simple PIDs in these games is tightly bounded from above by the PID's robustness of incompatibility against noise.  In Sec.~\ref{sec:experimental}, we discuss the experimental setup of nontransient guessing games and put forward a variant class of games that lowers the experimental requirement while also faithfully characterizing the incompatibility preorder between PIDs.

\section{Programmable Quantum Devices}
\label{sec:device}

In this section, we first review the traditional concept of measurement incompatibility formulated as a resource for programmable measurement devices~\cite{Buscemi-2020a}.  Then we extend the traditional framework by incorporating programmable instrument devices and a generalized concept of incompatibility for such devices.

\begin{figure*}[t]
\begin{minipage}{0.49\textwidth}
\subfloat[A general PID\@. \label{fig:PID-spatial}]{\includegraphics[scale=0.15]{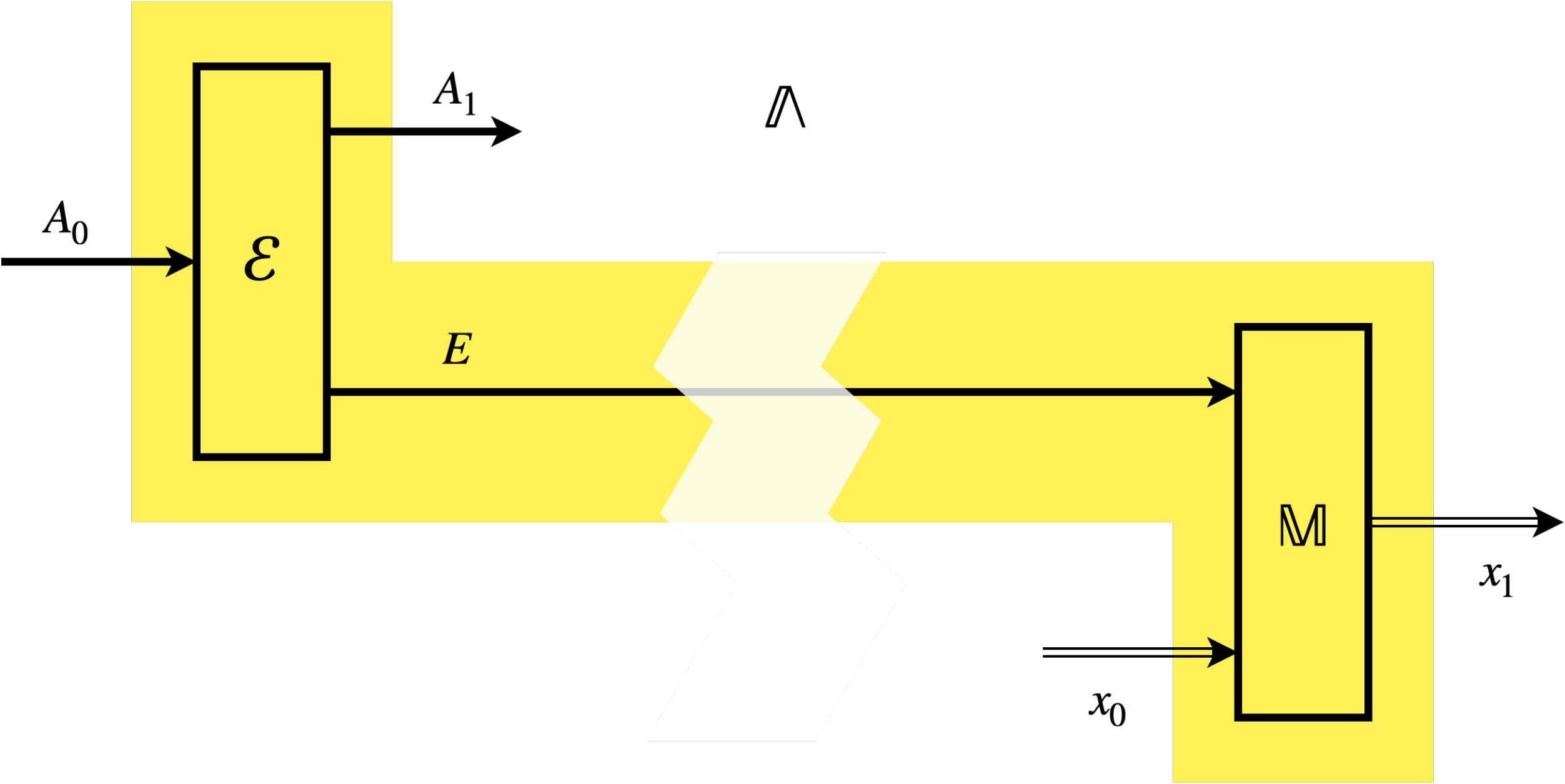}}
\end{minipage} \hfill
\begin{minipage}{0.49\textwidth}
\subfloat[A general PID\@. \label{fig:PID-temporal}]{\includegraphics[scale=0.15]{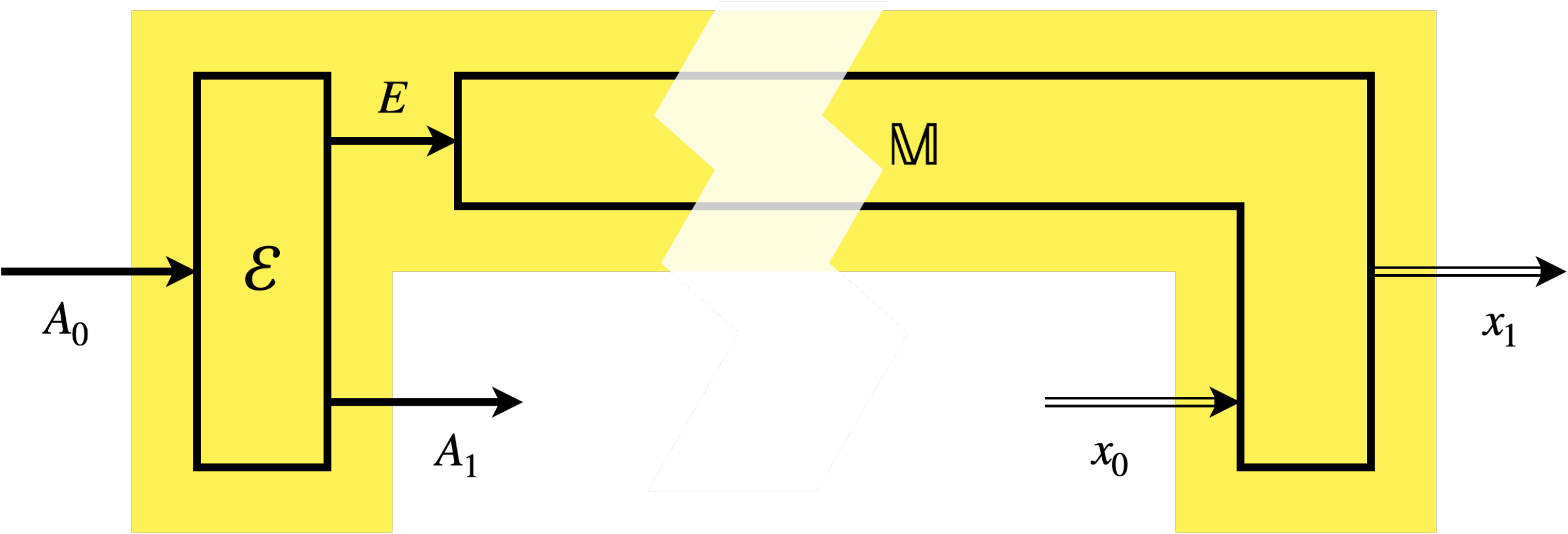}} \\
\subfloat[A simple PID\@. \label{fig:PID-simple}]{\includegraphics[scale=0.15]{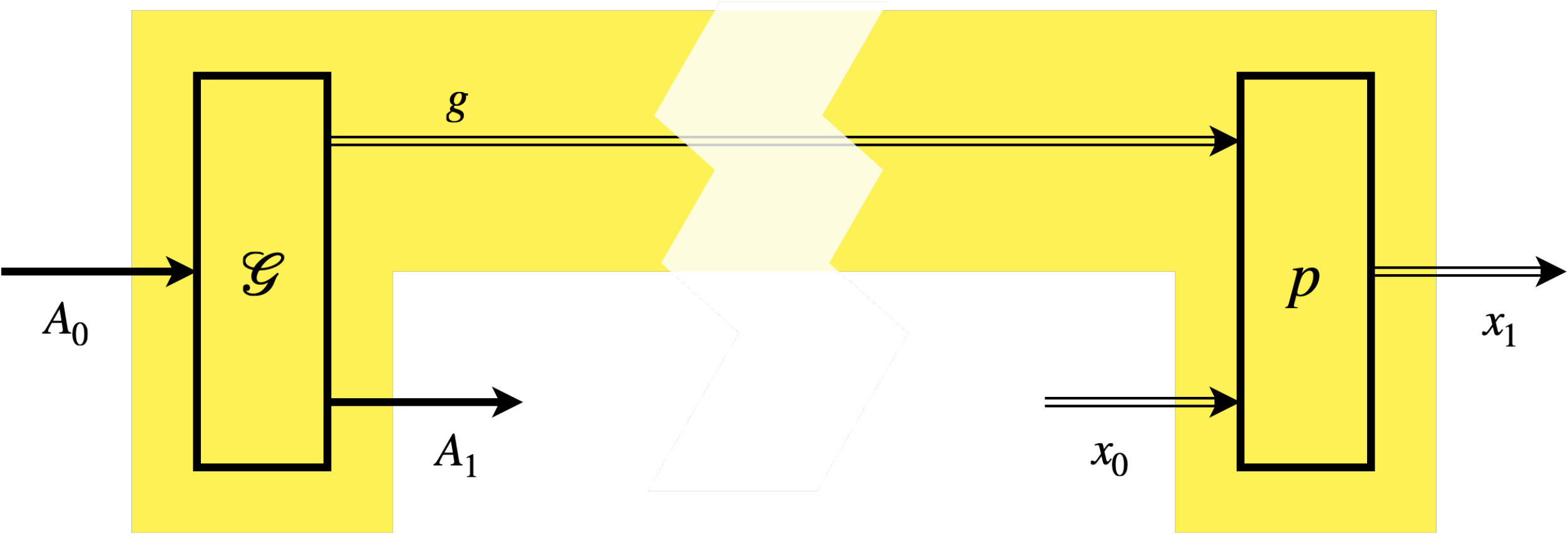}}
\end{minipage}
\caption{Decomposition of a general PID (Figs.~\ref{fig:PID-spatial} and \ref{fig:PID-temporal}) and that of a simple PID (Fig.~\ref{fig:PID-simple}).  Solid arrows stand for quantum systems and hollow arrows for classical systems.  Time flows from left to right.  The opaque rips indicate that the quantum (left) and classical (right) parts of the devices are temporally separated under the late-program assumption.  (a) A general PID $\pd{\Lambda}$ can be realized by connecting one output system $E$ of a broadcast channel $\m{E}^{A_0\to A_1E}$ to a PMD $\pd{M}$ using a quantum memory channel $\id^E$.  As such, the inner working of the PID can be understood as a process of channel steering (see Sec.~\ref{sec:steering}).  (b) The general PID can be represented in a different configuration, with the system $A_1$ displaced downwards and the quantum memory channel subsumed within the PMD $\pd{M}$.  (c) A simple PID can be realized with a ``mother'' instrument $\ins{G}$ and a classical channel $p$.  In ``steering'' terms, the simple PID implements an unsteerable channel assemblage.  Note that the simple PID has the same quantum delay time $\Delta t_\abb{D}\approx0$ between $A_0$ and $A_1$ as the general PID does.  However, the quantum memory of the general PID needs a much longer lifetime $\Delta t_\abb{QM}\gg\Delta t_\abb{D}$ in general to accept a late arriving classical program $x_0$, unlike the simple PID.}
\label{fig:PID}
\end{figure*}

\subsection{Programmable measurement devices}
\label{sec:PMD}

A \emph{programmable measurement device (PMD)}~\cite{Buscemi-2020a}, alias a \emph{multimeter}~\cite{Gour-2018b}, is a quantum measurement device capable of implementing a family of POVMs conditioned on a classical control signal.  A PMD $\pd{M}$ is mathematically represented by a collection of positive semidefinite operators $\pd{M}\equiv\{M_{x_1|x_0}\}_{x_0,x_1}$ such that $\sum_{x_1}M_{x_1|x_0}=\1$ for all $x_0$.  The classical input $x_0$ is known as the \emph{program}, which indicates the particular POVM $\{M_{x_1|x_0}\}_{x_1}$ to be performed on the quantum input.  The classical output $x_1$ labels the measurement outcome of the POVM\@.  PMDs represent the most general qc-to-c CPTP maps.

A PMD $\pd{M}$ is said to be \emph{simple} whenever it can be simulated with a ``mother'' POVM followed by some (controllable) classical postprocessing; namely, it implements a \emph{compatible} family of POVMs admitting the following decomposition:
\begin{align}
\label{eq:PMD-simple}
	M_{x_1|x_0}&=\sum_{g}p_{x_1|x_0,g}G_g\quad\forall x_0,x_1,
\end{align}
where $\{G_g\}_g$ is a POVM and $\{p_{x_1|x_0,g}\}_{x_0,x_1,g}$ is a classical channel (i.e., a conditional probability distribution).  A PMD not decomposable in the form of Eq.~\eqref{eq:PMD-simple} is \emph{nonsimple}, as the family of POVM it implements is \emph{incompatible}.

An advantage of studying quantum measurements in terms of PMDs is that it links measurement incompatibility and quantum memories in a physically motivated way.  As discussed in Sec.~\ref{sec:programmability}, a practical conception of programmability should take into account the unavoidable asynchronicity between the quantum input and the classical program, and so programmable devices should (ideally) allow the experimenter to issue the program at any desirable time~\cite{Buscemi-2020a}.  This temporal freedom can be simply captured by the late-program assumption, which we regard as a basic principle in the programmability context.  For a PMD to function as a qc-to-c box under this assumption, an internal quantum memory is generally needed to store the quantum input until the program is submitted to the classical system $X_0$.  However, if the PMD is controlling a compatible family of measurements, i.e., the PMD being simple, then no quantum memory is needed.  Instead, the ``mother'' POVM $\{G_g\}_g$ can be performed as soon as the quantum input arrives, and the outcome $g$ is stored in a classical memory until the program arrives.  Thus, the requirement of a quantum memory to implement a PMD is another way of characterizing measurement incompatibility.  It is then natural to identify simple PMDs as being ``free'' objects since they do not require quantum memories to implement, and a resource theory of PMD nonsimplicity is physically well justified.

\subsection{Programmable instrument devices}
\label{sec:PID}

Next we extend the theory of programmability to quantum instruments.  A quantum instrument is a generalized version of measurement that incorporates a quantum output representing the postmeasurement state.  The most straightforward generalization of a PMD is a \emph{multi-instrument}~\cite{Gour-2018b}, a device capable of implementing a collection of quantum instruments conditioned on a classical control signal.  A multi-instrument $\pd{\Lambda}$ is mathematically represented by a collection of completely positive (CP) maps $\pd{\Lambda}\equiv\{\Lambda_{x_1|x_0}\}_{x_0,x_1}$ such that $\sum_{x_1}\Lambda_{x_1|x_0}$ is trace preserving (TP) for all $x_0$.  To make the quantum input and output systems $A_0$ and $A_1$ explicit, we sometimes write $\pd{\Lambda}\equiv\{\Lambda_{x_1|x_0}^{A_0\to A_1}\}_{x_0,x_1}$.

Multiinstruments represent the most general qc-to-qc CPTP maps.  However, according to our conception of programmability, general multi-instruments are not suitable models for abstracting ``programmable'' quantum instruments.  As discussed in Sec.~\ref{sec:programmability} generally and in Sec.~\ref{sec:PMD} for PMDs, practical programmable instruments should withstand the late-program assumption; they should function while the classical program is free to arrive at $X_0$ anytime after the quantum input arrives at $A_0$.  By the same assumption, the program could even arrive \emph{after} the device is scheduled to dispense some quantum output at $A_1$.  For a device to be physically realizable in such a circumstance, there must be no signaling from the classical input $X_0$ to the quantum output $A_1$.  This motivates the following definition.

\begin{definition}
\label{def:PID}
A multi-instrument $\pd{\Lambda}\equiv\{\Lambda_{x_1|x_0}\}_{x_0,x_1}$ is called a \textbf{programmable instrument device} whenever it is nonsignaling from the classical input to the quantum output; namely, there exists a quantum channel $\Lambda$ such that
\begin{align}
\label{eq:PID} 
	\sum_{x_1}\Lambda_{x_1|x_0}&=\Lambda\quad\forall x_0.
\end{align}
We say that the PID $\pd{\Lambda}$ is \textbf{simple} (and otherwise \textbf{nonsimple}) whenever there exists a quantum instrument $\ins{G}\equiv\{\m{G}_g\}_g$ and a classical channel $p\equiv\{p_{x_1|x_0,g}\}_{x_0,x_1,g}$ such that
\begin{align}
\label{eq:PID-simple}
	\Lambda_{x_1|x_0}&=\sum_{g}p_{x_1|x_0,g}\m{G}_g\quad\forall x_0,x_1.
\end{align}
\end{definition}

The above definition of a programmable instrument device generalizes the definition of a PMD in a way that respects the late-program assumption.  Likewise, the concept of incompatibility in terms of device nonsimplicity is extended from POVMs to instruments.  Following an analogous argument that previously applies to PMDs, one can find that the difference between a nonsimple PID and a simple PID is precisely captured by whether the device needs a quantum memory with a non-negligible lifetime $\Delta t_\abb{QM}\gg\Delta t_\abb{D}\approx0$ to implement.  Accordingly, it is natural to identify nonsimple PIDs as resources in our theory of programmable instruments, whereas simple PIDs are free objects.

While our formulation of PIDs is motivated by the notion of programmability, the bipartite picture shown in Fig.~\ref{fig:spatial} can be helpful in understanding the internal structure of such devices.  We envision that Alice has the quantum input and output in her laboratory while Evan controls the classical input and output.  The nonsignaling condition (from Evan to Alice) in the definition of a PID is also known as ``semicausality''~\cite{Beckman-2001a}.  It has been proved that every semicausal map is ``semilocalizable''~\cite{Eggeling-2002a}, meaning that the map can be decomposed into local operations by Alice and Evan individually combined with one-way quantum communication from Alice to Evan, as shown in Fig.~\ref{fig:PID-spatial}.  Simple PIDs are then precisely those in which the one-way quantum communication can be replaced with one-way classical communication, as shown in Fig.~\ref{fig:PID-simple}.  In Secs.~\ref{sec:steering} and \ref{sec:SEM}, we provide a more formal statement (Proposition~\ref{prop:operational}) and related discussions regarding the internal structure of a PID\@.

\section{Free Simulations of Programmable Devices}
\label{sec:simulation}

In this section, we complete the construction of the resource theory of PID nonsimplicity by proposing a set of free transformations between PIDs.  Hereafter, we will refer to free transformations applied to programmable devices (either PIDs or PMDs) as \emph{free simulations} of the devices.

\subsection{Free simulations of PMDs}
\label{sec:simulation-PMD}

Before we introduce what constitutes a free simulation of PIDs, we first recall the free simulations of PMDs in the resource theory of PMD nonsimplicity~\cite{Buscemi-2020a}.  Later as we define the free simulations of PIDs, we must ensure that they reduce to the predefined simulations of PMDs when both the source PID and the target PID are PMDs.

\begin{definition}[\cite{Buscemi-2020a}]
\label{def:simulation-PMD}
Let $\pd{M}\equiv\{M_{x_1|x_0}\}_{x_0,x_1}$ and $\pd{N}\equiv\{N_{y_1|y_0}\}_{y_0,y_1}$ be two PMDs.  We say that $\pd{M}$ can \textbf{freely simulate} $\pd{N}$, denoted by $\pd{M}\succcurlyeq_\abb{M}\pd{N}$, whenever there exists a quantum instrument $\ins{K}\equiv\{\m{K}_k\}_k$ and two classical channels $p\equiv\{p_{x_0,l|y_0,k}\}_{x_0,y_0,k,l}$ and $q\equiv\{q_{y_1|x_1,l}\}_{x_1,y_1,l}$ such that
\begin{align}
	\label{eq:simulation-PMD}
	N_{y_1|y_0}&=\sum_{x_0,x_1,k,l}q_{y_1|x_1,l}p_{x_0,l|y_0,k}\m{K}_k^\dagger\left[M_{x_1|x_0}\right]\quad\forall y_0,y_1,
\end{align}
where $(\cdot)^\dagger$ denotes adjunction.  We call the transformation $\pd{M}\mapsto\pd{N}$ a \textbf{free simulation} of PMDs.
\end{definition}

\begin{figure}[t]
\includegraphics[scale=0.15]{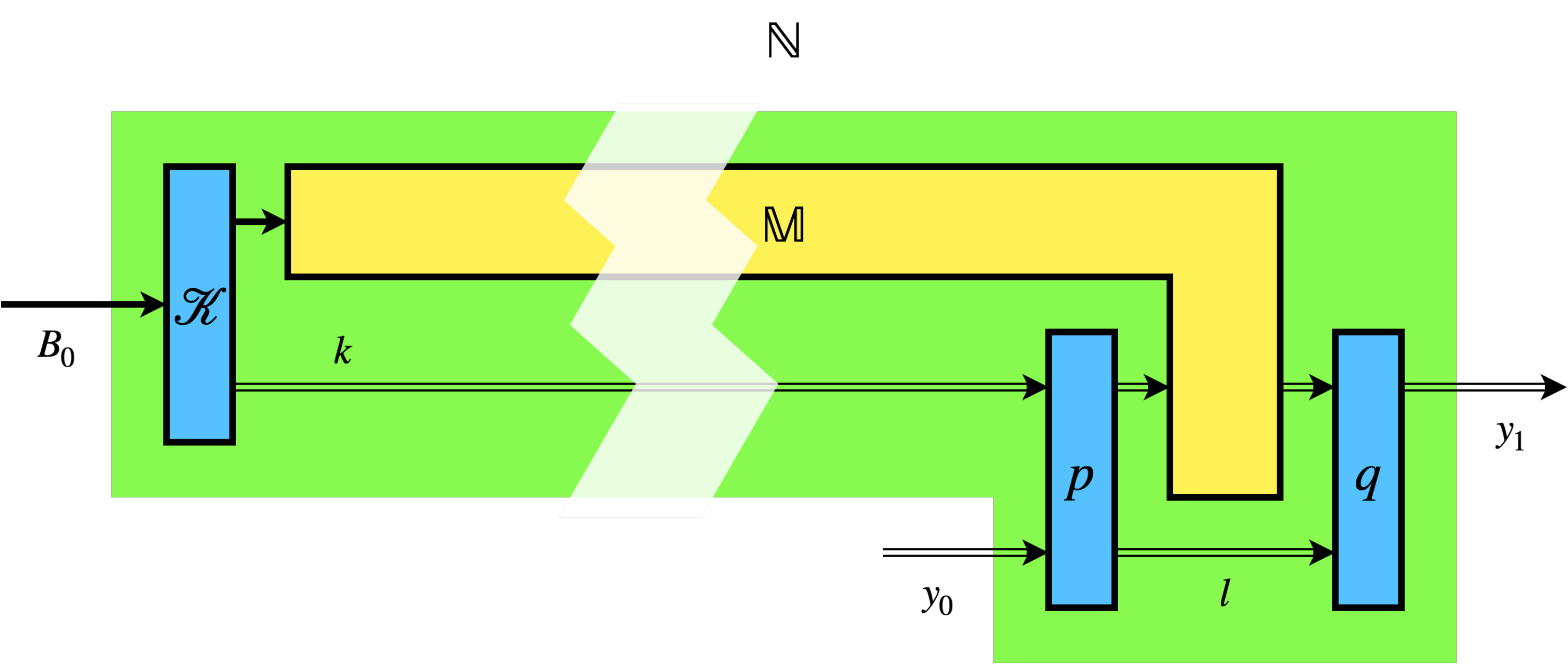}
\caption{The free simulation (blue) of a PMD $\pd{N}$ (green) using another PMD $\pd{M}$ (yellow) according to Eq.~\eqref{eq:simulation-PMD}.  It has been shown that nonsimple PMDs cannot be freely simulated by simple PMDs~\cite{Buscemi-2020a}.  Simulations in this form are precisely those that can be realized without additional quantum memories under the late-program assumption.}
\label{fig:simulation-PMD}
\end{figure}

The operational significance of Definition~\ref{def:simulation-PMD} is demonstrated by the fact that $\pd{M}\succcurlyeq_\abb{M}\pd{N}$ if and only if $\pd{M}$ can be physically transformed (i.e., via a quantum superchannel~\cite{Chiribella-2008a, Gour-2019a}) into $\pd{N}$ using no additional quantum memory~\cite{Buscemi-2020a}, as represented in Fig.~\ref{fig:simulation-PMD}.  Meanwhile, free simulations of PMDs have been shown to possess essential resource-theoretic properties including preserving PMD simplicity and being able to generate the entire set of simple PMDs~\cite{Buscemi-2020a}.  Thus they are formally qualified as the free transformations in a resource theory of PMD nonsimplicity, and the relation $\succcurlyeq_\abb{M}$ is a legitimate incompatibility preorder on the set of PMDs.

\subsection{Free simulations of PIDs}
\label{sec:simulation-PID}

Now we are ready to propose the free simulations for the resource theory of PID nonsimplicity.  In what follows, we first identify the complete class of PID transformations that do not require quantum memories to implement, and then we demonstrate its legitimacy as the set of free transformations from a resource-theoretic standpoint.  Since PIDs are semicausal quantum channels (i.e., quantum $2$-combs), transformations between them are supposed to be quantum $4$-combs~\cite{Chiribella-2009a}.

\begin{figure}[t]
\includegraphics[scale=0.15]{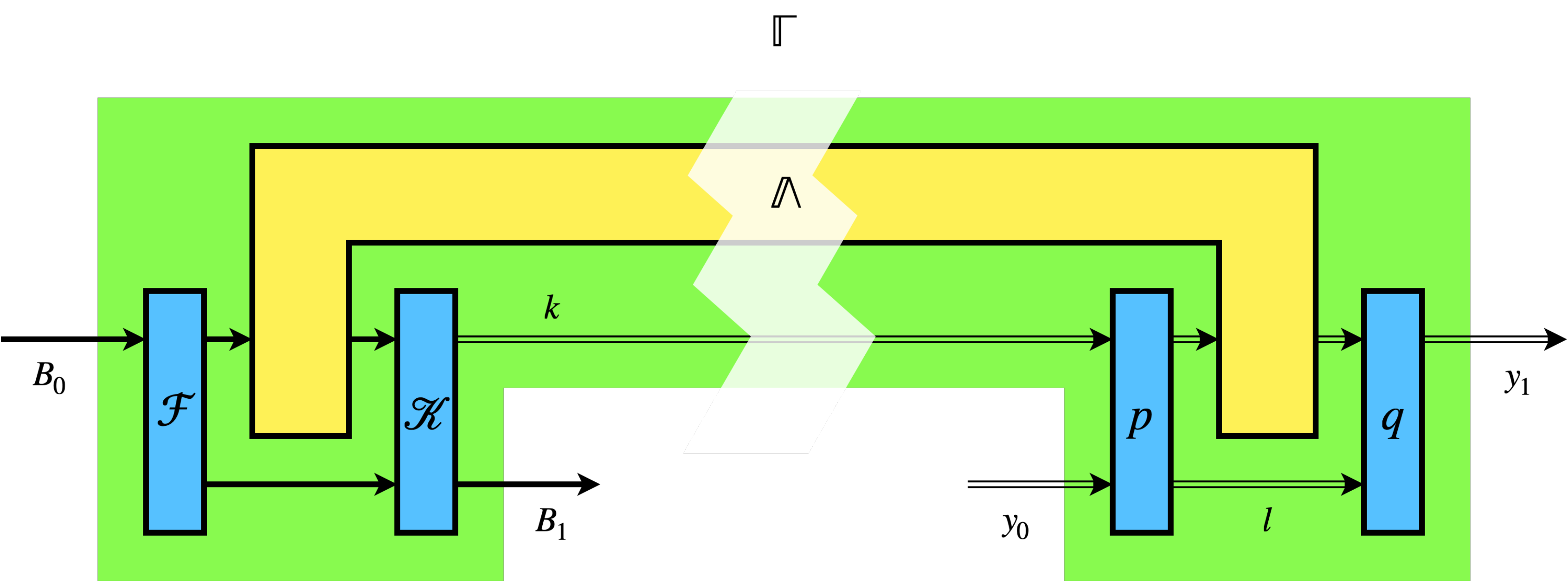}
\caption{The free simulation (blue) of a PID $\pd{\Gamma}$ (green) using another PID $\pd{\Lambda}$ (yellow) according to Eq.~\eqref{eq:simulation-PID}.  The simulation is composed of (i) pre-, post-, and side processing of the quantum part, (ii) pre-, post-, and side processing of the classical part, and (iii) an external classical memory connecting the quantum and classical parts.  Under the late-program assumption, free simulations of PIDs represent the most general transformations given that using any quantum memory with a lifetime exceeding $\Delta t_\abb{D}\approx0$ is forbidden.  This figure reduces to Fig.~\ref{fig:simulation-PMD} when the quantum output systems $A_1$ and $B_1$ are trivial.}
\label{fig:simulation-PID}
\end{figure}

Programmability of PIDs highlights the temporal separation between its quantum systems and classical systems, as displayed in Fig.~\ref{fig:PID}.  So quantum memories across this separation are the only resource that should be forbidden when simulating PIDs.  As a result, the experimenter should have the full ability to (i) have any physical process concatenated in sequence or appended in parallel to the quantum part or the classical part ``locally,'' and (ii) feed any side information generated by the quantum part into the classical part, as long as this information is stored in a classical memory before the classical program arrives, as represented in Fig.~\ref{fig:simulation-PID}.  Note that the quantum delay time $\Delta t_\abb{D}$ between the quantum systems $A_0$ and $A_1$ is assumed negligible and not regarded as a resource compared to the internal quantum memory lifetime $\Delta t_\abb{QM}$.  Hence, the side channel parallel to the quantum part need not be classical.  The formal definition of a free simulation of PIDs is given as follows.

\begin{definition}
\label{def:simulation-PID}
Let $\pd{\Lambda}\equiv\{\Lambda_{x_1|x_0}\}_{x_0,x_1}$ and $\pd{\Gamma}\equiv\{\Gamma_{y_1|y_0}\}_{y_0,y_1}$ be two PIDs.  We say that $\pd{\Lambda}$ can \textbf{freely simulate} $\pd{\Gamma}$, denoted by $\pd{\Lambda}\succcurlyeq_\abb{I}\pd{\Gamma}$, whenever there exists a quantum channel $\m{F}$, a quantum instrument $\ins{K}\equiv\{\m{K}_k\}_k$, and two classical channels $p\equiv\{p_{x_0,l|y_0,k}\}_{x_0,y_0,k,l}$ and $q\equiv\{q_{y_1|x_1,l}\}_{x_1,y_1,l}$ such that
\begin{align}
\label{eq:simulation-PID}
	\Gamma_{y_1|y_0}&=\sum_{x_0,x_1,k,l}q_{y_1|x_1,l}p_{x_0,l|y_0,k}\m{K}_k\circ\left(\Lambda_{x_1|x_0}\otimes\id\right)\circ\m{F} \notag\\
	&\quad\quad\forall y_0,y_1.
\end{align}
We call the transformation $\pd{\Lambda}\mapsto\pd{\Gamma}$ a \textbf{free simulation} of PIDs.
\end{definition}

As can be recognized from Eq.~\eqref{eq:simulation-PID} or Fig.~\ref{fig:simulation-PID}, free simulations of PIDs preserve the classical-to-quantum nonsignaling constraint, and thus they always map PIDs to PIDs.  One can conveniently verify that Definition~\ref{def:simulation-PID} reduces to Definition~\ref{def:simulation-PMD} (free simulations of PMDs) when the quantum output systems of both $\pd{\Lambda}$ and $\pd{\Gamma}$ are trivial (i.e., $1$ dimensional).

The following theorem demonstrates the legitimacy of identifying the free simulations of PIDs as the free transformations for PID nonsimplicity.  This implies that the relation $\succcurlyeq_\abb{I}$ is an incompatibility preorder on the set of PIDs.

\begin{theorem}
\label{thm:simulation-PID}
For fixed index sets where $x_0,x_1,y_0,y_1$ belong, the free simulations of PIDs have the following properties.
\begin{itemize}[leftmargin=1.5em]\setlength\itemsep{-0.2em}
	\item[(1)] Simplicity: a simple PID cannot freely simulate any nonsimple PIDs.
	\item[(2)] Reachability: any PID can freely simulate any simple PID\@.
	\item[(3)] Composability: the sequential or parallel composition of free simulations is a free simulation.
	\item[(4)] Closedness: the limit of a sequence of free simulations (if exists) is a free simulation.
	\item[(5)] Convexity: the probabilistic mixture of free simulations is a free simulation.
\end{itemize}
\end{theorem}

The proof of Theorem~\ref{thm:simulation-PID} is in Appendix~\ref{app:simulation-PID}.

Theorem~\ref{thm:simulation-PID}(1) and (2) and the sequential composability in Theorem~\ref{thm:simulation-PID}(3) guarantee that the free simulations meet the minimal requirements for qualifying as the free transformations for PID nonsimplicity~\cite{Chitambar-2019a}.  The parallel composability in Theorem~\ref{thm:simulation-PID}(3) implies that the resulting resource theory admits a tensor-product structure~\cite{Chitambar-2019a}.  Crucially, Theorem~\ref{thm:simulation-PID}(4) and (5) indicate that the resource theory of PID nonsimplicity has the nice mathematical property of being closed and convex.  Operationally, convexity means that the definition of free simulations of PIDs has implicitly included the use of shared randomness among the constituent physical units of a free simulation.

Before closing this section, we remark that our free simulations of PIDs constitute the \emph{complete} set of physical transformations that do not exploit quantum memories whose lifetime exceeds $\Delta t_\abb{D}\approx0$.  This fact can be demonstrated by invoking the theory of quantum networks~\cite{Chiribella-2009a}, combined with the observation that all input and output systems of $\pd{\Lambda}$ and $\pd{\Gamma}$ must be put in the present causal order in Fig.~\ref{fig:simulation-PID} (see Ref.~\cite[Theorem~8 and Fig.~11]{Chiribella-2009a}).  However, we leave it as an open question whether the set of free simulations considered here is the maximal set of transformations that do not generate PID nonsimplicity, or conversely, whether there exists a completely simplicity-preserving comb~\cite{Chitambar-2019a} that requires a quantum side memory with a non-negligible lifetime to implement.

\section{Relationship with Steering and Measurement Incompatibility}
\label{sec:relationship}

In this section, we expand on the relationship between PID nonsimplicity, steering, and PMD nonsimplicity.  We first clarify that each PID can be implemented through a process of channel steering and vice versa, and so a resource theory of the former implies that of the latter.  Persisting with the steering viewpoint, we then unfold some underlying connections between nonsimplicity of PIDs and of PMDs, or equivalently, between channel steering and measurement incompatibility, and ultimately we demonstrate that PMDs themselves can be cast as a measure of nonsimplicity for PIDs.

\subsection{PIDs as assemblages in channel steering}
\label{sec:steering}

\begin{table*}[t]
\begin{tabular}{c c c c c c c} \botrule
\makecell[bl]{\\[-2ex] Resource theories of \\ nonsimple \\ programmability} && \multicolumn{2}{c}{Programmable devices (nonsignaling assemblages)} && \multicolumn{2}{c}{Simple devices (compatible assemblages)} \\[0.5ex] \hline
\makecell[tl]{\\[-1.5ex] PSD nonsimplicity \\ \quad (EPR steering~\cite{Gallego-2015a})} && \makecell[t]{\\[-1.5ex] \includegraphics[scale=0.1]{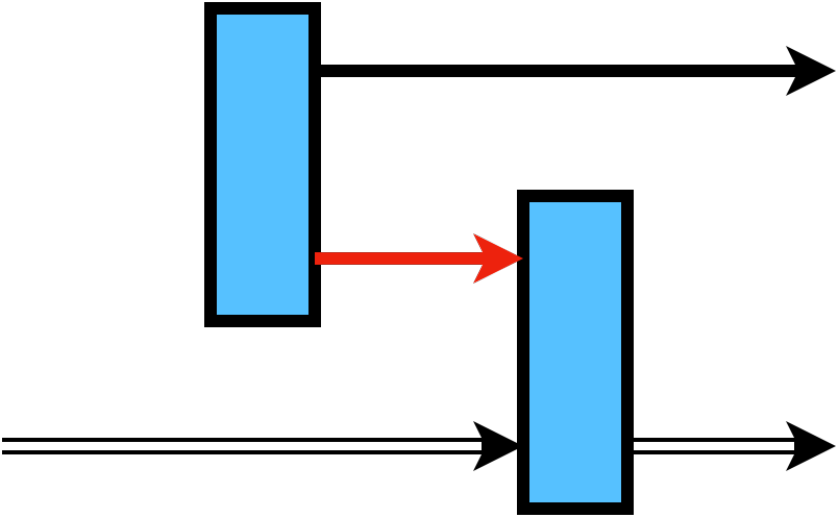}} & \makecell[t]{\\[-1.5ex] $\{\rho_{x_1|x_0}\}_{x_0,x_1}\colon\sum_{x_1}\rho_{x_1|x_0}=\rho\;\;\forall x_0$} && \makecell[t]{\\[-1ex] \includegraphics[scale=0.1]{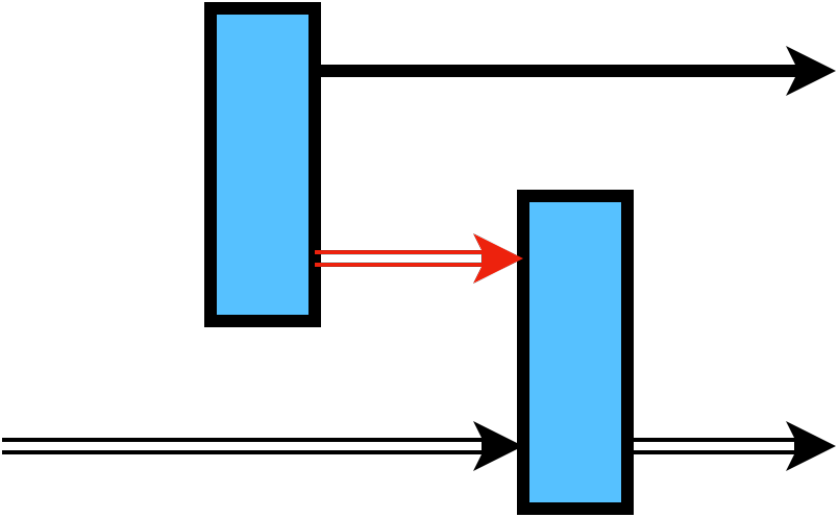}} & \makecell[t]{\\[-1.5ex] $\{\rho_{x_1|x_0}\}_{x_0,x_1}\colon\rho_{x_1|x_0}=\sum_{g}p_{x_1|x_0,g}\eta_g\;\;\forall x_0,x_1$} \\[6ex]
\makecell[tl]{\\[-1.5ex] PMD nonsimplicity \\ \quad \cite{Buscemi-2020a} (measurement \\ \quad incompatibility \\ \quad \cite{Pusey-2015a})} && \makecell[t]{\\[-1.5ex] \includegraphics[scale=0.1]{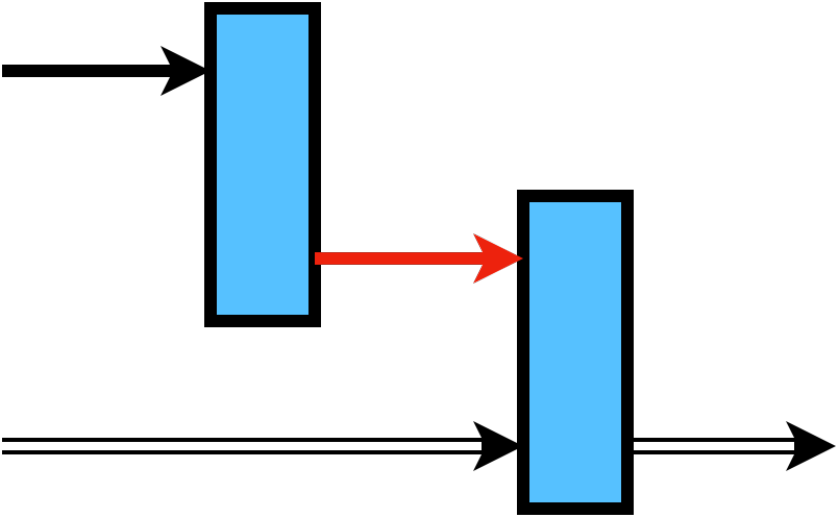}} & \makecell[t]{\\[-1.5ex] $\{M_{x_1|x_0}\}_{x_0,x_1}\colon\sum_{x_1}M_{x_1|x_0}=\1\;\;\forall x_0$} && \makecell[t]{\\[-1.5ex] \includegraphics[scale=0.1]{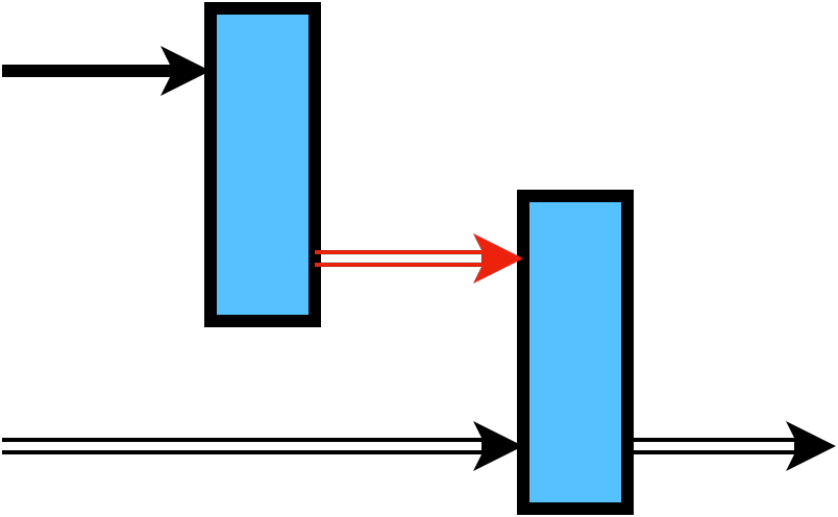}} & \makecell[t]{\\[-1.5ex] $\{M_{x_1|x_0}\}_{x_0,x_1}\colon M_{x_1|x_0}=\sum_{g}p_{x_1|x_0,g}G_g\;\;\forall x_0,x_1$} \\[9ex]
\makecell[tl]{\\[-1.5ex] PID nonsimplicity \\ \quad (channel steering) \\ \quad [this paper]} && \makecell[t]{\\[-1.5ex] \includegraphics[scale=0.1]{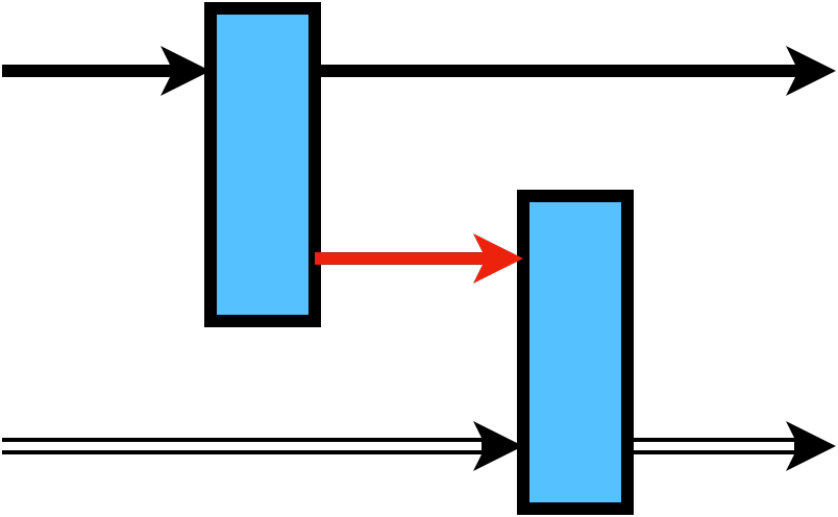}} & \makecell[t]{\\[-1.5ex] $\{\Lambda_{x_1|x_0}\}_{x_0,x_1}\colon\sum_{x_1}\Lambda_{x_1|x_0}=\Lambda\;\;\forall x_0$} && \makecell[t]{\\[-1.5ex] \includegraphics[scale=0.1]{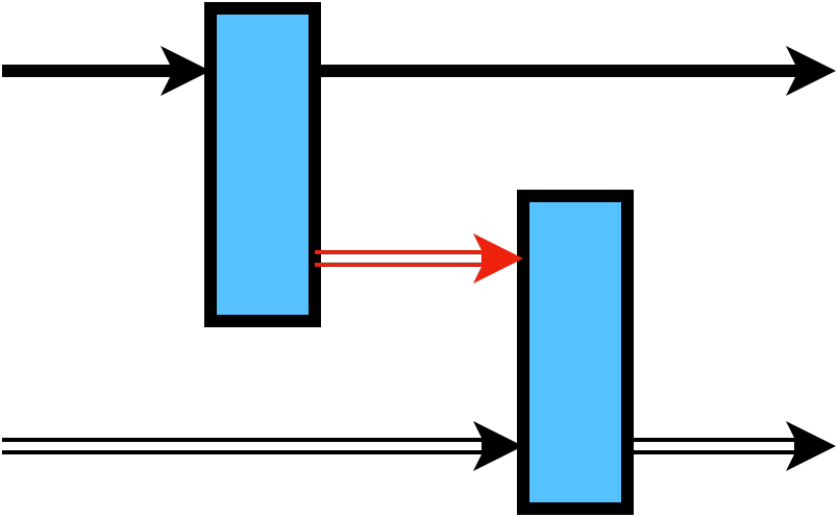}} & \makecell[t]{\\[-1.5ex] $\{\Lambda_{x_1|x_0}\}_{x_0,x_1}\colon\Lambda_{x_1|x_0}=\sum_{g}p_{x_1|x_0,g}\m{G}_g\;\;\forall x_0,x_1$} \\[7ex] \botrule
\end{tabular}
\caption{A comparison between programmable devices (i.e., general objects) and simple devices (i.e., free objects) in the resource theories of nonsimplicity of programmable source, measurement, and instrument devices (PSDs, PMDs, and PIDs, respectively).  Our resource theory of PID nonsimplicity, equivalently a resource theory of channel steering, is a generalized theory unifying both the resource theory of PSD nonsimplicity (i.e., EPR steering~\cite{Gallego-2015a}) and that of PMD nonsimplicity~\cite{Buscemi-2020a} (i.e., measurement incompatibility~\cite{Pusey-2015a}).  Specifically, PID nonsimplicity reduces to PSD nonsimplicity when the quantum input is trivial, and it reduces to PMD nonsimplicity when the quantum output is trivial.  Programmable devices in these theories are all subject to the nonsignaling constraint, and simple devices all implement compatible assemblages.  The compatibility in these free objects can universally be viewed as a consequence of the classicality of the internal memory (hollow red arrows), which should in general be a quantum memory (solid red arrows).}
\label{tab:relationship}
\end{table*}

As a dynamical generalization of the celebrated phenomenon of EPR steering~\cite{Schrodinger-1935a, Wiseman-2007a, Jones-2007a}, channel steering~\cite{Piani-2015b} provides a natural and effective way of understanding the internal structure of PIDs.  Referencing the bipartite picture in Fig.~\ref{fig:spatial}, the scenario of channel steering is described as follows.  Consider a broadcast channel $\m{E}^{A_0\to A_1E}$ with the systems $A_0$ and $A_1$ held by Alice and the system $E$ leaked to Evan.  Without any proactive interference, Evan can remotely ``steer'' the subchannel decomposition of Alice's marginal channel $\Lambda^{A_0\to A_1}\coloneq\tr_E\circ\m{E}^{A_0\to A_1E}$ by directing his system $E$ to a PMD $\pd{M}\equiv\{M_{x_1|x_0}\}_{x_0,x_1}$.  This steering process would lead to a family of instruments $\pd{\Lambda}\equiv\{\Lambda_{x_1|x_0}\}_{x_0,x_1}$ on Alice's side, typically known as a \emph{channel assemblage}~\cite{Piani-2015b}, defined by
\begin{align}
\label{eq:channel-steering}
	\Lambda_{x_1|x_0}^{A_0\to A_1}\left[\cdot\right]&\coloneq\tr_E\left[\left(\1^{A_1}\otimes M_{x_1|x_0}^E\right)\m{E}^{A_0\to A_1E}\left[\cdot\right]\right]\quad\forall x_0,x_1.
\end{align}
The assemblage $\pd{\Lambda}$ is said to be \emph{unsteerable} whenever it can be realized following Eq.~\eqref{eq:channel-steering} with $\m{E}^{A_0\to A_1E}$ being an \emph{incoherent extension} of its marginal channel $\Lambda^{A_0\to A_1}$~\cite{Piani-2015b}, namely, whenever there exists a quantum instrument $\ins{G}\equiv\{\m{G}_g\}_g$ and states $\epsilon_g$ for all $g$ such that
\begin{align}
\label{eq:incoherent-extension}
	\m{E}^{A_0\to A_1E}&=\sum_{g}\m{G}_g^{A_0\to A_1}\otimes\epsilon_g^E.
\end{align}
It follows as an observation that if the PMD $\pd{M}$ is simple, then the channel assemblage $\pd{\Lambda}$ induced by Eq.~\eqref{eq:channel-steering} must be unsteerable regardless of $\m{E}^{A_0\to A_1E}$~\cite{Piani-2015b}.  We introduce the following shorthand to denote a channel assemblage generated via steering.

\begin{definition}
\label{def:steering}
Let $\pd{\Lambda}\equiv\{\Lambda_{x_1|x_0}\}_{x_0,x_1}$ be a channel assemblage, $\m{E}^{A_0\to A_1E}$ a broadcast channel, and $\pd{M}\equiv\{M_{x_1|x_0}\}_{x_0,x_1}$ a PMD\@. We say that $(\m{E},\pd{M})$ is a \textbf{steering decomposition} of $\pd{\Lambda}$, denoted by $\pd{\Lambda}\leftarrowtail(\m{E},\pd{M})$, whenever Eq.~\eqref{eq:channel-steering} is satisfied, i.e., whenever $\pd{\Lambda}$ can be induced by a process of channel steering featuring $\m{E}$ and $\pd{M}$.
\end{definition}

Clearly, each channel assemblage $\pd{\Lambda}\leftarrowtail(\m{E},\pd{M})$ can be regarded as a PID, since by Eq.~\eqref{eq:channel-steering} the nonsignaling constraint in Eq.~\eqref{eq:PID} is satisfied.  Conversely, as we argued in Sec.~\ref{sec:PID}, due to the equivalence between semicausality and semilocalizability~\cite{Eggeling-2002a}, each PID $\pd{\Lambda}$ admits a steering decomposition $\pd{\Lambda}\leftarrowtail(\m{E},\pd{M})$, and therefore we can envision a process of channel steering going on within each PID, as shown in Fig.~\ref{fig:PID-spatial}.  Then we can observe from Fig.~\ref{fig:PID-simple} that simple PIDs are precisely those that implement unsteerable channel assemblages.  Likewise, the free simulations of PIDs in Definition~\ref{def:simulation-PID} correspond to transformations of channel assemblages that are realizable via Alice-to-Evan one-way local operations and classical communication (one-way LOCC).  In this way, the resource theory of PID nonsimplicity that we developed in Secs.~\ref{sec:PID} and \ref{sec:simulation-PID} can be equivalently interpreted as a resource theory of channel steering.

The scenario of channel steering reduces to that of EPR steering when the quantum input system $A_0$ is trivial~\cite{Wiseman-2007a, Jones-2007a}.  In this case, when Evan feeds his part of a bipartite state $\xi^{A_1E}$ into a PMD $\pd{M}$, a state assemblage $\{\rho_{x_1|x_0}\}_{x_0,x_1}$ is generated on Alice's side:
\begin{align}
\label{eq:EPR-steering}
	\rho_{x_1|x_0}^{A_1}&\coloneq\tr_E\left[\left(\1^{A_1}\otimes M_{x_1|x_0}^E\right)\xi^{A_1E}\right]\quad\forall x_0,x_1.
\end{align}
Accordingly, the assemblage $\{\rho_{x_1|x_0}\}_{x_0,x_1}$ is \emph{unsteerable} when it can be induced by a bipartite state $\xi^{A_1E}$ being separable:
\begin{align}
	\xi^{A_1E}=\sum_{g}\eta_g^{A_1}\otimes\epsilon_g^E,
\end{align}
where $\{\eta_g\}_g$ is a state ensemble and $\epsilon_g$ is a state for all $g$.  In parallel to Eq.~\eqref{eq:PID-simple}, an unsteerable state assemblage $\{\rho_{x_1|x_0}\}_{x_0,x_1}$ demonstrates compatibility by admitting a so-called ``local-hidden-state'' model~\cite{Wiseman-2007a}:
\begin{align}
	\rho_{x_1|x_0}^{A_1}&=\sum_{g}p_{x_1|x_0,g}\eta_g^{A_1}\quad\forall x_0,x_1,
\end{align}
where $\{p_{x_1|x_0,g}\}_{x_0,x_1,g}$ is a classical channel.  As in channel steering, if the PMD $\pd{M}$ is simple, then the induced state assemblage must be unsteerable.  Conversely, it has also been proved that all nonsimple PMDs are useful to generate \emph{steerable} state assemblages when $\xi^{A_1E}$ is a pure state of maximum Schmidt rank~\cite{Quintino-2014a, Uola-2014a}.

As a special case of channel steering, EPR steering can also be understood as a theory of nonsimplicity of \emph{programmable source devices (PSDs)} (alias nonsignaling \emph{multisources}~\cite{Gour-2018b}) in the programmability framework, and simple PSDs are those implementing unsteerable state assemblages.  As a result, our resource theory of PID nonsimplicity subsumes the existing resource theory of EPR steering with one-way LOCC as free operations~\cite{Gallego-2015a}.  A comparison between resource theories of nonsimplicity of different types of programmable devices is provided in Table~\ref{tab:relationship}.

\subsection{Steering-equivalence mapping}
\label{sec:SEM}

The concept of ``steering-equivalent observables'' (SEO) plays an essential role in studying the relationship between EPR steering and measurement incompatibility~\cite{Uola-2015a}.  Now we generalize this concept from state assemblages to PIDs (i.e., to channel assemblages), so as to connect PID nonsimplicity (i.e., channel steering) with PMD nonsimplicity (i.e., measurement incompatibility).  Given a PID $\pd{\Lambda}\equiv\{\Lambda_{x_1|x_0}^{A_0\to A_1}\}_{x_0,x_1}$, the nonsignaling constraint guarantees the existence of the following channel:
\begin{align}
	\Lambda^{A_0\to A_1}&\coloneq\sum_{x_1}\Lambda_{x_1|x_0}^{A_0\to A_1}\quad\forall x_0.
\end{align}
The Choi operator of $\Lambda_{x_1|x_0}^{A_0\to A_1}$ is given by
\begin{align}
	J_{\Lambda_{x_1|x_0}}^{A_0A_1}&\coloneq\left(\id^{A_0}\otimes\Lambda_{x_1|x_0}^{\rpl{A}_0\to A_1}\right)\left[\phi_+^{A_0\rpl{A}_0}\right]\quad\forall x_0,x_1,
\end{align}
where $\rpl{A}_0$ is a system isomorphic to $A_0$ and $\phi_+^{A_0\rpl{A}_0}\equiv\sum_{i,j}\op{i}{j}^{A_0}\otimes\op{i}{j}^{\rpl{A}_0}$.  The Choi operator of $\Lambda^{A_0\to A_1}$ is given by
\begin{align}
\label{eq:Choi}
	J_\Lambda^{A_0A_1}&\coloneq\sum_{x_1}J_{\Lambda_{x_1|x_0}}^{A_0A_1}\quad\forall x_0.
\end{align}
Let $A^*$ be a quantum system such that its associated Hilbert space, denoted by $\spa{H}^{A^*}$, is isomorphic to the support of $J_\Lambda^{A_0A_1}$.  In other words, it satisfies $\spa{H}^{A^*}\cong\supp(J_\Lambda^{A_0A_1})\subseteq\spa{H}^{A_0A_1}$, where $\spa{H}^{A_0A_1}\equiv\spa{H}^{A_0}\otimes\spa{H}^{A_1}$ is the composite Hilbert space associated with the systems $A_0$ and $A_1$.  By Choi's theorem, $J_{\Lambda_{x_1|x_0}}^{A_0A_1}$ is positive semidefinite since $\Lambda_{x_1|x_0}^{A_0\to A_1}$ is CP for all $x_0,x_1$.  By Eq.~\eqref{eq:Choi}, this implies $\supp(J_{\Lambda_{x_1|x_0}}^{A_0A_1})\subseteq\supp(J_\Lambda^{A_0A_1})\cong\spa{H}^{A^*}$, and therefore the image of $J_{\Lambda_{x_1|x_0}}^{A_0A_1}$ in the system $A^*$, denoted by $J_{\Lambda_{x_1|x_0}}^{A^*}$, is well defined for all $x_0,x_1$.  The following definition generalizes the steering-equivalent observes (SEO) defined on state assemblages to PIDs.

\begin{definition}
\label{def:SEM}
The \textbf{steering-equivalence mapping}, denoted by $\f{SEM}$, is a mapping from the set of PIDs to the set of PMDs, and it sends a PID $\pd{\Lambda}\equiv\{\Lambda_{x_1|x_0}^{A_0\to A_1}\}_{x_0,x_1}$ to a PMD $\f{SEM}(\pd{\Lambda})\equiv\{S_{x_1|x_0}^{A^*}\}_{x_0,x_1}$ where
\begin{align}
\label{eq:SEM}
	S_{x_1|x_0}^{A^*}&\coloneq(J_\Lambda^{A^*})^{-\frac{1}{2}}J_{\Lambda_{x_1|x_0}}^{A^*}(J_\Lambda^{A^*})^{-\frac{1}{2}}\quad\forall x_0,x_1.
\end{align}
\end{definition}

It can be conveniently verified that $\f{SEM}(\pd{\Lambda})$ is a valid PMD given any PID $\pd{\Lambda}$.  It is also apparent from Definition~\ref{def:SEM} that, by the Choi--Jamiołkowski isomorphism, a PID $\pd{\Lambda}$ is nonsimple if and only if the PMD $\f{SEM}(\pd{\Lambda})$ is nonsimple.  That is to say, the membership problem of steerable channel assemblages can be reduced to the membership problem of incompatible families of POVMs through $\f{SEM}$.  This generalizes Theorem~1 of Ref.~\cite{Uola-2015a}, which addresses the membership problem of steerable state assemblages via the SEO\@.  The SEO of a state assemblage has been shown to possess an operational interpretation as the transposed PMD that induces the state assemblage from a minimal state extension~\cite{Uola-2015a}.  We show in the following proposition that this type of operational interpretation remains effective for $\f{SEM}$ in the generalized scenario of channel steering.

\begin{proposition}
\label{prop:operational}
Let $\pd{\Lambda}\equiv\{\Lambda_{x_1|x_0}^{A_0\to A_1}\}_{x_0,x_1}$ be a PID, and let $\pd{S}\equiv\{S_{x_1|x_0}^{A^*}\}_{x_0,x_1}$ be a PMD such that $\pd{S}=\f{SEM}(\pd{\Lambda})$.  Then there exists an isometric channel $\m{V}^{A_0\to A_1A^*}$ such that
\begin{align}
\label{eq:operational}
	\Lambda_{x_1|x_0}^{A_0\to A_1}\left[\cdot\right]&=\tr_{A^*}\left[\left(\1^{A_1}\otimes(S_{x_1|x_0}^\top)^{A^*}\right)\m{V}^{A_0\to A_1A^*}\left[\cdot\right]\right] \notag\\
	&\quad\quad\forall x_0,x_1,
\end{align}
where $(\cdot)^\top$ denotes transposition under a fixed orthonormal basis.
\end{proposition}

The proof of Proposition~\ref{prop:operational} is in Appendix~\ref{app:operational}.

Proposition~\ref{prop:operational} indicates that given any PID $\pd{\Lambda}$, there exists a broadcast channel $\m{V}^{A_0\to A_1A^*}$ such that the steering decomposition $\pd{\Lambda}\leftarrowtail(\m{V},\f{SEM}(\pd{\Lambda})^\top)$ holds, where $\m{V}^{A_0\to A_1A^*}$ is an isometric dilation of $\Lambda^{A_0\to A_1}$ and $\f{SEM}(\pd{\Lambda})^\top$ is the elementwise transpose of the PMD $\f{SEM}(\pd{\Lambda})$.  In particular, this provides a formal and independent demonstration for the internal structure of a general PID as depicted in Fig.~\ref{fig:PID-spatial}, which we previously argued by invoking the equivalence between semicausality and semilocalizability.

On the other hand, the existence of a steering decomposition for any PID does not imply that such a decomposition is unique, and in fact it is not unique.  Despite this, there is a close relation between any steering decomposition of a given PID and the ``canonical'' steering decomposition specified in Proposition~\ref{prop:operational}, as given by the following proposition.

\begin{proposition}
\label{prop:canonicity}
Let $\pd{\Lambda}$ be a PID, $\m{E}^{A_0\to A_1E}$ a broadcast channel, and $\pd{M}\equiv\{M_{x_1|x_0}\}_{x_0,x_1}$ a PMD such that $\pd{\Lambda}\leftarrowtail(\m{E},\pd{M})$.  Then we have $\pd{M}^\top\succcurlyeq_\abb{M}\f{SEM}(\pd{\Lambda})$, where $\pd{M}^\top\equiv\{M_{x_1|x_0}^\top\}_{x_0,x_1}$.
\end{proposition}

The proof of Proposition~\ref{prop:canonicity} is in Appendix~\ref{app:canonicity}.

Proposition~\ref{prop:canonicity} indicates that $\f{SEM}(\pd{\Lambda})^\top$ is the least resourceful PMD that can be used to compose a given PID $\pd{\Lambda}$, because every other PMD $\pd{M}$ that suffices to do so must be convertible to $\f{SEM}(\pd{\Lambda})^\top$ via free simulations.  This is the reason why we refer to the decomposition $\pd{\Lambda}\leftarrowtail(\m{V},\f{SEM}(\pd{\Lambda})^\top)$ in Proposition~\ref{prop:operational} as the ``canonical'' or ``minimal'' decomposition for $\pd{\Lambda}$.  To understand this physically, we can think of any PID decomposition $\pd{\Lambda}\leftarrowtail(\m{E},\pd{M})$ as having a configuration as depicted in Fig.~\ref{fig:PID-temporal}, and the nonsimplicity of $\pd{\Lambda}$ is essentially attributed to the PMD $\pd{M}$, since that is where the quantum memory with a lifetime $\Delta t_\abb{QM}\gg\Delta t_\abb{D}\approx0$ resides.  Then Proposition~\ref{prop:canonicity} implies that when the decomposition $\pd{\Lambda}\leftarrowtail(\m{E},\pd{M})$ is canonical, namely, when $\m{E}^{A_0\to A_1E}=\m{V}^{A_0\to A_1A^*}$ is the minimal isometric dilation (i.e., the maximally coherent channel extension~\cite{Piani-2015b}), the physical resource within $\pd{M}$ is best utilized.

Built on the aforementioned results, we conclude this section with the following theorem.  We show that the mapping $\f{SEM}$ behaves as a nonsimplicity monotone of PIDs (equivalently, a steering monotone of channel assemblages), in the sense that it preserves the incompatibility preorder specified by the free simulations of programmable devices.  This means that the nonsimplicity of $\f{SEM}(\pd{\Lambda})$ is not only an indicator, but also a \emph{measure} of the nonsimplicity of $\pd{\Lambda}$.

\begin{theorem}
\label{thm:monotonicity}
The mapping $\f{SEM}$ is a faithful nonsimplicity monotone.  Formally, it has the following properties.
\begin{itemize}[leftmargin=1.5em]\setlength\itemsep{-0.2em}
\item[(1)] Faithfulness: $\pd{\Lambda}$ is a simple PID if and only if $\f{SEM}(\pd{\Lambda})$ is a simple PMD\@.
\item[(2)] Monotonicity: if $\pd{\Lambda}\succcurlyeq_\abb{I}\pd{\Gamma}$, then $\f{SEM}(\pd{\Lambda})\succcurlyeq_\abb{M}\f{SEM}(\pd{\Gamma})$.
\end{itemize}
\end{theorem}

The proof of Theorem~\ref{thm:monotonicity} is detailed in Appendices~\ref{app:faithfulness} and \ref{app:monotonicity}.  It is a proof by construction based on Propositions~\ref{prop:operational} and \ref{prop:canonicity}.

Conventionally, by ``resource monotones'' we allude to real-valued functions that are nonincreasing under free transformations~\cite{Chitambar-2019a}.  In Theorem~\ref{thm:monotonicity}, however, the term has a broader meaning of order-preserving mappings under free transformations, and such mappings can in general be between objects, as $\f{SEM}$ is.  A direct application of such generalized monotones is to convert resource monotones in one resource theory to resource monotones in another resource theory.  For instance, given any resource monotone $f$ for PMD nonsimplicity (i.e., measurement incompatibility), we immediately obtain an induced resource monotone $f\circ\f{SEM}$ for PID nonsimplicity (i.e., channel steering), regardless of what $f$ is and whether $f$ is real valued or not.  

On the other hand, we remark that object-valued monotones like $\f{SEM}$ are also interesting in their own rights, since they reveal fundamental connections between two different resource theories and may trigger insights into the physical nature of the resources involved.  As for $\f{SEM}$, with its operational interpretation established in Proposition~\ref{prop:operational}, Theorem~\ref{thm:monotonicity} indicates that a more resourceful PID must have a more resourceful internal PMD under the ``canonical'' steering decomposition.  This certainly supports our previous viewpoint of attributing the physical resource (i.e., a quantum memory with a non-negligible lifetime) in a PID to its internal PMD (as in Fig.~\ref{fig:PID-temporal}).

\section{Semi-Device-Independent Characterization with Nontransient Guessing Games}
\label{sec:game}

In this section, we demonstrate the operational significance of PID nonsimplicity in the scenario of guessing games.  We propose a class of guessing games with double temporal stages, and we call them \emph{nontransient guessing games}.  Just as nonlocal games feature spatial separation between different parties~\cite{Clauser-1969a, Buscemi-2012b}, nontransient games feature temporal separation between different stages, and therefore they are a suitable setting for characterizing correlations that exist across time, such as memory effect~\cite{Rosset-2018a} and programmability~\cite{Buscemi-2020a}.  We show that the winning probabilities of PIDs in the nontransient guessing games compose a complete set of incompatibility monotones, fully characterizing the incompatibility preorder between PIDs.  This also implies that every nonsimple PID provides a nontrivial advantage over simple PIDs in some guessing game.  Furthermore, we prove that this advantage is bounded from above by the robustness of incompatibility of the PID and that this bound is tight.  Finally, we comment on the limitations of nontransient guessing games in terms of experimental difficulties, and we propose a variant class of guessing games that overcomes such difficulties while also giving rise to a complete set of incompatibility monotones.

\subsection{A complete set of incompatibility monotones}
\label{sec:convertibility}

A nontransient guessing game is a parametrized interactive protocol between a player and a referee.  Although such a game will be utilized to test the player's PID, the game itself is actually ``semi-device-independent.''  That is, it requires that the referee's operations be faithfully executed, but does not make any assumptions about the player's device or strategy.

\begin{definition}
\label{def:game-1}
A \textbf{nontransient guessing game} between Alice (the player) and Bob (the referee) is specified by a bipartite POVM $\povm{M}\equiv\{M_{m,n}\}_{m,n}$, and it has two stages separated by a time interval $\Delta t\gg\Delta t_\abb{D}\approx0$.
\begin{enumerate}[leftmargin=1.5em]\setlength\itemsep{-0.2em}
	\item In the first stage, Bob sends Alice one half of a maximally entangled state $\varphi_+\equiv1/d\sum_{i,j}\op{i}{j}\otimes\op{i}{j}$, and then he asks Alice to submit a quantum system back to him.  Bob measures Alice's submitted system and the other half of $\varphi_+$ jointly according to the POVM $\povm{M}$ and obtains a tuple $(m,n)$ as the outcome.
	\item In the second stage (after $\Delta t$ has passed), Bob announces the index $m$ to Alice, and then he asks Alice to make a guess $n'$ at the other index $n$.  Alice wins the game whenever she guesses correctly, i.e., whenever $n'=n$.
\end{enumerate}
Throughout the game, Alice has no access to quantum memories with a lifetime larger than $\Delta t_\abb{D}\approx0$.
\end{definition}

We note that in Definition~\ref{def:game-1}, Alice's device and strategy are both uncharacterized.  To serve the purpose of testing PIDs, we now assume that Alice holds a PID $\pd{\Lambda}\equiv\{\Lambda_{x_1|x_0}\}_{x_0,x_1}$ in hand, which may count as an additional resource for her in the game.  It is also convenient for us to assume that the quantum delay time of $\pd{\Lambda}$ is no greater than the $\Delta t_\abb{D}$ specified in Definition~\ref{def:game-1}.

\begin{figure}[t]
\includegraphics[scale=0.15]{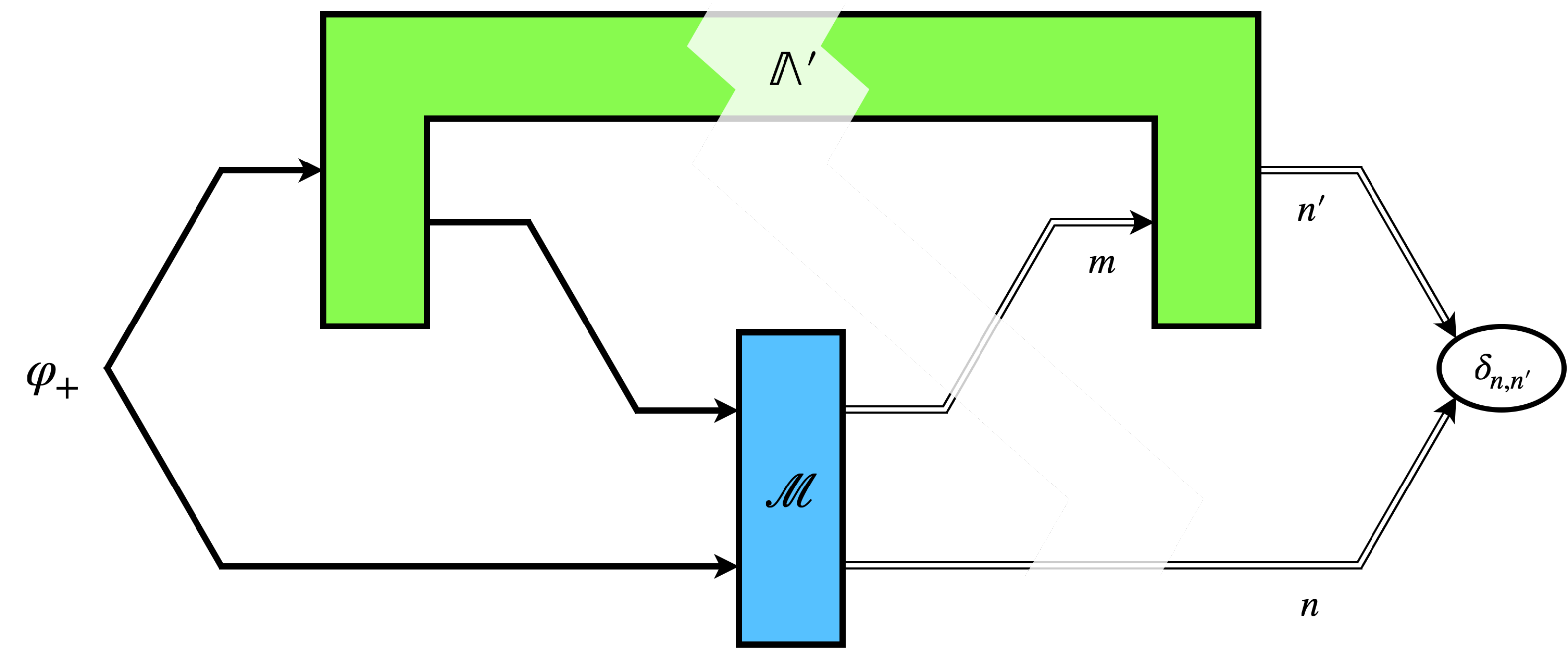}
\caption{A nontransient guessing game between Alice (the player) and Bob (the referee).  The setting of the game is specified by Bob's POVM $\povm{M}\equiv\{M_{m,n}\}_{m,n}$.  Alice's strategy to the game is represented by a PID $\pd{\Lambda}'$ (green), which is freely simulated by an actual PID $\pd{\Lambda}$ held in her hand.  She wins the game whenever she makes a correct guess at one of Bob's outcome indices $n$ given the other index $m$, i.e., whenever $n'=n$.}
\label{fig:game-1}
\end{figure}

We note that from Alice's perspective, the setting she is dealing with perfectly satisfies the late-program assumption; namely, the classical signal $m$ does not arrive until a significant time interval $\Delta t\gg\Delta t_\abb{D}\approx0$ after her quantum output is released.  Therefore, the most general strategy for Alice to follow is to use her PID $\pd{\Lambda}$ to simulate whatever PID $\pd{\Lambda}'$ she can and to insert $\pd{\Lambda}'$ into the open slots of the game, as illustrated in Fig.~\ref{fig:game-1}.  In addition, given that the quantum delay time of $\pd{\Lambda}$ is within $\Delta t_\abb{D}$ and that no quantum memory with a lifetime exceeding $\Delta t_\abb{D}$ is accessible (see Definition~\ref{def:game-1}), we can rest assured that Alice's simulation of $\pd{\Lambda}'$ using $\pd{\Lambda}$ is a free simulation, i.e., $\pd{\Lambda}\succcurlyeq_\abb{I}\pd{\Lambda}'$ (see Definition~\ref{def:simulation-PID}).  As a result, Alice's maximum winning probability in the nontransient guessing game specified by the POVM $\povm{M}$ is given by
\begin{align}
\label{eq:game-1}
	P_\abb{guess}(\pd{\Lambda};\povm{M})&\coloneq\max_{\pd{\Lambda}'\colon\pd{\Lambda}\succcurlyeq_\abb{I}\pd{\Lambda}'}\sum_{m,n}\tr\left[M_{m,n}\left(\id\otimes\Lambda_{n|m}'\right)\left[\varphi_+\right]\right].
\end{align}

By the transitivity of the incompatibility preorder $\succcurlyeq_\abb{I}$ [i.e., the sequential composability of free simulations in Theorem~\ref{thm:simulation-PID}(3)], we can observe that Alice's winning probability $P_\abb{guess}(\pd{\Lambda};\povm{M})$ is a nonsimplicity monotone with respect to $\pd{\Lambda}$ given any POVM $\povm{M}$.  The following theorem states that when considering all different POVMs, the winning probabilities $\{P_\abb{guess}(\pd{\Lambda};\povm{M})\}_{\povm{M}}$ compose a \emph{complete} set of nonsimplicity monotones, which faithfully reflects convertibility between PIDs under free simulations.

\begin{theorem}
\label{thm:convertibility}
Let $\pd{\Lambda}\equiv\{\Lambda_{x_1|x_0}\}_{x_0,x_1}$ and $\pd{\Gamma}\equiv\{\Gamma_{y_1|y_0}\}_{y_0,y_1}$ be two PIDs.  Then $\pd{\Lambda}\succcurlyeq_\abb{I}\pd{\Gamma}$ if and only if $P_\abb{guess}(\pd{\Lambda};\povm{M})\geq P_\abb{guess}(\pd{\Gamma};\povm{M})$ for every bipartite POVM $\povm{M}\equiv\{M_{m,n}\}_{m,n}$.
\end{theorem}

The ``only if'' part of Theorem~\ref{thm:convertibility} is evident from the monotonicity of winning probabilities as argued before.  The proof of the ``if'' part is detailed in Appendix~\ref{app:convertibility}.  It is a proof by construction, utilizing the closedness and convexity of the resource theory [Theorem~\ref{thm:simulation-PID}(4) and (5)], the hyperplane separation theorem~\cite{Boyd-2004a}, and a technique employed in Refs.~\cite{Piani-2015a, Lipka-Bartosik-2020a}.

A prominent implication of Theorem~\ref{thm:convertibility} is that each nontransient guessing game can be used as a certification of nonsimplicity of a PID, and the combination of all such certifications is sufficient to compose a faithful criterion for deciding nonsimplicity.  Specifically, since all simple PIDs are interconvertible via free simulations [Theorem~\ref{thm:simulation-PID}(2)], Theorem~\ref{thm:convertibility} implies that all simple PIDs give rise to the same winning probability in any fixed nontransient guessing game specified by $\povm{M}$, which equals
\begin{align}
\label{eq:game-1-simple}
	P_\abb{guess}^\abb{simple}(\povm{M})&\coloneq\max_{\pd{\Omega}\colon\abb{simple}}P_\abb{guess}(\pd{\Omega};\povm{M}) \\
	&=\max_{\pd{\Omega}\colon\abb{simple}}\sum_{m,n}\tr\left[M_{m,n}\left(\id\otimes\Omega_{n|m}\right)\left[\varphi_+\right]\right].
\end{align}
Therefore, as long as Alice's winning probability in this game is observed to be greater than $P_\abb{guess}^\abb{simple}(\povm{M})$, we can conclude with certainty that Alice holds some device that functions as a nonsimple PID\@.  Conversely, if Alice holds a nonsimple PID $\pd{\Lambda}$ and always follows an optimal strategy while playing the games, then there must exist a specific game that certifies the nonsimplicity of her device, i.e., a POVM $\opt{\povm{M}}$ such that $P_\abb{guess}(\pd{\Lambda};\opt{\povm{M}})>P_\abb{guess}^\abb{simple}(\opt{\povm{M}})$.

\subsection{Robustness of incompatibility as the supremum of game advantage}
\label{sec:robustness}

The \emph{robustness of resource}~\cite{Harrow-2003a, Brandao-2015a} is a well-studied resource measure that reflects how tolerant a resource is to an admixture of generic noise.  It is universally well defined in any resource theory, and it possesses many desirable properties as a resource measure, including faithfulness, monotonicity, and resource-dependent convexity (i.e., being a convex function when the set of free objects is convex)~\cite{Chitambar-2019a}.  For the resource theory of PID nonsimplicity, the robustness measure can be defined as follows.

\begin{definition}
\label{def:robustness}
The \textbf{robustness of incompatibility} of a PID $\pd{\Lambda}\equiv\{\Lambda_{x_1|x_0}\}_{x_0,x_1}$, denoted by $\f{RoI}(\pd{\Lambda})$, is defined as
\begin{subequations}
\label{eq:robustness}
\begin{align}
	\f{RoI}(\pd{\Lambda})&\coloneq\min_{r\geq0}r \\
	\textnormal{subject to:}&\quad\left\{\frac{\Lambda_{x_1|x_0}+r\Upsilon_{x_1|x_0}}{1+r}\right\}_{x_0,x_1}\textnormal{ is a simple PID}, \\
	&\quad\left\{\Upsilon_{x_1|x_0}\right\}_{x_0,x_1}\textnormal{ is a PID}.
\end{align}
\end{subequations}
\end{definition}

As we discussed before, Theorem~\ref{thm:convertibility} implies that every nonsimple PID can provide a nontrivial operational advantage over simple PIDs in some nontransient guessing game.  The following theorem shows that this advantage can be quantitatively characterized by the robustness of incompatibility of the PID\@.

\begin{theorem}
\label{thm:robustness}
Let $\pd{\Lambda}\equiv\{\Lambda_{x_1|x_0}\}_{x_0,x_1}$ be a PID\@.  Then
\begin{align}
	\sup_{\povm{M}\equiv\{M_{m,n}\}_{m,n}}\frac{P_\abb{guess}(\pd{\Lambda};\povm{M})}{P_\abb{guess}^\abb{simple}(\povm{M})}&=1+\f{RoI}(\pd{\Lambda}),
\end{align}
where the supremum is over all bipartite POVMs.
\end{theorem}

The proof of Theorem~\ref{thm:robustness} is detailed in Appendix~\ref{app:robustness}.  It consists of two parts, following a similar structure to the proofs of comparable results in Refs.~\cite{Piani-2015a, Uola-2019a, Takagi-2019b, Buscemi-2020a}.  The first part proves that the advantage provided by any PID $\pd{\Lambda}$ in any nontransient guessing game, in terms of the ratio $P_\abb{guess}(\pd{\Lambda};\povm{M})/P_\abb{guess}^\abb{simple}(\povm{M})$, can never exceed $1+\f{RoI}(\pd{\Lambda})$.  This is done via a slight reformulation of Eq.~\eqref{eq:robustness}.  The second part explicitly constructs a sequence of games (specified by a sequence of POVMs) that approaches the aforementioned robustness upper bound on the advantage arbitrarily close, thus showing that this upper bound is tight.  This is done by utilizing the dual conic program of Eq.~\eqref{eq:robustness}.  It is worth mentioning that while the construction of the sequence of POVMs may require an unbounded number of measurement outcomes, the dimensionality of the quantum systems to be measured is bounded.  We also note that the convexity of the resource theory of PID nonsimplicity [Theorem~\ref{thm:simulation-PID}(5)] plays a crucial role here, as it guarantees strong duality between the conic programs for the robustness of incompatibility.

\subsection{Constructing experimentally friendly incompatibility tests}
\label{sec:experimental}

We now take a closer look at the experimental setup of using nontransient guessing games to test PID nonsimplicity.  According to Definition~\ref{def:game-1} and as illustrated in Fig.~\ref{fig:game-1}, throughout the game procedure, no quantum memory with a lifetime larger than $\Delta t_\abb{D}\approx0$ is ever needed by Alice or Bob.  In this sense, nontransient guessing games are \emph{resource efficient}, as they do not consume any physical resource they are actually testing.  On the other hand, testing PID nonsimplicity or convertibility following the scheme proposed in Theorem~\ref{thm:convertibility} can still be costly to experiment.  This is because the scheme requires Bob to be able to implement infinitely many different POVMs reliably, which is hard to achieve in realistic settings.  Therefore, we are motivated to design nonsimplicity tests that are experimentally friendlier.  In what follows, we propose a new class of guessing games that also gives rise to a complete set of incompatibility monotones.  We call such games \emph{postinformation guessing games}, as they generalize the games after the same name in Ref.~\cite{Buscemi-2020a} for testing incompatibility between POVMs.  Compared to nontransient guessing games, postinformation guessing games are experimentally more convenient to realize.

\begin{definition}
\label{def:game-2}
A \textbf{postinformation guessing game} between Alice (the player) and Bob (the referee) is specified by a state ensemble $\ens{\varsigma}\equiv\{\sigma_{m,n,l}\}_{m,n,l}$ and a POVM $\povm{L}\equiv\{L_{l'}\}_{l'}$, and it has two stages separated by a time interval $\Delta t\gg\Delta t_\abb{D}\approx0$.
\begin{enumerate}[leftmargin=1.5em]\setlength\itemsep{-0.2em}
	\item In the first stage, Bob generates an index triple $(m,n,l)$ with probability $\tr[\sigma_{m,n,l}]$ and sends the state $\sigma_{m,n,l}/\tr[\sigma_{m,n,l}]$ to Alice without announcing $(m,n,l)$.  Then Bob asks Alice to submit a quantum system back to him.  Bob measures Alice's submitted system according to the POVM $\povm{L}$ and obtains an outcome $l'$.
	\item In the second stage (after $\Delta t$ has passed), Bob announces the index $j$ to Alice, and then he asks Alice to make a guess $n'$ at the index $n$.  Alice wins the game whenever she guesses correctly and in the meantime does not alter the index $l$, i.e., whenever $n'=n$ and $l'=l$.
\end{enumerate}
Throughout the game, Alice has no access to quantum memories with a lifetime larger than $\Delta t_\abb{D}\approx0$.
\end{definition}

\begin{figure}[t]
\includegraphics[scale=0.15]{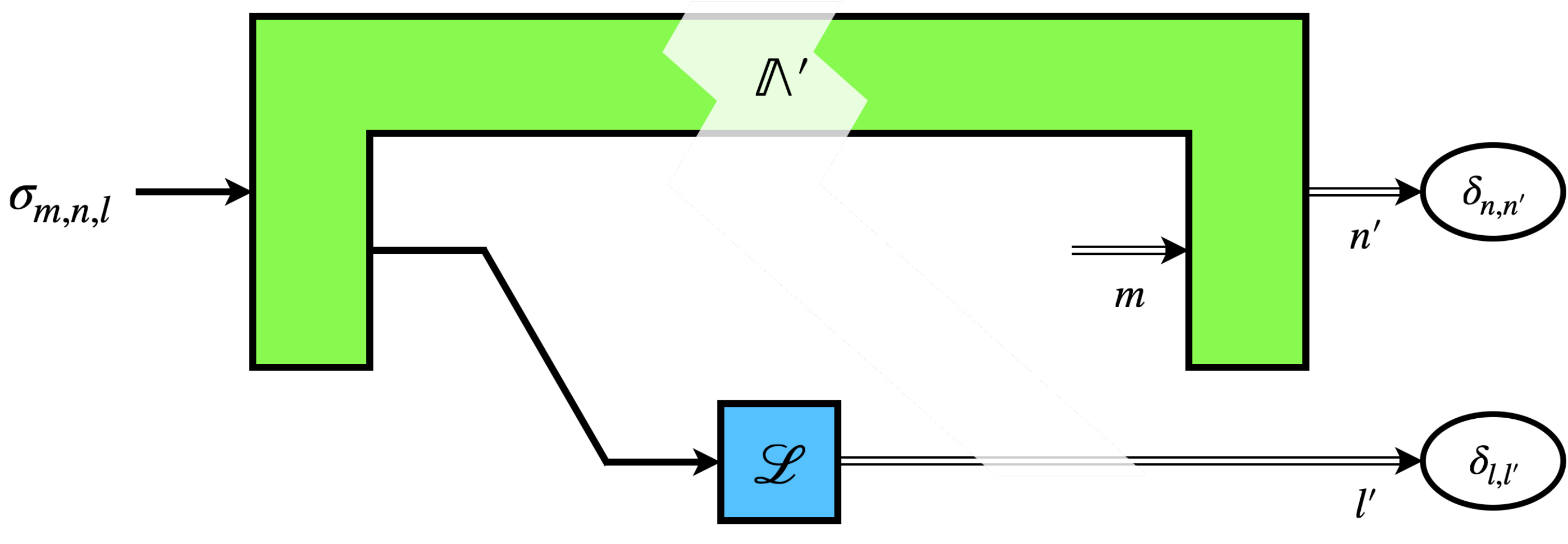}
\caption{A postinformation guessing game between Alice (the player) and Bob (the referee).  The setting of the game is specified by Bob's state ensemble $\ens{\varsigma}\equiv\{\sigma_{m,n,l}\}_{m,n,l}$ and POVM $\povm{L}\equiv\{L_{l'}\}_{l'}$.  Alice's strategy to the game is represented by a PID $\pd{\Lambda}'$ (green), which is freely simulated by an actual PID $\pd{\Lambda}$ held in her hand.  She wins the game whenever she makes a correct guess at one of Bob's indices $n$ while not altering the index $l$ given the index $m$, i.e., whenever $n'=n$ and $l'=l$.}
\label{fig:game-2}
\end{figure}

We note that Definition~\ref{def:game-2} is also semi-device-independent in the sense that Alice's device and strategy are uncharacterized.  As before, we now assume that Alice is assisted by a PID $\pd{\Lambda}\equiv\{\Lambda_{x_1|x_0}\}_{x_0,x_1}$ while playing the game and that the quantum delay time of $\pd{\Lambda}$ is no greater than the $\Delta t_\abb{D}$ specified in Definition~\ref{def:game-2}.  Since Alice has no access to quantum memories with a lifetime exceeding $\Delta t_\abb{D}$, her most general strategy is described by a PID $\pd{\Lambda}'$ such that $\pd{\Lambda}'\succcurlyeq_\abb{I}\pd{\Lambda}$, as illustrated in Fig.~\ref{fig:game-2}.  As a result, Alice's maximum winning probability in the postinformation guessing game specified by the state ensemble $\ens{\varsigma}$ and the POVM $\povm{L}$ is given by
\begin{align}
	\label{eq:game-2}
	P_\abb{guess}'(\pd{\Lambda};\ens{\varsigma},\povm{L})&\coloneq\max_{\pd{\Lambda}'\colon\pd{\Lambda}\succcurlyeq_\abb{I}\pd{\Lambda}'}\sum_{m,n,l}\tr\left[L_l\Lambda_{n|m}'\left[\sigma_{m,n,l}\right]\right].
\end{align}

The following proposition states that for a certain POVM $\povm{L}$, the winning probabilities $\{P_\abb{guess}'(\pd{\Lambda};\ens{\varsigma},\povm{L})\}_\ens{\varsigma}$ compose a complete set of nonsimplicity monotones with respect to $\pd{\Lambda}$ when considering different source ensembles.

\begin{proposition}
\label{prop:convertibility}
Let $\pd{\Lambda}\equiv\{\Lambda_{x_1|x_0}^{A_0\to A_1}\}_{x_0\in\idx{X}_0,x_1\in\idx{X}_1}$ and $\pd{\Gamma}\equiv\{\Gamma_{y_1|y_0}^{B_0\to B_1}\}_{y_0\in\idx{Y}_0,y_1\in\idx{Y}_1}$ be two PIDs, and let $\povm{L}\equiv\{L_{l'}^{B_1}\}_{l'\in\idx{L}}$ be an informationally complete POVM\@.  Then $\pd{\Lambda}\succcurlyeq_\abb{I}\pd{\Gamma}$ if and only if $P_\abb{guess}'(\pd{\Lambda};\ens{\varsigma},\povm{L})\geq P_\abb{guess}'(\pd{\Gamma};\ens{\varsigma},\povm{L})$ for every state ensemble $\ens{\varsigma}\equiv\{\sigma_{m,n,l}^{B_0}\}_{m\in\idx{Y}_0,n\in\idx{Y}_1,l\in\idx{L}}$.
\end{proposition}

The proof of Proposition~\ref{prop:convertibility} is in Appendix~\ref{app:experimental}.  

Proposition~\ref{prop:convertibility} generalizes Ref.~\cite[Theorem~1]{Buscemi-2020a} in the sense that it reduces to the latter when the index $l$ and Alice's quantum output are trivial.  Compared to Theorem~\ref{thm:convertibility}, Proposition~\ref{prop:convertibility} enables faithful tests of convertibility between PIDs under free simulations (and thus of nonsimplicity of a PID) with a much lower experimental requirement.  First, unlike in Theorem~\ref{thm:convertibility}, the tests in Proposition~\ref{prop:convertibility} do not demand infinitely many different POVMs on Bob's side.  Besides, only single-system operations are involved to carry out such tests, while entanglement distribution and bipartite measurements are avoided, reducing the experimental difficulty to a further extent.  Although Proposition~\ref{prop:convertibility} still requires an infinitude of different source ensembles, it is also true that one can bypass this challenge by using a single tomographically complete state ensemble and classical postprocessing to simulate all different source ensembles in these tests.  As a result, while implementing the tests in Proposition~\ref{prop:convertibility}, Bob can reuse his quantum hardware over and over again and only vary his classical postprocessing when the game parameters differ between tests.

\section{Conclusions and Discussions}
\label{sec:conclusions}

\subsection{Summary of results}
\label{sec:summary}

In this paper, we have conducted a resource-theoretic analysis of incompatibility between quantum instruments in terms of nonsimple programmability of quantum devices.  We have been physically motivated by the notion of programmability, which envisions certain quantum devices as objects that can be programmed at any time, i.e., regardless of when the quantum input arrives.  This naturally restricts the investigation to programmable instrument devices, which are classically controlled mechanisms that implement nonsignaling multi-instruments (Definition~\ref{def:PID}).  Every PID possesses two characteristic time intervals: (i) the quantum delay time $\Delta t_\abb{D}$, which quantifies how quickly the device produces its quantum output, and (ii) the lifetime of the internal quantum memory $\Delta t_\abb{QM}$, which quantifies how long the device is able to store some form of quantum information to influence the classical output.  To provide the experimenter with full temporal freedom on when the program can be issued, simple PIDs can have $\Delta t_\abb{QM}^\abb{simple}\leq\Delta t_\abb{D}$, whereas nonsimple PIDs must satisfy $\Delta t_\abb{QM}\gg\Delta t_\abb{D}$.  Quantum memories are thus the resource that enables nonsimple programmability.  To isolate the different memory demands between simple and nonsimple PIDs, we have assumed $\Delta t_\abb{D}\approx0$ for all PIDs so that only nonsimple PIDs require a built-in quantum memory with a non-negligible lifetime to implement.
	
In the resource theory of PID nonsimplicity, the experimenter is allowed to perform arbitrary auxiliary processing that does not depend on quantum memories with a non-negligible lifetime.  This restricts the allowable transformations between PIDs to a set of quantum combs called free simulations (Definition~\ref{def:simulation-PID}).  As nonsimplicity of PIDs is mathematically captured by incompatibility of the family of instruments being implemented, the ability of one PID to freely simulate another identifies an incompatibility preorder between these devices (Theorem~\ref{thm:simulation-PID}).

Every PID can be understood as a channel assemblage produced through a process of channel steering, and simple PIDs correspond to unsteerable channel assemblages.  So yet another way to frame this work is as a resource theory of channel steering.  From a practical point of view, channel steering offers a way of investigating properties of a given broadcast channel when the measurement device of one receiver is untrusted~\cite{Piani-2015a, Pusey-2015a}.  We have deepened the connections between PID nonsimplicity and channel steering by deriving for every PID a unique steering decomposition (Propositions~\ref{prop:operational}) and showing that this decomposition is ``canonical'' (Proposition~\ref{prop:canonicity}).  An essential ingredient of this decomposition, called the steering-equivalence mapping (Definition~\ref{def:SEM}), has subsequently been identified as an object-valued incompatibility monotone (Theorem~\ref{thm:monotonicity}).  This monotonicity result reflects a fundamental connection between the resource theory of PID nonsimplicity and that of measurement incompatibility, and consequently, any measure of incompatibility between POVMs previously studied in the literature~\cite{Heinosaari-2015a, Guerini-2017a, Oszmaniec-2019a, Uola-2019a} can now be used to quantify incompatibility between quantum instruments.

We have also proposed operational schemes for measuring and benchmarking nonsimplicity of PIDs by designing a class of games, called nontransient guessing games (Definition~\ref{def:game-1}).  These games have temporally separated stages in a way that resembles the spatially separated parties in nonlocal games, and therefore they are adept at characterizing correlations that exist across time.  We have shown that the maximum winning probability in any nontransient guessing game is a nonsimplicity monotone with respect to the player's PID, and the collection of all such winning probabilities under different game settings provides a complete criterion for judging whether a given PID can freely simulate another (Theorem~\ref{thm:convertibility}).  Since no assumption needs to be made about the player's device or strategy, nontransient guessing games also provide semi-device-independent certifications for PID nonsimplicity.  We have also established a tight upper bound on the operational advantage of a given PID over simple PIDs in nontransient guessing games in terms of a well-studied resource measure, namely the robustness against noise (Theorem~\ref{thm:robustness}).  This result endows the robustness of incompatibility with a clear operational meaning.  Considering the fact that testing PID convertibility using nontransient guessing games can be experimentally costly to implement, we have provided an alternative but experimentally friendlier scheme for such tests based on a class of so-called postinformation guessing games (Proposition~\ref{prop:convertibility}).

\subsection{Outlook}
\label{sec:outlook}

Our work leads to several directions for future research.  First, we have treated PID nonsimplicity as a dynamical resource distributed over quantum networks~\cite{Chiribella-2009a} rather than carried by quantum channels~\cite{Gour-2019b, Gour-2020a}.  The difference here is that, PIDs are quantum $2$-combs (i.e., quantum networks with two vertices), and so they can be manipulated by quantum $4$-combs~\cite{Chiribella-2009a}, whereas generic quantum channels can only be manipulated by quantum superchannels~\cite{Chiribella-2008a, Gour-2019a}.  A potential direction is to generalize the concept of incompatibility and its resource theory to more complex network layouts.  It would also be interesting to study resources other than incompatibility in the network setting~\cite{Milz-2021a, Milz-2022a}, and a generic framework for studying resources in networks has been lacking.

Second, we have investigated the relationships between a number of quantum correlations (see Table~\ref{tab:relationship}), and we have clarified that the presence of quantum memories (i.e., entanglement-nonbreaking channels) is a precondition for any of these correlations.  This is a qualitative remark, and one can continue to conduct a quantitative analysis on the pivotal role of quantum memories by asking what is the limit of nonclassicality in quantum correlations generated by \emph{unideal} quantum memories (i.e., nonidentity channels).  Following this line, one may further expect that a universal framework for studying various quantum correlations can be established based on the resource theory of quantum memories~\cite{Rosset-2018a}.

Finally, as we mentioned before, PIDs are quantum $2$-combs with the second vertex being classical, and thus they can also be interpreted as quantum superchannels transforming one POVM to another.  Interestingly, despite simple PIDs being a proper subset of general PIDs, any conversion between two single-party POVMs via general PIDs can always be realized via simple PIDs.  This indicates a vanishing operational distinction between general and simple PIDs in terms of converting single-party POVMs.  However, the distinction between general and simple PIDs becomes conspicuous when we consider convertibility between bipartite POVMs via \emph{partial} action of these PIDs.  This is well demonstrated by the nontransient guessing games, where a PID $\pd{\Lambda}'$ acts on one part of a bipartite POVM $\povm{M}$ (see Fig.~\ref{fig:game-1}), and the performance gap between general and simple PIDs is nontrivial (Theorem~\ref{thm:convertibility}).  This kind of interplay between simple PIDs and (single-party and bipartite) POVMs is somehow reminiscent of that between entanglement-breaking channels and (single-party and bipartite) states.  As for the latter, convertibility between single-party states via general channels is equivalent to convertibility via entanglement-breaking channels, whereas convertibility between bipartite states via partial action of general channels does not imply convertibility via partial action of entanglement-breaking channels.  Partial action of PIDs on bipartite POVMs is still not fully understood, and we leave the exploration for future work.

\section*{Acknowledgements}
\label{sec:acknowledgements}

We thank Vishal Singh and Mark M. Wilde for helpful comments on the presentation of the paper.  This work was supported by the Department of Energy (DOE) Q-NEXT: National Quantum Information Science Research Center.


\onecolumngrid
\appendix

\section{Properties of Free Simulations}
\label{app:simulation-PID}

In this Appendix, we demonstrate resource-theoretic properties possessed by free simulations of PIDs.  These properties include simplicity [Theorem~\ref{thm:simulation-PID}(1)], reachability [Theorem~\ref{thm:simulation-PID}(2)], composability [Theorem~\ref{thm:simulation-PID}(3)], closedness [Theorem~\ref{thm:simulation-PID}(4)], and convexity [Theorem~\ref{thm:simulation-PID}(5)].

\subsection{Proof of Theorem~\ref{thm:simulation-PID}(1)}
\label{app:simplicity}

Let $\pd{\Lambda}\equiv\{\Lambda_{x_1|x_0}^{A_0\to A_1}\}_{x_0,x_1}$ be a simple PID\@.  By Definition~\ref{def:PID}, there exists a quantum instrument $\{\m{G}_g^{A_0\to A_1}\}_g$ and a classical channel (i.e., a conditional probability distribution) $\{p_{x_1|x_0,g}\}_{x_0,x_1,g}$ such that
\begin{align}
	\Lambda_{x_1|x_0}^{A_0\to A_1}&=\sum_{g}p_{x_1|x_0,g}\m{G}_g^{A_0\to A_1}\quad\forall x_0,x_1.
\end{align}
Let $\pd{\Gamma}\equiv\{\Gamma_{y_1|y_0}^{B_0\to B_1}\}_{y_0,y_1}$ be a PID such that $\pd{\Lambda}\succcurlyeq_\abb{I}\pd{\Gamma}$.  By Definition~\ref{def:simulation-PID}, there exists a quantum channel $\m{F}'^{B_0\to A_0D}$, a quantum instrument $\{\m{K}_k'^{A_1D\to B_1}\}_k$, and two classical channels $\{p'_{x_0,l|y_0,k}\}_{x_0,y_0,k,l}$ and $\{q'_{y_1|x_1,l}\}_{x_1,y_1,l}$ such that
\begin{align}
	\Gamma_{y_1|y_0}^{B_0\to B_1}&=\sum_{x_0,x_1,k,l}q'_{y_1|x_1,l}p'_{x_0,l|y_0,k}\m{K}_k'^{A_1D\to B_1}\circ\left(\Lambda_{x_1|x_0}^{A_0\to A_1}\otimes\id^D\right)\circ\m{F}'^{B_0\to A_0D} \\
	&=\sum_{x_0,x_1,g,k,l}q'_{y_1|x_1,l}p_{x_1|x_0,g}p'_{x_0,l|y_0,k}\m{K}_k'^{A_1D\to B_1}\circ\left(\m{G}_g^{A_0\to A_1}\otimes\id^D\right)\circ\m{F}'^{B_0\to A_0D}\quad\forall y_0,y_1. \label{eq:a1:1}
\end{align}
Define a quantum instrument $\{\m{G}_{g,k}'^{B_0\to B_1}\}_{g,k}$ and a classical channel $\{p''_{y_1|y_0,g,k}\}_{y_0,y_1,g,k}$ as follows:
\begin{align}
	\m{G}_{g,k}'^{B_0\to B_1}&\coloneq\m{K}_k'^{A_1D\to B_1}\circ\left(\m{G}_g^{A_0\to A_1}\otimes\id^D\right)\circ\m{F}'^{B_0\to A_0D}\quad\forall g,k, \label{eq:a1:2}\\
	p''_{y_1|y_0,g,k}&\coloneq\sum_{x_0,x_1,l}q'_{y_1|x_1,l}p_{x_1|x_0,g}p'_{x_0,l|y_0,k}\quad\forall y_0,y_1,g,k. \label{eq:a1:3}
\end{align}
Then it follows from Eqs.~\eqref{eq:a1:1}--\eqref{eq:a1:3} that
\begin{align}
	\sum_{g,k}p''_{y_1|y_0,g,k}\m{G}_{g,k}'^{B_0\to B_1}&=\sum_{x_0,x_1,g,k,l}q'_{y_1|x_1,l}p_{x_1|x_0,g}p'_{x_0,l|y_0,k}\m{K}_k'^{A_1D\to B_1}\circ\left(\m{G}_g^{A_0\to A_1}\otimes\id^D\right)\circ\m{F}'^{B_0\to A_0D} \\
	&=\Gamma_{y_1|y_0}^{B_0\to B_1}\quad\forall y_0,y_1.
\end{align}
By Definition~\ref{def:PID}, this shows that the simulated PID $\pd{\Gamma}$ is simple.  This concludes the proof of Theorem~\ref{thm:simulation-PID}(1).

\subsection{Proof of Theorem~\ref{thm:simulation-PID}(2)}
\label{app:reachability}

Let $\pd{\Lambda}\equiv\{\Lambda_{x_1|x_0}^{A_0\to A_1}\}_{x_0,x_1}$ be a PID, and let $\pd{\Gamma}\equiv\{\Gamma_{y_1|y_0}^{B_0\to B_1}\}_{y_0,y_1}$ be a simple PID\@.  By Definition~\ref{def:PID}, there exists a quantum instrument $\{\m{G}_g^{B_0\to B_1}\}_g$ and a classical channel $\{p_{y_1|y_0,g}\}_{y_0,y_1,g}$ such that
\begin{align}
\label{eq:a2:1}
	\Gamma_{y_1|y_0}^{B_0\to B_1}&=\sum_{g}p_{y_1|y_0,g}\m{G}_g^{B_0\to B_1}\quad\forall y_0,y_1.
\end{align}
Let $\delta$ denote the classical identity channel, which satisfies $\delta_{b|a}\coloneq1$ if $a=b$ and $\delta_{b|a}\coloneq0$ if $a\neq b$.  Define a quantum channel $\m{F}'^{B_0\to A_0D}$, a quantum instrument $\{\m{K}_g'^{A_1D\to B_1}\}_g$, and two classical channels $\{p'_{x_0,l|y_0,g}\}_{x_0,y_0,g,l}$ and $\{q'_{y_1|x_1,l}\}_{x_1,y_1,s}$ as follows:
\begin{align}
	\m{F}'^{B_0\to A_0D}&\coloneq\op{0}{0}^{A_0}\otimes\id^{B_0\to D}, \label{eq:a2:2}\\
	\m{K}_g'^{A_1D\to B_1}&\coloneq\m{G}_g^{D\to B_1}\circ\tr_{A_1}\quad\forall g, \label{eq:a2:3}\\
	p'_{x_0,l|y_0,g}&\coloneq\delta_{x_0|0}p_{l|y_0,g}\quad\forall x_0,y_0,g,l, \label{eq:a2:4}\\
	q'_{y_1|x_1,l}&\coloneq\delta_{y_1|l}\quad\forall x_1,y_1,l. \label{eq:a2:5}
\end{align}
Then it follows from Eqs.~\eqref{eq:a2:1}--\eqref{eq:a2:5} that
\begin{align}
	&\sum_{x_0,x_1,g,l}q'_{y_1|x_1,l}p'_{x_0,l|y_0,g}\m{K}_g'^{A_1D\to B_1}\circ\left(\Lambda_{x_1|x_0}^{A_0\to A_1}\otimes\id^D\right)\circ\m{F}'^{B_0\to A_0D} \notag\\
	&\quad=\sum_{x_0,x_1,g,l}\delta_{y_1|l}\delta_{x_0|0}p_{l|y_0,g}\m{G}_g^{D\to B_1}\circ\tr_{A_1}\circ\left(\Lambda_{x_1|x_0}^{A_0\to A_1}\left[\op{0}{0}^{A_0}\right]\otimes\id^{B_0\to D}\right) \\
	&\quad=\sum_{g}p_{y_1|y_0,g}\m{G}_g^{B_0\to B_1} \\
	&\quad=\Gamma_{y_1|y_0}^{B_0\to B_1}\quad\forall y_0,y_1.
\end{align}
By Definition~\ref{def:simulation-PID}, this shows $\pd{\Lambda}\succcurlyeq_\abb{I}\pd{\Gamma}$.  This concludes the proof of Theorem~\ref{thm:simulation-PID}(2).

\subsection{Proof of Theorem~\ref{thm:simulation-PID}(3)}
\label{app:composability}

First, we prove sequential composability.  Consider a free simulation $\pd{\Lambda}\mapsto\pd{\Gamma}$, where $\pd{\Lambda}\equiv\{\Lambda_{x_1|x_0}^{A_0\to A_1}\}_{x_0,x_1}$ and $\pd{\Gamma}\equiv\{\Gamma_{y_1|y_0}^{B_0\to B_1}\}_{y_0,y_1}$ are two PIDs.  By Definition~\ref{def:simulation-PID}, the free simulation can be represented by a quantum channel $\m{F}^{B_0\to A_0D}$, a quantum instrument $\{\m{K}_k^{A_1D\to B_1}\}_k$, and two classical channels $\{p_{x_0,l|y_0,k}\}_{x_0,y_0,k,l}$ and $\{q_{y_1|x_1,l}\}_{x_1,y_1,l}$ such that
\begin{align}
\label{eq:a3:1}
	\Gamma_{y_1|y_0}^{B_0\to B_1}&=\sum_{x_0,x_1,k,l}q_{y_1|x_1,l}p_{x_0,l|y_0,k}\m{K}_k^{A_1D\to B_1}\circ\left(\Lambda_{x_1|x_0}^{A_0\to A_1}\otimes\id^D\right)\circ\m{F}^{B_0\to A_0D}\quad\forall y_0,y_1.
\end{align}
Consider another free simulation $\pd{\Gamma}\mapsto\pd{\Psi}$, where $\pd{\Psi}\equiv\{\Psi_{z_1|z_0}^{C_0\to C_1}\}_{z_0,z_1}$ is a PID\@.  By Definition~\ref{def:simulation-PID}, this free simulation can be represented by a quantum channel $\m{F}'^{C_0\to B_0E}$, a quantum instrument $\{\m{K}_{k'}'^{B_1E\to C_1}\}_{k'}$, and two classical channels $\{p'_{y_0,l'|z_0,k'}\}_{y_0,z_0,k',l'}$ and $\{q'_{z_1|y_1,l'}\}_{y_1,z_1,l'}$ such that
\begin{align}
\label{eq:a3:2}
	\Psi_{z_1|z_0}^{C_0\to C_1}&=\sum_{y_0,y_1,k',l'}q'_{z_1|y_1,l'}p'_{y_0,l'|z_0,k'}\m{K}_{k'}'^{B_1E\to C_1}\circ\left(\Gamma_{y_1|y_0}^{B_0\to B_1}\otimes\id^E\right)\circ\m{F}'^{C_0\to B_0E}\quad\forall z_0,z_1.
\end{align}
Combining Eqs.~\eqref{eq:a3:1} and \eqref{eq:a3:2}, the sequential composition of the above two free simulations can be described as $\pd{\Lambda}\mapsto\pd{\Psi}$ such that
\begin{align}
\label{eq:a3:3}
	\Psi_{z_1|z_0}^{C_0\to C_1}&=\sum_{x_0,x_1,y_0,y_1,k,k',l,l'}q'_{z_1|y_1,l'}q_{y_1|x_1,l}p_{x_0,l|y_0,k}p'_{y_0,l'|z_0,k'}\m{K}_{k'}'^{B_1E\to C_1}\circ\left(\m{K}_k^{A_1D\to B_1}\otimes\id^E\right) \notag\\
	&\quad\circ\left(\Lambda_{x_1|x_0}^{A_0\to A_1}\otimes\id^{DE}\right)\circ\left(\m{F}^{B_0\to A_0D}\otimes\id^E\right)\circ\m{F}'^{C_0\to B_0E}\quad\forall z_0,z_1.
\end{align}
Define a quantum channel $\m{F}''^{C_0\to A_0DE}$, a quantum instrument $\{\m{K}_{k,k'}''^{A_1DE\to C_1}\}_{k,k'}$, and two classical channels $\{p''_{x_0,l,l'|z_0,k,k'}\}_{x_0,z_0,k,k',l,l'}$ and $\{q''_{z_1|x_1,l,l'}\}_{x_1,z_1,l,l'}$ as follows:
\begin{align}
	\m{F}''^{C_0\to A_0DE}&\coloneq\left(\m{F}^{B_0\to A_0D}\otimes\id^E\right)\circ\m{F}'^{C_0\to B_0E}, \label{eq:a3:4}\\
	\m{K}_{k,k'}''^{A_1DE\to C_1}&\coloneq\m{K}_{k'}'^{B_1E\to C_1}\circ\left(\m{K}_k^{A_1D\to B_1}\otimes\id^E\right)\quad\forall k,k', \label{eq:a3:5}\\
	p''_{x_0,l,l'|z_0,k,k'}&\coloneq\sum_{y_0}p_{x_0,l|y_0,k}p'_{y_0,l'|z_0,k'}\quad\forall x_0,z_0,k,k',l,l', \label{eq:a3:6}\\
	q''_{z_1|x_1,l,l'}&\coloneq\sum_{y_1}q'_{z_1|y_1,l'}q_{y_1|x_1,l}\quad\forall x_1,z_1,l,l'. \label{eq:a3:7}
\end{align}
Then it follows from Eqs.~\eqref{eq:a3:3}--\eqref{eq:a3:7} that
\begin{align}
	&\sum_{x_0,x_1,k,k',l,l'}q''_{z_1|x_1,l,l'}p''_{x_0,l,l'|z_0,k,k'}\m{K}_{k,k'}''^{A_1DE\to C_1}\circ\left(\Lambda_{x_1|x_0}^{A_0\to A_1}\otimes\id^{DE}\right)\circ\m{F}''^{C_0\to A_0DE} \notag\\
	&\quad=\sum_{x_0,x_1,y_0,y_1,k,k',l,l'}q'_{z_1|y_1,l'}q_{y_1|x_1,l}p_{x_0,l|y_0,k}p'_{y_0,l'|z_0,k'}\m{K}_{k'}'^{B_1E\to C_1}\circ\left(\m{K}_k^{A_1D\to B_1}\otimes\id^E\right) \notag\\
	&\quad\quad\circ\left(\Lambda_{x_1|x_0}^{A_0\to A_1}\otimes\id^{DE}\right)\circ\left(\m{F}^{B_0\to A_0D}\otimes\id^E\right)\circ\m{F}'^{C_0\to B_0E} \\
	&\quad=\Psi_{z_1|z_0}^{C_0\to C_1}\quad\forall z_0,z_1.
\end{align}
By Definition~\ref{def:simulation-PID}, this shows that the sequential composition $\pd{\Lambda}\mapsto\pd{\Psi}$ of the two free simulations is a free simulation.

Next, we prove parallel composability.  Consider a free simulation $\pd{\Lambda}\mapsto\pd{\Gamma}$, where $\pd{\Lambda}\equiv\{\Lambda_{x_1|x_0}^{A_0\to A_1}\}_{x_0,x_1}$ and $\pd{\Gamma}\equiv\{\Gamma_{y_1|y_0}^{B_0\to B_1}\}_{y_0,y_1}$ are two PIDs.  By Definition~\ref{def:simulation-PID}, the free simulation can be represented by a quantum channel $\m{F}^{B_0\to A_0D}$, a quantum instrument $\{\m{K}_k^{A_1D\to B_1}\}_k$, and two classical channels $\{p_{x_0,l|y_0,k}\}_{x_0,y_0,k,l}$ and $\{q_{y_1|x_1,l}\}_{x_1,y_1,l}$ such that
\begin{align}
	\label{eq:a3:8}
	\Gamma_{y_1|y_0}^{B_0\to B_1}&=\sum_{x_0,x_1,k,l}q_{y_1|x_1,l}p_{x_0,l|y_0,k}\m{K}_k^{A_1D\to B_1}\circ\left(\Lambda_{x_1|x_0}^{A_0\to A_1}\otimes\id^D\right)\circ\m{F}^{B_0\to A_0D}\quad\forall y_0,y_1.
\end{align}
Consider another free simulation $\pd{\Lambda}'\mapsto\pd{\Gamma}'$, where $\pd{\Lambda}'\equiv\{\Lambda_{x_1'|x_0'}'^{A_0'\to A_1'}\}_{x_0',x_1'}$ and $\pd{\Gamma}'\equiv\{\Gamma_{y_1'|y_0'}'^{B_0'\to B_1'}\}_{y_0',y_1'}$ are two PIDs.  By Definition~\ref{def:simulation-PID}, this free simulation can be represented by a quantum channel $\m{F}'^{B_0'\to A_0'D'}$, a quantum instrument $\{\m{K}_{k'}'^{B_1'D'\to A_1'}\}_{k'}$, and two classical channels $\{p'_{x_0',l'|y_0',k'}\}_{x_0',y_0',k',l'}$ and $\{q'_{y_1'|x_1',l'}\}_{x_1',y_1',l'}$ such that
\begin{align}
	\label{eq:a3:9}
	\Gamma_{y_1'|y_0'}'^{B_0'\to B_1'}&=\sum_{x_0',x_1',k',l'}q'_{y_1'|x_1',l'}p'_{x_0',l'|y_0',k'}\m{K}_{k'}'^{A_1'D'\to B_1'}\circ\left(\Lambda_{x_1'|x_0'}'^{A_0'\to A_1'}\otimes\id^{D'}\right)\circ\m{F}'^{B_0'\to A_0'D'}\quad\forall y_0',y_1'.
\end{align}
Combining Eqs.~\eqref{eq:a3:8} and \eqref{eq:a3:9}, the parallel composition of the above two free simulations can be described as $\pd{\Lambda}''\mapsto\pd{\Gamma}''$, where $\pd{\Lambda}''\equiv\{\Lambda_{x_1,x_1'|x_0,x_0'}''^{A_0A_0'\to A_1A_1'}\}_{x_0,x_0',x_1,x_1'}$ and $\pd{\Gamma}''\equiv\{\Gamma_{y_1,y_1'|y_0,y_0'}''^{B_0B_0'\to B_1B_1'}\}_{y_0,y_0',y_1,y_1'}$ are two PIDs, such that
\begin{align}
	\label{eq:a3:10}
	\Gamma_{y_1,y_1'|y_0,y_0'}''^{B_0B_0'\to B_1B_1'}&=\sum_{x_0,x_0',x_1,x_1',k,k',l,l'}q_{y_1|x_1,l}q'_{y_1'|x_1',l'}p_{x_0,l|y_0,k}p'_{x_0',l'|y_0',k'}\left(\m{K}_k^{A_1D\to B_1}\otimes\m{K}_{k'}'^{A_1'D'\to B_1'}\right) \notag\\
	&\quad\circ\left(\Lambda_{x_1,x_1'|x_0,x_0'}''^{A_0A_0'\to A_1A_1'}\otimes\id^{DD'}\right)\circ\left(\m{F}^{B_0\to A_0D}\otimes\m{F}'^{B_0'\to A_0'D'}\right)\quad\forall y_0,y_0',y_1,y_1'.
\end{align}
Define a quantum channel $\m{F}''^{B_0B_0'\to A_0DA_0'D'}$, a quantum instrument $\{\m{K}_{k,k'}''^{A_1A_1'DD'\to B_1B_1'}\}_{k,k'}$, and two classical channels $\{p''_{x_0,x_0',l,l'|y_0,y_0',k,k'}\}_{x_0,x_0',y_0,y_0',k,k',l,l'}$ and $\{q''_{y_1,y_1'|x_1,x_1',l,l'}\}_{x_1,x_1',y_1,y_1',l,l'}$ as follows:
\begin{align}
	\m{F}''^{B_0B_0'\to A_0A_0'DD'}&\coloneq\m{F}^{B_0\to A_0D}\otimes\m{F}'^{B_0'\to A_0'D'}, \label{eq:a3:11}\\
	\m{K}_{k,k'}''^{A_1A_1'DD'\to B_1B_1'}&\coloneq\m{K}_k^{A_1D\to B_1}\otimes\m{K}_{k'}'^{A_1'D'\to B_1'}\quad\forall k,k', \label{eq:a3:12}\\
	p''_{x_0,x_0',l,l'|y_0,y_0',k,k'}&\coloneq p_{x_0,l|y_0,k}p'_{x_0',l'|y_0',k'}\quad\forall x_0,x_0',y_0,y_0',k,k',l,l', \label{eq:a3:13}\\
	q''_{y_1,y_1'|x_1,x_1',l,l'}&\coloneq q_{y_1|x_1,l}q'_{y_1'|x_1',l'}\quad\forall x_1,x_1',y_1,y_1',l,l'. \label{eq:a3:14}
\end{align}
Then it follows from Eqs.~\eqref{eq:a3:10}--\eqref{eq:a3:14} that
\begin{align}
	&\sum_{x_0,x_1,k,k',l,l'}q''_{y_1,y_1'|x_1,x_1',l,l'}p''_{x_0,x_0',l,l'|y_0,y_0',k,k'}\m{K}_{k,k'}''^{A_1A_1'DD'\to B_1B_1'}\circ\left(\Lambda_{x_1,x_1'|x_0,x_0'}''^{A_0A_0'\to A_1A_1'}\otimes\id^{DD'}\right)\circ\m{F}''^{B_0B_0'\to A_0A_0'DD'} \notag\\
	&\quad=\sum_{x_0,x_0',x_1,x_1',k,k',l,l'}q_{y_1|x_1,l}q'_{y_1'|x_1',l'}p_{x_0,l|y_0,k}p'_{x_0',l'|y_0',k'}\left(\m{K}_k^{A_1D\to B_1}\otimes\m{K}_{k'}'^{A_1'D'\to B_1'}\right) \notag\\
	&\quad\quad\circ\left(\Lambda_{x_1,x_1'|x_0,x_0'}''^{A_0A_0'\to A_1A_1'}\otimes\id^{DD'}\right)\circ\left(\m{F}^{B_0\to A_0D}\otimes\m{F}'^{B_0'\to A_0'D'}\right) \\
	&\quad=\Gamma_{y_1,y_1'|y_0,y_0'}''^{B_0B_0'\to B_1B_1'}\quad\forall y_0,y_0',y_1,y_1'.
\end{align}
By Definition~\ref{def:simulation-PID}, this shows that the parallel composition $\pd{\Lambda}''\mapsto\pd{\Gamma}''$ of the two free simulations is a free simulation.  This concludes the proof of Theorem~\ref{thm:simulation-PID}(3).

\subsection{Proof of Theorem~\ref{thm:simulation-PID}(4)}
\label{app:closedness}

Consider a free simulation $\pd{\Lambda}\mapsto\pd{\Gamma}$, where $\pd{\Lambda}\equiv\{\Lambda_{x_1|x_0}^{A_0\to A_1}\}_{x_0\in\idx{X}_0,x_1\in\idx{X}_1}$ and $\pd{\Gamma}\equiv\{\Gamma_{y_1|y_0}^{B_0\to B_1}\}_{y_0\in\idx{Y}_0,y_1\in\idx{Y}_1}$ are two PIDs.  By Definition~\ref{def:simulation-PID}, the free simulation can be represented by a quantum channel $\m{F}^{B_0\to A_0D}$, a quantum instrument $\{\m{K}_k^{A_1D\to B_1}\}_k$, and two classical channels $\{p_{x_0,l|y_0,k}\}_{x_0,y_0,k,l}$ and $\{q_{y_1|x_1,l}\}_{x_1,y_1,l}$ such that
\begin{align}
	\label{eq:a4:1}
	\Gamma_{y_1|y_0}^{B_0\to B_1}&=\sum_{x_0,x_1,k,l}q_{y_1|x_1,l}p_{x_0,l|y_0,k}\m{K}_k^{A_1D\to B_1}\circ\left(\Lambda_{x_1|x_0}^{A_0\to A_1}\otimes\id^D\right)\circ\m{F}^{B_0\to A_0D}\quad\forall y_0,y_1.
\end{align}
By Ref.~\cite[Theorem~1(4)]{Gour-2019a}, the dimensionality of the system $D$ can be bounded by the product of the dimensionalities of the systems $A_0$ and $B_0$.  Since every classical channel can be decomposed into a probabilistic mixture of deterministic classical channels, there exists a conditional probability distribution $\{q'_{l'|l}\}_{l,l'}$ such that
\begin{align}
	\label{eq:a4:2}
	q_{y_1|x_1,l}&=\sum_{l'\in\idx{L}'}\delta_{y_1|l'(x_1)}q'_{l'|l}\quad\forall x_1,y_1,l,
\end{align}
where $\idx{L}'\coloneq\idx{Y}_1^{\idx{X}_1}$ is the finite set of all functions from $\idx{X}_1$ to $\idx{Y}_1$.  Define a classical channel $\{p'_{x_0,l'|y_0,k}\}_{x_0,y_0,k,l'}$ as follows:
\begin{align}
	\label{eq:a4:3}
	p'_{x_0,l'|y_0,k}&\coloneq\sum_{l}q'_{l'|l}p_{x_0|y_0,k}\quad\forall x_0,y_0,k,l'.
\end{align}
Likewise, there exists a conditional probability distribution $\{p''_{k'|k}\}_{k,k'}$ such that
\begin{align}
	\label{eq:a4:4}
	p_{x_0,l'|y_0,k}'&=\sum_{k'\in\idx{K}'}\delta_{x_0,l'|k'(y_0)}p''_{k'|k}\quad\forall x_0,y_0,k,l',
\end{align}
where $\idx{K}'\coloneq(\idx{X}_0\otimes\idx{L}')^{\idx{Y}_0}$ is the finite set of all functions from $\idx{Y}_0$ to $\idx{X}_0\otimes\idx{L}'$.  Define a quantum instrument $\{\m{K}_{k'}'^{A_1D\to B_1}\}_{k'}$ and two classical channels $\{p'''_{x_0,l'|y_0,k'}\}_{x_0,y_0,k',l'}$ and $\{q''_{y_1|x_1,l'}\}_{x_1,y_1,l'}$ as follows:
\begin{align}
	\m{K}_{k'}'^{A_1D\to B_1}&\coloneq\sum_{k}p''_{k'|k}\m{K}_k^{A_1D\to B_1}\quad\forall k', \label{eq:a4:5}\\
	p'''_{x_0,l'|y_0,k'}&\coloneq\delta_{x_0,l'|k'(y_0)}\quad\forall x_0,y_0,k',l', \label{eq:a4:6}\\
	q''_{y_1|x_1,l'}&\coloneq\delta_{y_1|l'(x_1)}\quad\forall x_1,y_1,l'. \label{eq:a4:7}
\end{align}
Then it follows from Eqs.~\eqref{eq:a4:1}--\eqref{eq:a4:7} that
\begin{align}
	&\sum_{x_0,x_1,k',l'}q''_{y_1|x_1,l'}p'''_{x_0,l'|y_0,k'}\m{K}_{k'}'^{A_1D\to B_1}\circ\left(\Lambda_{x_1|x_0}^{A_0\to A_1}\otimes\id^D\right)\circ\m{F}^{B_0\to A_0D} \label{eq:a4:8}\\
	&\quad=\sum_{x_0,x_1,k',l'}\delta_{y_1|l'(x_1)}\delta_{x_0,l'|k'(y_0)}\m{K}_{k'}'^{A_1D\to B_1}\circ\left(\Lambda_{x_1|x_0}^{A_0\to A_1}\otimes\id^D\right)\circ\m{F}^{B_0\to A_0D} \\
	&\quad=\sum_{x_0,x_1,k,k',l'}\delta_{y_1|l'(x_1)}\delta_{x_0,l'|k'(y_0)}p''_{k'|k}\m{K}_k^{A_1D\to B_1}\circ\left(\Lambda_{x_1|x_0}^{A_0\to A_1}\otimes\id^D\right)\circ\m{F}^{B_0\to A_0D} \\
	&\quad=\sum_{x_0,x_1,k,l'}\delta_{y_1|l'(x_1)}p'_{x_0,l'|y_0,k}\m{K}_k^{A_1D\to B_1}\circ\left(\Lambda_{x_1|x_0}^{A_0\to A_1}\otimes\id^D\right)\circ\m{F}^{B_0\to A_0D} \\
	&\quad=\sum_{x_0,x_1,k,l,l'}\delta_{y_1|l'(x_1)}q'_{l'|l}p_{x_0|y_0,k}\m{K}_k^{A_1D\to B_1}\circ\left(\Lambda_{x_1|x_0}^{A_0\to A_1}\otimes\id^D\right)\circ\m{F}^{B_0\to A_0D} \\
	&\quad=\sum_{x_0,x_1,k,l}q_{y_1|x_1,l}p_{x_0|y_0,k}\m{K}_k^{A_1D\to B_1}\circ\left(\Lambda_{x_1|x_0}^{A_0\to A_1}\otimes\id^D\right)\circ\m{F}^{B_0\to A_0D} \\
	&\quad=\Gamma_{y_1|y_0}^{B_0\to B_1}\quad\forall y_0,y_1.
\end{align}
By Definition~\ref{def:simulation-PID}, this shows that every free simulation can be realized using side channels of bounded size, and therefore the set of free simulations is closed.  This concludes the proof of Theorem~\ref{thm:simulation-PID}(4).

\subsection{Proof of Theorem~\ref{thm:simulation-PID}(5)}
\label{app:convexity}

Consider an ensemble of free simulations labeled by a random index $i$.  The $i$th free simulation is applied with probability $p_i\geq0$ such that $\sum_{i}p_i=1$.  By Definition~\ref{def:simulation-PID}, the $i$th free simulation can be represented by a quantum channel $\m{F}_{(i)}'^{B_0\to A_0D}$, a quantum instrument $\{\m{K}_{k|i}'^{A_1D\to B_1}\}_k$, and two classical channels $\{p'_{x_0,l|y_0,k,i}\}_{x_0,y_0,k,l}$ and $\{q'_{y_1|x_1,l,i}\}_{x_1,y_1,l}$.  The probabilistic mixture of these free simulations is then described by  $\pd{\Lambda}\mapsto\pd{\Gamma}$, where $\pd{\Lambda}\equiv\{\Lambda_{x_1|x_0}^{A_0\to A_1}\}_{x_0,x_1}$ and $\pd{\Gamma}\equiv\{\Gamma_{y_1|y_0}^{B_0\to B_1}\}_{y_0,y_1}$ are two PIDs, such that
\begin{align}
\label{eq:a5:1}
	\Gamma_{y_1|y_0}^{B_0\to B_1}&=\sum_{i}p_i\sum_{x_0,x_1,k,l}q'_{y_1|x_1,l,i}p'_{x_0,l|y_0,k,i}\m{K}_{k|i}'^{A_1D\to B_1}\circ\left(\Lambda_{x_1|x_0}^{A_0\to A_1}\otimes\id^D\right)\circ\m{F}_{(i)}'^{B_0\to A_0D}\quad\forall y_0,y_1.
\end{align}
Define a quantum channel $\m{F}''^{B_0\to A_0DK}$, a quantum instrument $\{\m{K}_{k,k'}''^{A_1DK\to B_1}\}_{k,k'}$, and a classical channel $\{p''_{x_0,l,l'|y_0,k,k'}\}_{x_0,y_0,k,l,k',l'}$ as follows:
\begin{align}
	\m{F}''^{B_0\to A_0DK}&\coloneq\sum_{i}p_i\m{F}_{(i)}'^{B_0\to A_0D}\otimes\op{i}{i}^K, \label{eq:a5:2}\\
	\m{K}_{k,k'}''^{A_1DK\to B_1}\left[\cdot\right]&\coloneq\m{K}_{k|k'}'^{A_1D\to B_1}\left[\left(\1^{A_1D}\otimes\bra{k'}^K\right)\left[\cdot\right]\left(\1^{A_1D}\otimes\ket{k'}^K\right)\right]\quad\forall k,k', \label{eq:a5:3}\\
	p''_{x_0,l,l'|y_0,k,k'}&\coloneq p'_{x_0,l|y_0,k,k'}\delta_{l'|k'}\quad\forall x_0,y_0,k,l,k',l'. \label{eq:a5:4}
\end{align}
Then it follows from Eqs.~\eqref{eq:a5:1}--\eqref{eq:a5:4} that
\begin{align}
	&\sum_{x_0,x_1,k,l,k',l'}q'_{y_1|x_1,l,l'}p''_{x_0,l,l'|y_0,k,k'}\m{K}_{k,k'}''^{A_1DK\to B_1}\circ\left(\Lambda_{x_1|x_0}^{A_0\to A_1}\otimes\id^{DK}\right)\circ\m{F}''^{B_0\to A_0DK} \notag\\
	&\quad=\sum_{x_0,x_1,k,l,k',l'}q'_{y_1|x_1,l,l'}p'_{x_0,s|y_0,k,k'}\delta_{l'|k'}\m{K}_{k|k'}''^{A_1DK\to B_1}\circ\left(\Lambda_{x_1|x_0}^{A_0\to A_1}\otimes\id^{DK}\right)\circ\left(\sum_{i}p_i\m{F}_{(i)}'^{B_0\to A_0D}\otimes\ip{k'}{i}\ip{i}{k'}^K\right) \\
	&\quad=\sum_{i}p_i\sum_{x_0,x_1,k,l}q'_{y_1|x_1,l,i}p'_{x_0,l|y_0,k,i}\m{K}_{k|i}'^{A_1D\to B_1}\circ\left(\Lambda_{x_1|x_0}^{A_0\to A_1}\otimes\id^D\right)\circ\m{F}_{(i)}'^{B_0\to A_0D} \\
	&\quad=\Gamma_{y_1|y_0}^{B_0\to B_1}\quad\forall y_0,y_1.
\end{align}
By Definition~\ref{def:simulation-PID}, this shows that the probabilistic mixture $\pd{\Lambda}\mapsto\pd{\Psi}$ of the ensemble of free simulations is a free simulation.  This concludes the proof of Theorem~\ref{thm:simulation-PID}(5).

\section{Properties of the Steering-Equivalence Mapping}
\label{app:SEM}

In this Appendix, we demonstrate useful properties possessed by the steering-equivalence mapping.  These properties include an operational interpretation (Proposition~\ref{prop:operational}), canonicity (Proposition~\ref{prop:canonicity}), faithfulness [Theorem~\ref{thm:monotonicity}(1)], and monotonicity [Theorem~\ref{thm:monotonicity}(2)].

\subsection{Proof of Proposition~\ref{prop:operational}}
\label{app:operational}

Let $\pd{\Lambda}\equiv\{\Lambda_{x_1|x_0}^{A_0\to A_1}\}_{x_0,x_1}$ be a PID\@.  Let $\Lambda^{A_0\to A_1}\coloneq\sum_{x_1}\Lambda_{x_1|x_0}^{A_0\to A_1}$ be the marginal channel of $\pd{\Lambda}$ from $A_0$ to $A_1$.  Let $r\coloneq\rk(J_\Lambda^{A_0A_1})$.  The Choi operator $J_\Lambda^{A_0A_1}$ has a spectral decomposition as follows:
\begin{align}
	J_\Lambda^{A_0A_1}&=\sum_{i=0}^{r-1}a_i\op{\alpha_i}{\alpha_i}^{A_0A_1},
\end{align}
where $a_i>0$ is a positive real number for all $i$ and $\{\ket{\alpha_i}^{A_0A_1}\}_i$ is an orthonormal set of vectors.  Let $A^*$ be a system such that $\spa{H}^{A^*}\cong\supp(J_\Lambda^{A_0A_1})\subseteq\spa{H}^{A_0A_1}$.  Let $\ket{\alpha_i}^{A^*}$ be the image of $\ket{\alpha_i}^{A_0A_1}$ in $A^*$.  Let $\ket{\cc{\alpha_i}}^{A^*}$ be the complex conjugate of $\ket{\alpha_i}^{A^*}$ under a fixed orthonormal basis.  Then $\{\ket{\cc{\alpha_i}}^{A^*}\}_i$ is an orthonormal basis of $\spa{H}^{A^*}$.  By Ref.~\cite[Eq.~(2.2.36)]{Khatri-2020a}, there exists an operator $V^{A_0\to A_1A^*}$ such that
\begin{align}
	\left(\1^{A_0}\otimes V^{\rpl{A}_0\to A_1A^*}\right)\ket{\phi_+}^{A_0\rpl{A}_0}&=\sum_{i=0}^{r-1}\sqrt{a_i}\ket{\alpha_i}^{A_0A_1}\ket{\cc{\alpha_i}}^{A^*}.
\end{align}
Define a linear map $\m{V}^{A_0\to A_1A^*}$ as follows:
\begin{align}
	\m{V}^{A_0\to A_1A^*}\left[\cdot\right]&\coloneq V^{A_0\to A_1A^*}\left[\cdot\right](V^\dagger)^{A_1A^*\to A_0}.
\end{align}
We note that $\m{V}^{A_0\to A_1A^*}$ is an isometric dilation of $\Lambda^{A_0\to A_1}$, as can be verified by marginalizing its Choi operator $J_\m{V}^{A_0A_1A^*}$:
\begin{align}
	\tr_{A^*}\left[J_\m{V}^{A_0A_1A^*}\right]&=\tr_{A^*}\left[\left(\1^{A_0}\otimes V^{\rpl{A}_0\to A_1A^*}\right)\phi_+^{A_0\rpl{A}_0}\left(\1^{A_0}\otimes(V^\dagger)^{A_1A^*\to\rpl{A}_0}\right)\right] \\
	&=\tr_{A^*}\left[\sum_{i,j=0}^{r-1}\sqrt{a_ia_j}\op{\alpha_i}{\alpha_j}^{A_0A_1}\otimes\op{\cc{\alpha_i}}{\cc{\alpha_j}}^{A^*}\right] \\
	&=\sum_{i=0}^{r-1}a_i\op{\alpha_i}{\alpha_i}^{A_0A_1} \\*
	&=J_\Lambda^{A_0A_1}.
\end{align}
Let $\pd{S}\equiv\{S_{x_1|x_0}^{A^*}\}_{x_0,x_1}$ be a PMD such that $\pd{S}=\f{SEM}(\pd{\Lambda})$.  By Definition~\ref{def:SEM},
\begin{align}
	S_{x_1|x_0}^{A^*}&=(J_\Lambda^{A^*})^{-\frac{1}{2}}J_{\Lambda_{x_1|x_0}}^{A^*}(J_\Lambda^{A^*})^{-\frac{1}{2}}\quad\forall x_0,x_1.
\end{align}
Define an isometric channel $\m{W}^{A^*\to A_0A_1}$ as follows:
\begin{align}
	\m{W}^{A^*\to A_0A_1}\left[\cdot\right]&\coloneq\sum_{i,j=0}^{r-1}\bra{\alpha_i}\left[\cdot\right]\ket{\alpha_j}^{A^*}\op{\alpha_i}{\alpha_j}^{A_0A_1}.
\end{align}
It follows that
\begin{align}
	\tr_{A^*}\left[\left(\1^{A_0A_1}\otimes(S_{x_1|x_0}^\top)^{A^*}\right)J_\m{V}^{A_0A_1A^*}\right]&=\sum_{i,j=0}^{r-1}\sqrt{a_ia_j}\bra{\cc{\alpha_j}}S_{x_1|x_0}^\top\ket{\cc{\alpha_i}}^{A^*}\op{\alpha_i}{\alpha_j}^{A_0A_1} \\
	&=\sum_{i,j=0}^{r-1}\sqrt{a_ia_j}\bra{\alpha_i}S_{x_1|x_0}\ket{\alpha_j}^{A^*}\op{\alpha_i}{\alpha_j}^{A_0A_1} \\
	&=\left(\sum_{i=0}^{r-1}\sqrt{a_i}\ket{\alpha_i}^{A_0A_1}\bra{\alpha_i}^{A^*}\right)S_{x_1|x_0}^{A^*}\left(\sum_{j=0}^{r-1}\sqrt{a_j}\ket{\alpha_j}^{A^*}\bra{\alpha_j}^{A_0A_1}\right) \\
	&=\m{W}^{A^*\to A_0A_1}\left[(J_\Lambda^{A^*})^\frac{1}{2}S_{x_1|x_0}^{A^*}(J_\Lambda^{A^*})^\frac{1}{2}\right] \\
	&=\m{W}^{A^*\to A_0A_1}\left[J_{\Lambda_{x_1|x_0}}^{A^*}\right] \\
	&=J_{\Lambda_{x_1|x_0}}^{A_0A_1}\quad\forall x_0,x_1.
\end{align}
By the Choi--Jamiołkowski isomorphism, we can conclude that
\begin{align}
	\Lambda_{x_1|x_0}^{A_0\to A_1}\left[\cdot\right]&=\tr_{A^*}\left[\left(\1^{A_1}\otimes(S_{x_1|x_0}^\top)^{A^*}\right)\m{V}^{A_0\to A_1A^*}\left[\cdot\right]\right]\quad\forall x_0,x_1.
\end{align}
This concludes the proof of Proposition~\ref{prop:operational}.

\subsection{Proof of Proposition~\ref{prop:canonicity}}
\label{app:canonicity}

Let $\pd{\Lambda}\equiv\{\Lambda_{x_1|x_0}^{A_0\to A_1}\}_{x_0,x_1}$ be a PID\@.  Let $\Lambda^{A_0\to A_1}\coloneq\sum_{x_1}\Lambda_{x_1|x_0}^{A_0\to A_1}$ be the marginal channel of $\pd{\Lambda}$ from $A_0$ to $A_1$.  Let $\m{E}^{A_0\to A_1E}$ be a broadcast channel, and let $\pd{M}\equiv\{M_{x_1|x_0}^E\}_{x_0,x_1}$ be a PMD such that $\pd{\Lambda}\leftarrowtail(\m{E},\pd{M})$.  By Definition~\ref{def:steering},
\begin{align}
\label{eq:b2:1}
	\Lambda_{x_1|x_0}^{A_0\to A_1}\left[\cdot\right]&=\tr_E\left[\left(\1^{A_1}\otimes M_{x_1|x_0}^E\right)\m{E}^{A_0\to A_1E}\left[\cdot\right]\right]\quad\forall x_0,x_1.
\end{align}
Let $\m{V}^{A_0\to A_1EF}$ be an isometric dilation of $\m{E}^{A_0\to A_1E}$.  Let $V^{A_0\to A_1EF}$ be the isometry operator such that
\begin{align}
	\m{V}^{A_0\to A_1EF}\left[\cdot\right]&\coloneq V^{A_0\to A_1EF}\left[\cdot\right](V^\dagger)^{A_1EF\to A_0}.
\end{align}
Define a PMD $\pd{N}\equiv\{N_{x_1|x_0}^{EF}\}_{x_0,x_1}$ as follows:
\begin{align}
\label{eq:b2:2}
	N_{x_1|x_0}^{EF}&\coloneq M_{x_1|x_0}^E\otimes\1^F\quad\forall x_0,x_1.
\end{align}
It follows from Eqs.~\eqref{eq:b2:1} and \eqref{eq:b2:2} that
\begin{align}
	\tr_{EF}\left[\left(\1^{A_1}\otimes N_{x_1|x_0}^{EF}\right)\m{V}^{A_0\to A_1EF}\left[\cdot\right]\right]&=\tr_{E}\left[\left(\1^{A_1}\otimes M_{x_1|x_0}^E\right)\tr_F\circ\m{V}^{A_0\to A_1EF}\left[\cdot\right]\right] \\
	&=\tr_{E}\left[\left(\1^{A_1}\otimes M_{x_1|x_0}^E\right)\m{E}^{A_0\to A_1E}\left[\cdot\right]\right] \\
	&=\Lambda_{x_1|x_0}^{A_0\to A_1}\left[\cdot\right]\quad\forall x_0,x_1. \label{eq:b2:3}
\end{align}
The vector $(\1^{A_0}\otimes V^{\rpl{A}_0\to A_1EF})\ket{\phi_+}^{A_0\rpl{A}_0}$ has a Schmidt decomposition as follows:
\begin{align}
\label{eq:b2:4}
	\left(\1^{A_0}\otimes V^{\rpl{A}_0\to A_1EF}\right)\ket{\phi_+}^{A_0\rpl{A}_0}&=\sum_{i=0}^{r-1}\sqrt{a_i}\ket{\alpha_i}^{A_0A_1}\ket{\beta_i}^{EF},
\end{align}
where $a_i>0$ is a positive real number for all $i$ and $\{\ket{\alpha_i}^{A_0A_1}\}_i$ and $\{\ket{\beta_i}^{EF}\}_i$ are two orthonormal sets of vectors.  Let $A^*$ be a system such that $\spa{H}^{A^*}\cong\supp(J_\Lambda^{A_0A_1})\subseteq\spa{H}^{A_0A_1}$.  Let $\ket{\alpha_i}^{A^*}$ be the image of $\ket{\alpha_i}^{A_0A_1}$ in $A^*$.  Then $\{\ket{\alpha_i}^{A^*}\}_i$ is an orthonormal basis of $\spa{H}^{A^*}$.  Let $\ket{\cc{\beta_i}}^{EF}$ be the complex conjugate of $\ket{\beta_i}^{EF}$ under a fixed orthonormal basis.  Define an isometric channel $\m{W}^{A^*\to EF}$ as follows:
\begin{align}
	\m{W}^{A^*\to EF}\left[\cdot\right]&=\sum_{i,j=0}^{r-1}\bra{\alpha_i}\left[\cdot\right]\ket{\alpha_j}^{A^*}\op{\cc{\beta_i}}{\cc{\beta_j}}^{EF}.
\end{align}
Let $J_\m{V}^{A^*EF}$ denote the image of the Choi operator $J_\m{V}^{A_0A_1EF}$ of $\m{V}^{A_0\to A_1EF}$ in the composite system $A^*EF$.  Then
\begin{align}
	(J_\Lambda^{A^*})^\frac{1}{2}(\m{W}^\dagger)^{EF\to A^*}\left[(N_{x_1|x_0}^\top)^{EF}\right](J_\Lambda^{A^*})^\frac{1}{2}&=\sum_{i,j=0}^{r-1}\sqrt{a_ia_j}\op{\alpha_i}{\alpha_i}^{A^*}(\m{W}^\dagger)^{EF\to A^*}\left[(N_{x_1|x_0}^\top)^{EF}\right]\op{\alpha_j}{\alpha_j}^{A^*} \\
	&=\sum_{i,j=0}^{r-1}\sqrt{a_ia_j}\bra{\cc{\beta_i}}N_{x_1|x_0}^\top\ket{\cc{\beta_j}}^{EF}\op{\alpha_i}{\alpha_j}^{A^*} \\
	&=\sum_{i,j=0}^{r-1}\sqrt{a_ia_j}\bra{\beta_j}N_{x_1|x_0}\ket{\beta_i}^{EF}\op{\alpha_i}{\alpha_j}^{A^*} \\
	&=\tr_{EF}\left[\left(\1^{A^*}\otimes\left(N_{x_1|x_0}\right)^{EF}\right)\left(\sum_{i,j=0}^{r-1}\sqrt{a_ia_j}\op{\alpha_i}{\alpha_j}^{A^*}\otimes\op{\beta_i}{\beta_j}^{EF}\right)\right] \\
	&=\tr_{EF}\left[\left(\1^{A^*}\otimes N_{x_1|x_0}^{EF}\right)J_\m{V}^{A^*EF}\right] \label{eq:b2:5}\\
	&=J_{\Lambda_{x_1|x_0}}^{A^*}\quad\forall x_0,x_1. \label{eq:b2:6}
\end{align}
Here Eq.~\eqref{eq:b2:5} follows from Eq.~\eqref{eq:b2:4} and the isomorphism between $\spa{H}^{A^*}$ and $\supp(J_\Lambda^{A_0A_1})$; Eq.~\eqref{eq:b2:6} follows from Eq.~\eqref{eq:b2:3} and the same isomorphism.  It follows that
\begin{align}
	(J_\Lambda^{A^*})^{-\frac{1}{2}}J_{\Lambda_{x_1|x_0}}^{A^*}(J_\Lambda^{A^*})^{-\frac{1}{2}}&=(\m{W}^\dagger)^{EF\to A^*}\left[(N_{x_1|x_0}^\top)^{EF}\right]\quad\forall x_0,x_1.
\end{align}
By Definitions~\ref{def:simulation-PMD} and \ref{def:SEM}, this implies $\pd{N}^\top\succcurlyeq_\abb{M}\f{SEM}(\pd{\Lambda})$.  Since Eq.~\eqref{eq:b2:2} implies $\pd{M}^\top\succcurlyeq_\abb{M}\pd{N}^\top$, by the transitivity of the preorder $\succcurlyeq_\abb{M}$~\cite{Buscemi-2020a}, we have $\pd{M}^\top\succcurlyeq_\abb{M}\f{SEM}(\pd{\Lambda})$.  This concludes the proof of Proposition~\ref{prop:canonicity}.

\subsection{Proof of Theorem~\ref{thm:monotonicity}(1)}
\label{app:faithfulness}

By the Choi--Jamiołkowski isomorphism, a PID $\pd{\Lambda}\equiv\{\Lambda_{x_1|x_0}^{A_0\to A_1}\}_{x_0,x_1}$ is simple if and only if its corresponding assemblage of Choi states $\big\{J_{\Lambda_{x_1|x_0}}^{A_0A_1}/d_{A_0}\big\}_{x_0,x_1}$ is an unsteerable state assemblage~\cite{Piani-2015b}.  The state assemblage $\big\{J_{\Lambda_{x_1|x_0}}^{A_0A_1}/d_{A_0}\big\}_{x_0,x_1}$ is unsteerable if and only if its steering-equivalent observables are compatible~\cite[Theorem~1]{Uola-2015a}.  By Definition~\ref{def:SEM} and Ref.~\cite[Eq.~(5)]{Uola-2015a}, $\f{SEM}(\pd{\Lambda})$ is the steering-equivalent observables of the state assemblage $\big\{J_{\Lambda_{x_1|x_0}}^{A_0A_1}/d_{A_0}\big\}_{x_0,x_1}$.  Therefore, $\pd{\Lambda}$ is simple if and only if $\f{SEM}(\pd{\Lambda})$ is simple.  This concludes the proof of Theorem~\ref{thm:monotonicity}(1).

\subsection{Proof of Theorem~\ref{thm:monotonicity}(2)}
\label{app:monotonicity}

Let $\pd{\Lambda}\equiv\{\Lambda_{x_1|x_0}^{A_0\to A_1}\}_{x_0,x_1}$ and $\pd{\Gamma}\equiv\{\Gamma_{y_1|y_0}^{B_0\to B_1}\}_{y_0,y_1}$ be two PIDs such that $\pd{\Lambda}\succcurlyeq_\abb{I}\pd{\Gamma}$.  By Definition~\ref{def:simulation-PID}, there exists a quantum channel $\m{F}^{B_0\to A_0D}$, a quantum instrument $\{\m{K}_k^{A_1D\to B_1}\}_k$, and two classical channels $\{p_{x_0,l|y_0,k}\}_{x_0,y_0,k,l}$ and $\{q_{y_1|x_1,l}\}_{x_1,y_1,l}$ such that
\begin{align}
\label{eq:b3:1}
	\Gamma_{y_1|y_0}^{B_0\to B_1}&=\sum_{x_0,x_1,k,l}q_{y_1|x_1,l}p_{x_0,l|y_0,k}\m{K}_k^{A_1D\to B_1}\circ\left(\Lambda_{x_1|x_0}^{A_0\to A_1}\otimes\id^{C}\right)\circ\m{F}^{B_0\to A_0D}\quad\forall y_0,y_1.
\end{align}
Let $\Lambda^{A_0\to A_1}\coloneq\sum_{x_1}\Lambda_{x_1|x_0}^{A_0\to A_1}$ be the marginal channel of $\pd{\Lambda}$ from $A_0$ to $A_1$.  Let $A^*$ be a system such that $\spa{H}^{A^*}\cong\supp(J_\Lambda^{A_0A_1})\subseteq\spa{H}^{A_0A_1}$.  Let $\pd{S}\equiv\{S_{x_1|x_0}^{A^*}\}$ be a PMD such that $\pd{S}=\f{SEM}(\pd{\Lambda})$.  By Proposition~\ref{prop:operational}, there exists an isometric channel $\m{V}^{A_0\to A_1A^*}$ such that
\begin{align}
\label{eq:b3:2}
	\Lambda_{x_1|x_0}^{A_0\to A_1}\left[\cdot\right]&=\tr_{A^*}\left[\left(\1^{A_1}\otimes(S_{x_1|x_0}^\top)^{A^*}\right)\m{V}^{A_0\to A_1A^*}\left[\cdot\right]\right]\quad\forall x_0,x_1.
\end{align}
Define a quantum channel $\m{E}^{B_0\to B_1A^*K}$ as follows:
\begin{align}
\label{eq:b3:3}
	\m{E}^{B_0\to B_1A^*K}&=\sum_{k}\left(\m{K}_k^{A_1D\to B_1}\otimes\id^{A^*}\right)\circ\left(\m{V}^{A_0\to A_1A^*}\otimes\id^D\right)\circ\m{F}^{B_0\to A_0D}\otimes\op{k}{k}^K.
\end{align}
Define a quantum instrument $\ins{K'}\equiv\{\m{K}_{k'}'^{A^*K\to A^*}\}_{k'}$ as follows:
\begin{align}
	\m{K}_{k'}'^{A^*K\to A^*}\left[\cdot\right]&=\left(\1^{A^*}\otimes\bra{k'}^K\right)\left[\cdot\right]\left(\1^{A^*}\otimes\ket{k'}^K\right)\quad\forall k'.
\end{align}
Define a PMD $\pd{N}\equiv\{N_{y_1|y_0}^{A^*K}\}_{y_0,y_1}$ as follows:
\begin{align}
	N_{y_1|y_0}^{A^*K}&\coloneq\sum_{x_0,x_1,k',l}q_{y_1|x_1,l}p_{x_0,l|y_0,k'}(\m{K}_{k'}'^\dagger)^{A^*\to A^*K}\left[S_{x_1|x_0}^{A^*}\right] \label{eq:b3:4}\\
	&=\sum_{x_0,x_1,k',l}q_{y_1|x_1,l}p_{x_0,l|y_0,k'}S_{x_1|x_0}^{A^*}\otimes\op{k'}{k'}^K\quad\forall y_0,y_1. \label{eq:b3:5}
\end{align}
By Definition~\ref{def:simulation-PMD}, Eq.~\eqref{eq:b3:4} implies $\pd{S}\succcurlyeq_\abb{M}\pd{N}$.  It follows from Eqs.~\eqref{eq:b3:1}--\eqref{eq:b3:3}, and \eqref{eq:b3:5} that
\begin{align}
	&\tr_{A^*K}\left[\left(\1^{B_1}\otimes(N_{y_1|y_0}^\top)^{A^*K}\right)\m{E}^{B_0\to B_1A^*K}\left[\cdot\right]\right] \notag\\
	&\quad=\sum_{x_0,x_1,k,l,k'}q_{y_1|x_1,l}p_{x_0,l|y_0,k'}\tr_{A^*}\left[\left(\1^{B_1}\otimes(S_{x_1|x_0}^\top)^{A^*}\right)\left(\m{K}_k^{A_1D\to B_1}\otimes\id^{A^*}\right)\right. \notag\\
	&\quad\quad\left.\circ\left(\m{V}^{A_0\to A_1A^*}\otimes\id^D\right)\circ\m{F}^{B_0\to A_0D}\left[\cdot\right]\otimes\ip{k'}{k}\ip{k}{k'}^K\right] \\
	&\quad=\sum_{x_0,x_1,k,l}q_{y_1|x_1,l}p_{x_0,l|y_0,k}\m{K}_k^{A_1D\to B_1}\circ\tr_{A^*}\left[\left(\1^{A_1D}\otimes(S_{x_1|x_0}^\top)^{A^*}\right)\left(\m{V}^{A_0\to A_1A^*}\otimes\id^D\right)\circ\m{F}^{B_0\to A_0D}\left[\cdot\right]\right] \\
	&\quad=\sum_{x_0,x_1,k,l}q_{y_1|x_1,l}p_{x_0,l|y_0,k}\m{K}_k^{A_1D\to B_1}\circ\left(\Lambda_{x_1|x_0}^{A_0\to A_1}\otimes\id^D\right)\circ\m{F}^{B_0\to A_0D}\left[\cdot\right] \\
	&\quad=\Gamma_{y_1|y_0}^{B_0\to B_1}\left[\cdot\right]\quad\forall y_0,y_1.
\end{align}
By Definition~\ref{def:steering}, this implies $\pd{\Gamma}\leftarrowtail(\m{E},\pd{N}^\top)$, where $\pd{N}^\top\equiv\{(N_{y_1|y_0}^\top)^{A^*K}\}_{y_0,y_1}$.  Then by Proposition~\ref{prop:canonicity}, we have $\pd{N}\succcurlyeq_\abb{M}\f{SEM}(\pd{\Gamma})$.  Since $\f{SEM}(\pd{\Lambda})=\pd{S}\succcurlyeq_\abb{M}\pd{N}$, by the transitivity of the preorder $\succcurlyeq_\abb{M}$~\cite{Buscemi-2020a}, we have $\f{SEM}(\pd{\Lambda})\succcurlyeq_\abb{M}\f{SEM}(\pd{\Gamma})$.  This concludes the proof of Theorem~\ref{thm:monotonicity}(2).

\section{Semi-Device-Independent Characterization}
\label{app:game}

In this Appendix, we demonstrate the semi-device-independent characterization of PID nonsimplicity with guessing games.  This includes providing a complete set of incompatibility monotones (Theorem~\ref{thm:convertibility}) and an operational interpretation of the robustness of incompatibility (Theorem~\ref{thm:robustness}) based on nontransient guessing games, as well as a complete set of incompatibility monotones (Proposition~\ref{prop:convertibility}) based on postinformation guessing games.

\subsection{Proof of Theorem~\ref{thm:convertibility}}
\label{app:convertibility}

Let $\pd{\Lambda}\equiv\{\Lambda_{x_1|x_0}^{A_0\to A_1}\}_{x_0\in\idx{X}_0,x_1\in\idx{X}_1}$ be a PID\@.  Let $\povm{M}\equiv\{M_{m,n}^{C_0C_1}\}_{m\in\idx{M},n\in\idx{N}}$ be a bipartite POVM\@.  By Eq.~\eqref{eq:game-1}, Alice's maximum winning probability in the nontransient guessing game specified by $\povm{M}$ equals
\begin{align}
	P_\abb{guess}(\pd{\Lambda};\povm{M})&\coloneq\frac{1}{d_{C_0}}\max_{\pd{\Lambda}'\colon\pd{\Lambda}\succcurlyeq_\abb{I}\pd{\Lambda}'}\sum_{m\in\idx{M},n\in\idx{N}}\tr\left[M_{m,n}^{C_0C_1}\left(\id^{C_0}\otimes\Lambda_{n|m}'^{\rpl{C}_0\to C_1}\right)\left[\phi_+^{C_0\rpl{C}_0}\right]\right] \\
	&=\frac{1}{d_{C_0}}\max_{\pd{\Lambda}'\colon\pd{\Lambda}\succcurlyeq_\abb{I}\pd{\Lambda}'}\sum_{m\in\idx{M},n\in\idx{N}}\tr\left[M_{m,n}^{C_0C_1}J_{\Lambda_{n|m}'}^{C_0C_1}\right]. \label{eq:c1:1}
\end{align}
First, we prove the necessity of the convertibility conditions in Theorem~\ref{thm:convertibility}.  Let $\pd{\Lambda}\equiv\{\Lambda_{x_1|x_0}^{A_0\to A_1}\}_{x_0\in\idx{X}_0,x_1\in\idx{X}_1}$ and $\pd{\Gamma}\equiv\{\Gamma_{y_1|y_0}^{B_0\to B_1}\}_{y_0\in\idx{Y}_0,y_1\in\idx{Y}_1}$ be two PIDs such that $\pd{\Lambda}\succcurlyeq_\abb{I}\pd{\Gamma}$.  Let $\povm{M}\equiv\{M_{m,n}^{C_0C_1}\}_{m\in\idx{M},n\in\idx{N}}$ be a bipartite POVM\@.  By Theorem~\ref{thm:simulation-PID}(3), the preorder $\succcurlyeq_\abb{I}$ is transitive, thus
\begin{align}
	P_\abb{guess}(\pd{\Lambda};\povm{M})&=\frac{1}{d_{C_0}}\max_{\pd{\Lambda}'\colon\pd{\Lambda}\succcurlyeq_\abb{I}\pd{\Lambda}'}\sum_{m\in\idx{M},n\in\idx{N}}\tr\left[M_{m,n}^{C_0C_1}J_{\Lambda_{n|m}'}^{C_0C_1}\right] \\
	&\geq\frac{1}{d_{C_0}}\max_{\pd{\Lambda}'\colon\pd{\Gamma}\succcurlyeq_\abb{I}\pd{\Lambda}'}\sum_{m\in\idx{M},n\in\idx{N}}\tr\left[M_{m,n}^{C_0C_1}J_{\Lambda_{n|m}'}^{C_0C_1}\right] \\
	&=P_\abb{guess}(\pd{\Gamma};\povm{M}).
\end{align}
Next, we prove the sufficiency of the convertibility conditions in Theorem~\ref{thm:convertibility} by contradiction.  Let $\pd{\Lambda}\equiv\{\Lambda_{x_1|x_0}^{A_0\to A_1}\}_{x_0\in\idx{X}_0,x_1\in\idx{X}_1}$ and $\pd{\Gamma}\equiv\{\Gamma_{y_1|y_0}^{B_0\to B_1}\}_{y_0\in\idx{Y}_0,y_1\in\idx{Y}_1}$ be two PIDs such that $P_\abb{guess}(\pd{\Lambda};\povm{M})\geq P_\abb{guess}(\pd{\Gamma};\povm{M})$ for every bipartite POVM $\povm{M}\equiv\{M_{m,n}^{C_0C_1}\}_{m\in\idx{M},n\in\idx{N}}$.  We assume $\pd{\Lambda}\not\succcurlyeq_\abb{I}\pd{\Gamma}$.  This means that
\begin{align}
\label{eq:c1:2}
	\pd{\Gamma}&\notin\left\{\pd{\Lambda}'\equiv\left\{\Lambda_{y_1|y_0}'^{B_0\to B_1}\right\}_{y_0\in\idx{Y}_0,y_1\in\idx{Y}_1}\colon\pd{\Lambda}\succcurlyeq_\abb{I}\pd{\Lambda}'\right\}.
\end{align}
By Theorem~\ref{thm:simulation-PID}(4) and (5), the set of PIDs on the right-hand side of Eq.~\eqref{eq:c1:2} is closed and convex.  By the hyperplane separation theorem~\cite{Boyd-2004a}, there exists a set of Hermiticity-preserving linear maps $\{\m{O}_{y_0,y_1}^{B_0\to B_1}\}_{y_0\in\idx{Y}_0,y_1\in\idx{Y}_1}$ and a positive real number $\varepsilon>0$ such that
\begin{align}
	\left\langle\left\{\m{O}_{y_0,y_1}^{B_0\to B_1}\right\}_{y_0\in\idx{Y}_0,y_1\in\idx{Y}_1},\left\{\Gamma_{y_1|y_0}^{B_0\to B_1}\right\}_{y_0\in\idx{Y}_0,y_1\in\idx{Y}_1}\right\rangle&>\max_{\pd{\Lambda}'\colon\pd{\Lambda}\succcurlyeq_\abb{I}\pd{\Lambda}'}\left\langle\left\{\m{O}_{y_0,y_1}^{B_0\to B_1}\right\}_{y_0\in\idx{Y}_0,y_1\in\idx{Y}_1},\left\{\Lambda_{y_1|y_0}'^{B_0\to B_1}\right\}_{y_0\in\idx{Y}_0,y_1\in\idx{Y}_1}\right\rangle+\varepsilon.
\end{align}
The above equation can be interpreted based on the Hilbert--Schmidt inner product as follows:
\begin{align}
\label{eq:c1:3}
	\sum_{y_0\in\idx{Y}_0,y_1\in\idx{Y}_1}\tr\left[J_{\m{O}_{y_0,y_1}}^{B_0B_1}J_{\Gamma_{y_1|y_0}}^{B_0B_1}\right]&>\max_{\pd{\Lambda}'\colon\pd{\Lambda}\succcurlyeq_\abb{I}\pd{\Lambda}'}\sum_{y_0\in\idx{Y}_0,y_1\in\idx{Y}_1}\tr\left[J_{\m{O}_{y_0,y_1}}^{B_0B_1}J_{\Lambda_{y_1|y_0}'}^{B_0B_1}\right]+\varepsilon.
\end{align}
Let $c\coloneq\max_{y_0\in\idx{Y}_0,y_1\in\idx{Y}_1}\|J_{\m{O}_{y_0,y_1}}^{B_0B_1}\|_\infty$.  Define a bipartite POVM $\opt{\povm{M}}\equiv\{\opt{M}_{m,n}^{B_0B_1}\}_{m\in\idx{Y}_0,n\in\idx{N}}$ such that $\idx{N}\supset\idx{Y}_1$ as follows:
\begin{align}
\label{eq:c1:4}
	\opt{M}_{m,n}^{B_0B_1}&\coloneq\begin{cases}
		\frac{1}{2c\left|\idx{Y}_0\right|\left|\idx{Y}_1\right|}\left(J_{\m{O}_{m,n}}^{B_0B_1}+c\1^{B_0B_1}\right)\quad\forall m\in\idx{Y}_0,n\in\idx{Y}_1, \\
		\frac{1}{\left|\idx{Y}_0\right|\left(\left|\idx{N}\right|-\left|\idx{Y}_1\right|\right)}\left(\1^{B_0B_1}-\sum_{y_0\in\idx{Y}_0,y_1\in\idx{Y}_1}\opt{M}_{y_0,y_1}^{B_0B_1}\right)\quad\forall m\in\idx{Y}_0,n\in\idx{N}\setminus\idx{Y}_1.
	\end{cases}
\end{align}
It can be verified that $\opt{\povm{M}}$ is a valid POVM, as $\opt{M}_{m,n}^{B_0B_1}\geq0$ for all $m\in\idx{Y}_0,n\in\idx{N}$ and $\sum_{m\in\idx{Y}_0,n\in\idx{N}}\opt{M}_{m,n}^{B_0B_1}=\1^{B_0B_1}$.  Then
\begin{align}
	P_\abb{guess}(\pd{\Gamma};\opt{\povm{M}})&=\frac{1}{d_{B_0}}\max_{\pd{\Gamma}'\colon\pd{\Gamma}\succcurlyeq_\abb{I}\pd{\Gamma}'}\sum_{m\in\idx{Y}_0,n\in\idx{N}}\tr\left[\opt{M}_{m,n}^{B_0B_1}J_{\Gamma_{n|m}'}^{B_0B_1}\right] \\
	&\geq\frac{1}{d_{B_0}}\sum_{y_0\in\idx{Y}_0,y_1\in\idx{Y}_1}\tr\left[\opt{M}_{y_0,y_1}^{B_0B_1}J_{\Gamma_{y_1|y_0}}^{B_0B_1}\right] \\
	&=\frac{1}{2cd_{B_0}\left|\idx{Y}_0\right|\left|\idx{Y}_1\right|}\left(\sum_{y_0\in\idx{Y}_0,y_1\in\idx{Y}_1}\tr\left[J_{\m{O}_{y_0,y_1}}^{B_0B_1}J_{\Gamma_{y_1|y_0}}^{B_0B_1}\right]+c\sum_{y_0\in\idx{Y}_0,y_1\in\idx{Y}_1}\tr\left[J_{\Gamma_{y_1|y_0}}^{B_0B_1}\right]\right) \label{eq:c1:5}\\
	&=\frac{1}{2cd_{B_0}\left|\idx{Y}_0\right|\left|\idx{Y}_1\right|}\sum_{y_0\in\idx{Y}_0,y_1\in\idx{Y}_1}\tr\left[J_{\m{O}_{y_0,y_1}}^{B_0B_1}J_{\Gamma_{y_1|y_0}}^{B_0B_1}\right]+\frac{1}{2\left|\idx{Y}_1\right|} \\
	&>\frac{1}{2cd_{B_0}\left|\idx{Y}_0\right|\left|\idx{Y}_1\right|}\left(\max_{\pd{\Lambda}'\colon\pd{\Lambda}\succcurlyeq_\abb{I}\pd{\Lambda}'}\sum_{y_0\in\idx{Y}_0,y_1\in\idx{Y}_1}\tr\left[J_{\m{O}_{y_0,y_1}}^{B_0B_1}J_{\Lambda_{y_1|y_0}'}^{B_0B_1}\right]+\varepsilon\right)+\frac{1}{2\left|\idx{Y}_1\right|} \label{eq:c1:6}\\
	&=\frac{1}{2cd_{B_0}\left|\idx{Y}_0\right|\left|\idx{Y}_1\right|}\max_{\pd{\Lambda}'\colon\pd{\Lambda}\succcurlyeq_\abb{I}\pd{\Lambda}'}\left(\sum_{y_0\in\idx{Y}_0,y_1\in\idx{Y}_1}\tr\left[J_{\m{O}_{y_0,y_1}}^{B_0B_1}J_{\Lambda_{y_1|y_0}'}^{B_0B_1}\right]+c\sum_{y_0\in\idx{Y}_0,y_1\in\idx{Y}_1}\tr\left[J_{\Lambda_{y_1|y_0}'}^{B_0B_1}\right]\right)+\frac{\varepsilon}{2cd_{B_0}\left|\idx{Y}_0\right|\left|\idx{Y}_1\right|} \\
	&=\frac{1}{d_{B_0}}\max_{\pd{\Lambda}'\colon\pd{\Lambda}\succcurlyeq_\abb{I}\pd{\Lambda}'}\sum_{y_0\in\idx{Y}_0,y_1\in\idx{Y}_1}\tr\left[\opt{M}_{y_0,y_1}^{B_0B_1}J_{\Lambda_{y_1|y_0}'}^{B_0B_1}\right]+\frac{\varepsilon}{2cd_{B_0}\left|\idx{Y}_0\right|\left|\idx{Y}_1\right|} \label{eq:c1:7}\\
	&\geq\frac{1}{d_{B_0}}\max_{\pd{\Lambda}''\colon\pd{\Lambda}\succcurlyeq_\abb{I}\pd{\Lambda}''}\sum_{m\in\idx{Y}_0,n\in\idx{Y}_1}\tr\left[\opt{M}_{m,n}^{B_0B_1}J_{\Lambda_{n|m}''}^{B_0B_1}\right]+\frac{\varepsilon}{2cd_{B_0}\left|\idx{Y}_0\right|\left|\idx{Y}_1\right|} \label{eq:c1:8}\\
	&=\frac{1}{d_{B_0}}\max_{\pd{\Lambda}''\colon\pd{\Lambda}\succcurlyeq_\abb{I}\pd{\Lambda}''}\left(\sum_{m\in\idx{Y}_0,n\in\idx{N}}\tr\left[\opt{M}_{m,n}^{B_0B_1}J_{\Lambda_{n|m}''}^{B_0B_1}\right]-\sum_{m\in\idx{Y}_0,n\in\idx{N}\setminus\idx{Y}_1}\tr\left[\opt{M}_{m,n}^{B_0B_1}J_{\Lambda_{n|m}''}^{B_0B_1}\right]\right)+\frac{\varepsilon}{2cd_{B_0}\left|\idx{Y}_0\right|\left|\idx{Y}_1\right|} \\
	&\geq\frac{1}{d_{B_0}}\max_{\pd{\Lambda}''\colon\pd{\Lambda}\succcurlyeq_\abb{I}\pd{\Lambda}''}\left(\sum_{m\in\idx{Y}_0,n\in\idx{N}}\tr\left[\opt{M}_{m,n}^{B_0B_1}J_{\Lambda_{n|m}''}^{B_0B_1}\right]-\frac{1}{\left|\idx{Y}_0\right|\left(\left|\idx{N}\right|-\left|\idx{Y}_1\right|\right)}\sum_{m\in\idx{Y}_0,n\in\idx{N}\setminus\idx{Y}_1}\tr\left[J_{\Lambda_{n|m}''}^{B_0B_1}\right]\right)+\frac{\varepsilon}{2cd_{B_0}\left|\idx{Y}_0\right|\left|\idx{Y}_1\right|} \label{eq:c1:9}\\
	&\geq\frac{1}{d_{B_0}}\max_{\pd{\Lambda}''\colon\pd{\Lambda}\succcurlyeq_\abb{I}\pd{\Lambda}''}\left(\sum_{m\in\idx{Y}_0,n\in\idx{N}}\tr\left[\opt{M}_{m,n}^{B_0B_1}J_{\Lambda_{n|m}''}^{B_0B_1}\right]-\frac{1}{\left|\idx{Y}_0\right|\left(\left|\idx{N}\right|-\left|\idx{Y}_1\right|\right)}\sum_{m\in\idx{Y}_0,n\in\idx{N}}\tr\left[J_{\Lambda_{n|m}''}^{B_0B_1}\right]\right)+\frac{\varepsilon}{2cd_{B_0}\left|\idx{Y}_0\right|\left|\idx{Y}_1\right|} \\
	&=\frac{1}{d_{B_0}}\max_{\pd{\Lambda}''\colon\pd{\Lambda}\succcurlyeq_\abb{I}\pd{\Lambda}''}\sum_{m\in\idx{Y}_0,n\in\idx{N}}\tr\left[\opt{M}_{m,n}^{B_0B_1}J_{\Lambda_{n|m}''}^{B_0B_1}\right]-\frac{1}{\left|\idx{N}\right|-\left|\idx{Y}_1\right|}+\frac{\varepsilon}{2cd_{B_0}\left|\idx{Y}_0\right|\left|\idx{Y}_1\right|} \label{eq:c1:10}\\
	&=P_\abb{guess}(\pd{\Lambda};\opt{\povm{M}})-\frac{1}{\left|\idx{N}\right|-\left|\idx{Y}_1\right|}+\frac{\varepsilon}{2cd_{B_0}\left|\idx{Y}_0\right|\left|\idx{Y}_1\right|}. \label{eq:c1:11}
\end{align}
Here Eqs.~\eqref{eq:c1:5}, \eqref{eq:c1:7}, and \eqref{eq:c1:9} follow from Eq.~\eqref{eq:c1:4}; Ineq.~\eqref{eq:c1:6} follows from Ineq.~\eqref{eq:c1:3}.  Note that the PID in Ineq.~\eqref{eq:c1:8} has the form $\pd{\Lambda}''\equiv\{\Lambda_{n|m}''^{B_0\to B_1}\}_{m\in\idx{Y}_0,n\in\idx{N}}$ in contrast to the PID $\pd{\Lambda}'\equiv\{\Lambda_{y_1|y_0}'^{B_0\to B_1}\}_{y_0\in\idx{Y}_0,y_1\in\idx{Y}_1}$ in Eq.~\eqref{eq:c1:7}, and Ineq.~\eqref{eq:c1:8} follows from the fact that the freedom to make a guess $n$ outside the set of target outcomes $\idx{Y}_1$ does not increase the maximum winning probability.  Choose $\idx{N}\supseteq\idx{Y}_1$ to be an arbitrary index set such that
\begin{align}
	\left|\idx{N}\right|&>\left|\idx{Y}_1\right|+\frac{2cd_{B_0}\left|\idx{Y}_0\right|\left|\idx{Y}_1\right|}{\varepsilon}.
\end{align}
Then Eq.~\eqref{eq:c1:11} implies $P_\abb{guess}(\pd{\Lambda};\opt{\povm{M}})<P_\abb{guess}(\pd{\Gamma};\opt{\povm{M}})$, which contradicts the assumption that $P_\abb{guess}(\pd{\Lambda};\povm{M})\geq P_\abb{guess}(\pd{\Gamma};\povm{M})$ for every $\povm{M}$.  Therefore, we must have $\pd{\Lambda}\succcurlyeq_\abb{I}\pd{\Gamma}$.  This concludes the proof of Theorem~\ref{thm:convertibility}.

\subsection{Proof of Theorem~\ref{thm:robustness}}
\label{app:robustness}

Let $\pd{\Lambda}\equiv\{\Lambda_{x_1|x_0}^{A_0\to A_1}\}_{x_0\in\idx{X}_0,x_1\in\idx{X}_1}$ be a PID\@.  By Definition~\ref{def:robustness}, the robustness of incompatibility of $\pd{\Lambda}$ equals
\begin{subequations}
\label{eq:c2:1}
\begin{align}
	\f{RoI}(\pd{\Lambda})&\coloneq\min_{r\geq0}r \\
	\textnormal{subject to:}&\quad\left\{\frac{\Lambda_{x_1|x_0}+r\Upsilon_{x_1|x_0}}{1+r}\right\}_{x_0,x_1}\textnormal{ is a simple PID}, \label{eq:c2:2}\\
	&\quad\left\{\Upsilon_{x_1|x_0}\right\}_{x_0,x_1}\textnormal{ is a PID}. \label{eq:c2:3}
\end{align}
\end{subequations}
Since the set of simple PIDs is closed and convex and has a nonzero volume in the space of PIDs, the optimal solution to Program~\eqref{eq:c2:1} exists.  Let $\opt{\pd{\Upsilon}}\equiv\{\opt{\Upsilon}_{x_1|x_0}^{A_0\to A_1}\}_{x_0\in\idx{X}_0,x_1\in\idx{X}_1}$ be an optimal solution to Program~\eqref{eq:c2:1}, which is a PID\@.  Define a PID $\opt{\pd{\Omega}}\equiv\{\opt{\Omega}_{x_1|x_0}^{A_0\to A_1}\}_{x_0\in\idx{X}_0,x_1\in\idx{X}_1}$ as follows:
\begin{align}
\label{eq:c2:4}
	\opt{\Omega}_{x_1|x_0}^{A_0\to A_1}&\coloneq\frac{\Lambda_{x_1|x_0}^{A_0\to A_1}+\f{RoI}(\pd{\Lambda})\opt{\Upsilon}_{x_1|x_0}^{A_0\to A_1}}{1+\f{RoI}(\pd{\Lambda})}\quad\forall x_0,x_1.
\end{align}
By Eq.~\eqref{eq:c2:2}, $\opt{\pd{\Omega}}$ is simple.  Let $\s{C}$ denote the cone generated by the set of unsteerable state assemblages in $A_0A_1$:
\begin{align}
	\s{C}&\coloneq\left\{\left\{\omega_{x_1|x_0}^{A_0A_1}\right\}_{x_0\in\idx{X}_0,x_1\in\idx{X}_1}\colon\omega_{x_1|x_0}^{A_0A_1}=\sum_{g}p_{x_1|x_0,g}\eta_g^{A_0A_1},\;\eta_g^{A_0A_1}\geq0,\;p_{x_1|x_0,g}\geq0,\;\sum_{x_1}p_{x_1|x_0,g}=1\quad\forall x_0,x_1,g\right\}.
\end{align}
We note that the cone $\s{C}$ is generating with respect to the space of state ensembles in $A_0A_1$.  Let $\s{C}^*$ denote the dual cone of $\s{C}$:
\begin{align}
	\s{C}^*&\coloneq\left\{\left\{\kappa_{x_1|x_0}^{A_0A_1}\right\}_{x_0\in\idx{X}_0,x_1\in\idx{X}_1}\colon\sum_{x_0,x_1}\tr\left[\kappa_{x_1|x_0}^{A_0A_1}\omega_{x_1|x_0}^{A_0A_1}\right]\geq0\quad\forall\left\{\omega_{x_1|x_0}^{A_0A_1}\right\}_{x_0,x_1}\in\cone(\s{U}^{A_0A_1})\right\}.
\end{align}
Substituting $\omega_{x_1|x_0}^{A_0A_1}$ for $J_{\Lambda_{x_1|x_0}}^{A_0A_1}+rJ_{\Upsilon_{x_1|x_0}}^{A_0A_1}$, Program~\eqref{eq:c2:1} can be reformulated as a conic program as follows:
\begin{subequations}
\label{eq:c2:5}
\begin{align}
	\f{RoI}(\pd{\Lambda})&=\frac{1}{d_{A_0}\left|\idx{X}_0\right|}\min_{\left\{\omega_{x_1|x_0}^{A_0A_1}\right\}_{x_0,x_1}}\sum_{x_0,x_1}\tr\left[\omega_{x_1|x_0}^{A_0A_1}\right]-1 \\
	\textnormal{subject to:}&\quad\omega_{x_1|x_0}^{A_0A_1}-J_{\Lambda_{x_1|x_0}}^{A_0A_1}\geq0\quad\forall x_0,x_1, \\
	&\quad d_{A_0}\sum_{x_1}\tr_{A_1}\left[\omega_{x_1|x_0}^{A_0A_1}\right]=\sum_{x_1}\tr\left[\omega_{x_1|x_0}^{A_0A_1}\right]\1^{A_0}\quad\forall x_0, \\
	&\quad\left\{\omega_{x_1|x_0}^{A_0A_1}\right\}_{x_0,x_1}\in\s{C}.
\end{align}
\end{subequations}
Invoking the theory of conic programming duality~\cite{Boyd-2004a}, since $\omega_{x_1|x_0}^{A_0A_1}=2d_{A_0}\1^{A_0A_1}$ for all $x_0,x_1$ is a strictly feasible solution to Program~\eqref{eq:c2:5}, by Slater's condition, strong duality holds, and the optimal solution to the dual program exists.  The dual program of Program~\eqref{eq:c2:5} is given by
\begin{subequations}
\label{eq:c2:6}
\begin{align}
	\f{RoI}(\pd{\Lambda})&=\frac{1}{d_{A_0}\left|\idx{X}_0\right|}\max_{\left\{\alpha_{x_1|x_0}^{A_0A_1}\right\}_{x_0,x_1}}\sum_{x_0,x_1}\tr\left[\alpha_{x_1|x_0}^{A_0A_1}J_{\Lambda_{x_1|x_0}}^{A_0A_1}\right]-1 \label{eq:c2:7}\\
	\textnormal{subject to:}&\quad\alpha_{x_1|x_0}^{A_0A_1}\geq0\quad\forall x_0,x_1, \\
	&\quad\sum_{x_0}\tr\left[\beta_{x_0}^{A_0}\right]=d_{A_0}\left|\idx{X}_0\right|, \label{eq:c2:8}\\
	&\quad\left\{\beta_{x_0}^{A_0}\otimes\1^{A_1}-\alpha_{x_1|x_0}^{A_0A_1}\right\}_{x_0,x_1}\in\s{C}^*. \label{eq:c2:9}
\end{align}
\end{subequations}
Let $\{\opt{\alpha}_{x_1|x_0}^{A_0A_1}\}_{x_0,x_1}$, $\{\opt{\beta}_{x_1|x_0}^{A_0}\}_{x_0,x_1}$ be an optimal solution to Program~\eqref{eq:c2:6}.  First, we prove an upper bound on the game advantage in terms of the robustness of incompatibility.  Let $\povm{M}\equiv\{M_{m,n}^{C_0C_1}\}_{m\in\idx{M},n\in\idx{N}}$ be a bipartite POVM\@.  Let $\opt{\pd{\Lambda}}'\equiv\{\opt{\Lambda}_{n|m}'^{C_0\to C_1}\}_{m\in\idx{M},n\in\idx{N}}$ be an optimal solution to the maximization in Eq.~\eqref{eq:c1:1}, namely, the PID simulated by $\pd{\Lambda}$ under Alice's optimal strategy.  Let $\opt{\pd{\Upsilon}}'\equiv\{\opt{\Upsilon}_{n|m}'^{C_0\to C_1}\}_{m\in\idx{M},n\in\idx{N}}$ and $\opt{\pd{\Omega}}'\equiv\{\opt{\Omega}_{n|m}'^{C_0\to C_1}\}_{m\in\idx{M},n\in\idx{N}}$ denote the PIDs obtained by applying the same simulation strategy to $\opt{\pd{\Upsilon}}$ and $\opt{\pd{\Omega}}$, respectively.  Then
\begin{align}
	P_\abb{guess}(\pd{\Lambda};\povm{M})&=\frac{1}{d_{C_0}}\sum_{m\in\idx{M},n\in\idx{N}}\tr\left[M_{m,n}^{C_0C_1}J_{\opt{\Lambda}_{n|m}'}^{C_0C_1}\right] \\
	&=\frac{1}{d_{C_0}}\sum_{m\in\idx{M},n\in\idx{N}}\tr\left[M_{m,n}^{C_0C_1}\left(\left(1+\f{RoI}(\pd{\Lambda})\right)J_{\opt{\Omega}_{n|m}'}^{C_0C_1}-\f{RoI}(\pd{\Lambda})J_{\opt{\Upsilon}_{n|m}'}^{C_0C_1}\right)\right] \label{eq:c2:10}\\
	&\leq\frac{1+\f{RoI}(\pd{\Lambda})}{d_{C_0}}\sum_{m\in\idx{M},n\in\idx{N}}\tr\left[M_{m,n}^{C_0C_1}J_{\opt{\Omega}_{n|m}'}^{C_0C_1}\right] \\
	&\leq\frac{1+\f{RoI}(\pd{\Lambda})}{d_{C_0}}\max_{\pd{\Omega}\colon\opt{\pd{\Omega}}'\succcurlyeq_\abb{I}\pd{\Omega}}\sum_{m\in\idx{M},n\in\idx{N}}\tr\left[M_{m,n}^{C_0C_1}J_{\Omega_{n|m}}^{C_0C_1}\right] \\
	&=\frac{1+\f{RoI}(\pd{\Lambda})}{d_{C_0}}\max_{\pd{\Omega}\colon\abb{simple}}\sum_{m\in\idx{M},n\in\idx{N}}\tr\left[M_{m,n}^{C_0C_1}J_{\Omega_{n|m}}^{C_0C_1}\right] \label{eq:c2:11}\\
	&=\left(1+\f{RoI}(\pd{\Lambda})\right)P_\abb{guess}^\abb{simple}(\povm{M}).
\end{align}
Here Eq.~\eqref{eq:c2:10} follows from Eq.~\eqref{eq:c2:4} and the linearity of free simulations; Eq.~\eqref{eq:c2:11} follows from $\opt{\pd{\Omega}}'$ being simple and Theorem~\ref{thm:simulation-PID}(1) and (2).  This implies that
\begin{align}
\label{eq:c2:12}
	\sup_{\povm{M}}\frac{P_\abb{guess}(\pd{\Lambda};\povm{M})}{P_\abb{guess}^\abb{simple}(\povm{M})}&\leq1+\f{RoI}(\pd{\Lambda}).
\end{align}
Next, we show that Ineq.~\eqref{eq:c2:12} can be equalized by an infinite sequence of POVMs on $A_0A_1$.  Let $c\coloneq\|\sum_{x_0\in\idx{X}_0,x_1\in\idx{X}_1}\opt{\alpha}_{x_1|x_0}^{A_0A_1}\|_\infty$.  Define a bipartite POVM $\opt{\povm{M}}\equiv\{\opt{M}_{m,n}^{A_0A_1}\}_{m\in\idx{X}_0,n\in\idx{N}}$ such that $\idx{N}\supset\idx{X}_1$ as follows:
\begin{align}
\label{eq:c2:13}
	\opt{M}_{m,n}^{A_0A_1}&\coloneq\begin{cases}
		\frac{1}{c}\opt{\alpha}_{n|m}^{A_0A_1}\quad\forall m\in\idx{X}_0,n\in\idx{X}_1, \\
		\frac{1}{\left|\idx{X}_0\right|\left(\left|\idx{N}\right|-\left|\idx{X}_1\right|\right)}\left(\1^{A_0A_1}-\sum_{x_0\in\idx{X}_0,x_1\in\idx{X}_1}\opt{M}_{x_0,x_1}^{A_0A_1}\right)\quad\forall m\in\idx{X}_0,n\in\idx{N}\setminus\idx{X}_1.
	\end{cases}
\end{align}
It can be verified that $\opt{\povm{M}}$ is a valid POVM, as $\opt{M}_{m,n}^{A_0A_1}\geq0$ for all $m\in\idx{X}_0,n\in\idx{N}$ and $\sum_{m\in\idx{X}_0,n\in\idx{N}}\opt{M}_{m,n}^{A_0A_1}=\1^{A_0A_1}$.  Then
\begin{align}
	P_\abb{guess}(\pd{\Lambda},\opt{\povm{M}})&=\frac{1}{d_{A_0}}\max_{\pd{\Lambda}'\colon\pd{\Lambda}\succcurlyeq_\abb{I}\pd{\Lambda}'}\sum_{m\in\idx{X}_0,n\in\idx{N}}\tr\left[\opt{M}_{m,n}^{A_0A_1}J_{\Lambda_{n|m}'}^{A_0A_1}\right] \\
	&\geq\frac{1}{d_{A_0}}\sum_{x_0\in\idx{X}_0,x_1\in\idx{X}_1}\tr\left[\opt{M}_{x_0,x_1}^{A_0A_1}J_{\Lambda_{x_1|x_0}}^{A_0A_1}\right] \\
	&=\frac{1}{d_{A_0}c}\sum_{x_0\in\idx{X}_0,x_1\in\idx{X}_1}\tr\left[\opt{\alpha}_{x_1|x_0}^{A_0A_1}J_{\Lambda_{x_1|x_0}}^{A_0A_1}\right] \\
	&=\frac{\left|\idx{X}_0\right|\left(1+\f{RoI}(\pd{\Lambda})\right)}{c}. \label{eq:c2:14}
\end{align}
Here Eq.~\eqref{eq:c2:14} is by the definition of $\{\opt{\alpha}_{x_1|x_0}^{A_0A_1}\}_{x_0,x_1}$ and follows from Eq.~\eqref{eq:c2:7}.  In addition,
\begin{align}
	P_\abb{guess}^\abb{simple}(\opt{\povm{M}})&=\frac{1}{d_{A_0}}\max_{\pd{\Omega}\colon\abb{simple}}\sum_{m\in\idx{X}_0,n\in\idx{N}}\tr\left[\opt{M}_{m,n}^{A_0A_1}J_{\Omega_{n|m}}^{A_0A_1}\right] \\
	&=\frac{1}{d_{A_0}}\max_{\pd{\Omega}\colon\abb{simple}}\left(\sum_{m\in\idx{X}_0,n\in\idx{X}_1}\tr\left[\opt{M}_{m,n}^{A_0A_1}J_{\Omega_{n|m}}^{A_0A_1}\right]+\sum_{m\in\idx{X}_0,n\in\idx{N}\setminus\idx{X}_1}\tr\left[\opt{M}_{m,n}^{A_0A_1}J_{\Omega_{n|m}}^{A_0A_1}\right]\right) \\
	&\leq\frac{1}{d_{A_0}}\max_{\pd{\Omega}\colon\abb{simple}}\left(\sum_{m\in\idx{X}_0,n\in\idx{X}_1}\tr\left[\opt{M}_{m,n}^{A_0A_1}J_{\Omega_{n|m}}^{A_0A_1}\right]+\frac{1}{\left|\idx{X}_0\right|\left(\left|\idx{N}\right|-\left|\idx{X}_1\right|\right)}\sum_{m\in\idx{X}_0,n\in\idx{N}\setminus\idx{X}_1}\tr\left[J_{\Omega_{n|m}}^{A_0A_1}\right]\right) \label{eq:c2:15}\\
	&\leq\frac{1}{d_{A_0}}\max_{\pd{\Omega}\colon\abb{simple}}\left(\sum_{m\in\idx{X}_0,n\in\idx{X}_1}\tr\left[\opt{M}_{m,n}^{A_0A_1}J_{\Omega_{n|m}}^{A_0A_1}\right]+\frac{1}{\left|\idx{X}_0\right|\left(\left|\idx{N}\right|-\left|\idx{X}_1\right|\right)}\sum_{m\in\idx{X}_0,n\in\idx{N}}\tr\left[J_{\Omega_{n|m}}^{A_0A_1}\right]\right) \\
	&=\frac{1}{d_{A_0}}\max_{\pd{\Omega}\colon\abb{simple}}\sum_{m\in\idx{X}_0,n\in\idx{X}_1}\tr\left[\opt{M}_{m,n}^{A_0A_1}J_{\Omega_{n|m}}^{A_0A_1}\right]+\frac{1}{\left|\idx{N}\right|-\left|\idx{X}_1\right|} \label{eq:c2:16}\\
	&\leq\frac{1}{d_{A_0}}\max_{\pd{\Omega}'\colon\abb{simple}}\sum_{x_0\in\idx{X}_0,x_1\in\idx{X}_1}\tr\left[\opt{M}_{x_0,x_1}^{A_0A_1}J_{\Omega_{x_1|x_0}'}^{A_0A_1}\right]+\frac{1}{\left|\idx{N}\right|-\left|\idx{X}_1\right|} \label{eq:c2:17}\\
	&=\frac{1}{d_{A_0}c}\max_{\pd{\Omega}'\colon\abb{simple}}\sum_{x_0\in\idx{X}_0,x_1\in\idx{X}_1}\tr\left[\opt{\alpha}_{x_1|x_0}^{A_0A_1}J_{\Omega_{x_1|x_0}'}^{A_0A_1}\right]+\frac{1}{\left|\idx{N}\right|-\left|\idx{X}_1\right|} \label{eq:c2:18}\\
	&=\frac{1}{d_{A_0}c}\max_{\pd{\Omega}'\colon\abb{simple}}\sum_{x_0\in\idx{X}_0,x_1\in\idx{X}_1}\left(\tr\left[\left(\opt{\beta}_{x_0}^{A_0}\otimes\1^{A_1}\right)J_{\Omega_{x_1|x_0}'}^{A_0A_1}\right]-\tr\left[\left(\opt{\beta}_{x_0}^{A_0}\otimes\1^{A_1}-\opt{\alpha}_{x_1|x_0}^{A_0A_1}\right)J_{\Omega_{x_1|x_0}'}^{A_0A_1}\right]\right)+\frac{1}{\left|\idx{N}\right|-\left|\idx{X}_1\right|} \\
	&\leq\frac{1}{d_{A_0}c}\max_{\pd{\Omega}'\colon\abb{simple}}\sum_{x_0\in\idx{X}_0,x_1\in\idx{X}_1}\tr\left[\left(\opt{\beta}_{x_0}^{A_0}\otimes\1^{A_1}\right)J_{\Omega_{x_1|x_0}'}^{A_0A_1}\right]+\frac{1}{\left|\idx{N}\right|-\left|\idx{X}_1\right|} \label{eq:c2:19}\\
	&=\frac{1}{d_{A_0}c}\sum_{x_0\in\idx{X}_0}\tr\left[\opt{\beta}_{x_0}^{A_0}\right]+\frac{1}{\left|\idx{N}\right|-\left|\idx{X}_1\right|} \\
	&=\frac{\left|\idx{X}_0\right|}{c}+\frac{1}{\left|\idx{N}\right|-\left|\idx{X}_1\right|}. \label{eq:c2:20}
\end{align}
Here Eqs.~\eqref{eq:c2:15} and \eqref{eq:c2:18} follow from Eq.~\eqref{eq:c2:13}; Ineq.~\eqref{eq:c2:19} follows from Eq.~\eqref{eq:c2:8}; Eq.~\eqref{eq:c2:20} follows from Eq.~\eqref{eq:c2:9}.  Note that the PID in Ineq.~\eqref{eq:c2:17} has the form $\pd{\Omega}'\equiv\{\Omega_{x_1|x_0}'^{A_0\to A_1}\}_{x_0\in\idx{X}_0,x_1\in\idx{X}_1}$ in contrast to the PID $\pd{\Omega}\equiv\{\Omega_{n|m}^{A_0\to A_1}\}_{m\in\idx{X}_0,n\in\idx{N}}$ in Eq.~\eqref{eq:c2:16}, and Ineq.~\eqref{eq:c2:17} follows from the fact that the optimal strategy in Eq.~\eqref{eq:c2:16} necessarily makes a guess $n$ within the set of target outcomes $\idx{Y}_1$.  Then it follows from Eqs.~\eqref{eq:c2:14} and \eqref{eq:c2:20} that
\begin{align}
	\sup_{\povm{M}}\frac{P_\abb{guess}(\pd{\Lambda};\povm{M})}{P_\abb{guess}^\abb{simple}(\povm{M})}&\geq\lim_{|\idx{N}|\to\infty}\frac{P_\abb{guess}(\pd{\Lambda};\opt{\povm{M}})}{P_\abb{guess}^\abb{simple}(\opt{\povm{M}})} \\
	&=\lim_{|\idx{N}|\to\infty}\frac{1+\f{RoI}(\pd{\Lambda})}{1+\frac{c}{\left|\idx{X}_0\right|\left(\left|\idx{N}\right|-\left|\idx{X}_1\right|\right)}} \\
	&=1+\f{RoI}(\pd{\Lambda}). \label{eq:c2:21}
\end{align}
Combining Eqs.~\eqref{eq:c2:12} and \eqref{eq:c2:21}, we can conclude that
\begin{align}
	\sup_{\povm{M}}\frac{P_\abb{guess}(\pd{\Lambda};\povm{M})}{P_\abb{guess}^\abb{simple}(\povm{M})}&=1+\f{RoI}(\pd{\Lambda}).
\end{align}
This concludes the proof of Theorem~\ref{thm:robustness}.

\subsection{Proof of Proposition~\ref{prop:convertibility}}
\label{app:experimental}

First, we prove the necessity of the convertibility conditions in Proposition~\ref{prop:convertibility}.  Let $\pd{\Lambda}\equiv\{\Lambda_{x_1|x_0}^{A_0\to A_1}\}_{x_0\in\idx{X}_0,x_1\in\idx{X}_1}$ and $\pd{\Gamma}\equiv\{\Gamma_{y_1|y_0}^{A_0\to A_1}\}_{y_0\in\idx{Y}_0,y_1\in\idx{Y}_1}$ be two PIDs such that $\pd{\Lambda}\succcurlyeq_\abb{I}\pd{\Gamma}$.  Let $\povm{L}\equiv\{L_{l'}^{B_1}\}_{l'\in\idx{L}}$ be a POVM, and let $\ens{\varsigma}\equiv\{\sigma_{m,n,l}^{B_0}\}_{m\in\idx{Y}_0,n\in\idx{Y}_1,l\in\idx{L}}$ be a state ensemble.  By Theorem~\ref{thm:simulation-PID}(3), the preorder $\succcurlyeq_\abb{I}$ is transitive, thus
\begin{align}
	P_\abb{guess}'(\pd{\Lambda};\ens{\varsigma},\povm{L})&=\max_{\pd{\Lambda}'\colon\pd{\Lambda}\succcurlyeq_\abb{I}\pd{\Lambda}'}\sum_{m,n,l}\tr\left[L_l^{B_1}\Lambda_{n|m}'^{B_0\to B_1}\left[\sigma_{m,n,l}^{B_0}\right]\right] \\
	&\geq\max_{\pd{\Lambda}'\colon\pd{\Gamma}\succcurlyeq_\abb{I}\pd{\Lambda}'}\sum_{m,n,l}\tr\left[L_l^{B_1}\Lambda_{n|m}'^{B_0\to B_1}\left[\sigma_{m,n,l}^{B_0}\right]\right] \\
	&=P_\abb{guess}'(\pd{\Gamma};\ens{\varsigma},\povm{L}).
\end{align}
Next, we prove the sufficiency of the convertibility conditions in Proposition~\ref{prop:convertibility} by contradiction.  Let $\pd{\Lambda}\equiv\{\Lambda_{x_1|x_0}^{A_0\to A_1}\}_{x_0\in\idx{X}_0,x_1\in\idx{X}_1}$ be two PIDs, and let $\povm{L}\equiv\{L_{l'}^{B_1}\}_{l'\in\idx{L}}$ be an informationally complete POVM such that $P_\abb{guess}'(\pd{\Lambda};\ens{\varsigma},\povm{L})\geq P_\abb{guess}'(\pd{\Lambda};\ens{\varsigma},\povm{L})$ for every state ensemble $\ens{\varsigma}\equiv\{\sigma_{m,n,l}^{B_0}\}_{m\in\idx{Y}_0,n\in\idx{Y}_1,l\in\idx{L}}$.  We assume $\pd{\Lambda}\not\succcurlyeq_\abb{I}\pd{\Gamma}$.  Following the argument that leads to Eq.~\eqref{eq:c1:3}, there exists a set of Hermiticity-preserving linear maps $\{\m{O}_{y_0,y_1}^{B_0\to B_1}\}_{y_0\in\idx{Y}_0,y_1\in\idx{Y}_1}$ such that
\begin{align}
\label{eq:c3:1}
	\sum_{y_0,y_1}\tr\left[J_{\m{O}_{y_0,y_1}}^{B_0B_1}J_{\Gamma_{y_1|y_0}}^{B_0B_1}\right]&>\max_{\pd{\Lambda}'\colon\pd{\Lambda}\succcurlyeq_\abb{I}\pd{\Lambda}'}\sum_{y_0,y_1}\tr\left[J_{\m{O}_{y_0,y_1}}^{B_0B_1}J_{\Lambda_{y_1|y_0}'}^{B_0B_1}\right].
\end{align}
Since $\povm{L}$ is an informationally complete POVM, there exists a set of Hermitian operators $\{\mu_{y_0,y_1,l'}^{B_0}\}_{y_0\in\idx{Y}_0,y_1\in\idx{Y}_1,l'\in\idx{L}}$ such that
\begin{align}
	J_{\m{O}_{y_0,y_1}}^{B_0B_1}&=\sum_{l'}\mu_{y_0,y_1,l'}^{B_0}\otimes L_{l'}^{B_1}\quad\forall y_0,y_1.
\end{align}
Let $c\coloneq\max_{y_0\in\idx{Y}_0,y_1\in\idx{Y}_1}\|\mu_{y_0,y_1}^{B_0}\|_\infty$ and $c'\coloneq\sum_{m\in\idx{Y}_0,n\in\idx{Y}_1,l\in\idx{L}}\tr[\mu_{m,n,l}^{B_0}]$.  Define a state ensemble $\opt{\ens{\varsigma}}\equiv\{\opt{\sigma}_{m,n,l}^{B_0}\}_{m\in\idx{Y}_0,n\in\idx{Y}_1,l\in\idx{L}}$ as follows:
\begin{align}
\label{eq:c3:2}
	\opt{\sigma}_{m,n,l}^{B_0}&\coloneq\frac{1}{c'+cd_{B_0}\left|\idx{Y}_0\right|\left|\idx{Y}_1\right|\left|\idx{L}\right|}\left((\mu_{m,n,l}^\top)^{B_0}+c\1^{B_0}\right)\quad\forall m,n,l.
\end{align}
It can be verified that $\opt{\ens{\varsigma}}$ is a valid state ensemble, as $\opt{\sigma}_{m,n,l}^{B_0}\geq0$ for all $m\in\idx{Y}_0,n\in\idx{Y}_0,l\in\idx{L}$ and $\sum_{m,n,l}\tr[\opt{\sigma}_{m,n,l}^{B_0}]=1$.  Then
\begin{align}
	P_\abb{guess}'(\pd{\Lambda};\opt{\ens{\varsigma}},\povm{L})&=\max_{\pd{\Lambda}'\colon\pd{\Lambda}\succcurlyeq_\abb{I}\pd{\Lambda}'}\sum_{m,n,l}\tr\left[L_l^{B_1}\Gamma_{n|m}'^{B_0\to B_1}\left[\opt{\sigma}_{m,n,l}^{B_0}\right]\right] \\
	&=\frac{1}{c'+cd_{B_0}\left|\idx{Y}_0\right|\left|\idx{Y}_1\right|\left|\idx{L}\right|}\max_{\pd{\Lambda}'\colon\pd{\Lambda}\succcurlyeq_\abb{I}\pd{\Lambda}'}\left(\sum_{m,n,l}\tr\left[L_l^{B_1}\Lambda_{n|m}'^{B_0\to B_1}\left[(\mu_{m,n,l}^\top)^{B_0}\right]\right]+c\sum_{m,n,l}\tr\left[L_l^{B_1}\Lambda_{n|m}'^{B_0\to B_1}\left[\1^{B_0}\right]\right]\right) \label{eq:c3:3}\\
	&=\frac{1}{c'+cd_{B_0}\left|\idx{Y}_0\right|\left|\idx{Y}_1\right|\left|\idx{L}\right|}\max_{\pd{\Lambda}'\colon\pd{\Lambda}\succcurlyeq_\abb{I}\pd{\Lambda}'}\left(\sum_{m,n,l}\tr\left[\left(\mu_{m,n,l}^{B_0}\otimes L_l^{B_1}\right)\left(\id^{B_0}\otimes\Lambda_{n|m}'^{\rpl{B}_0\to B_1}\right)\left[\phi_+^{B_0\rpl{B}_0}\right]\right]+cd_{B_0}\left|\idx{Y}_0\right|\right) \\
	&=\frac{1}{c'+cd_{B_0}\left|\idx{Y}_0\right|\left|\idx{Y}_1\right|\left|\idx{L}\right|}\left(\max_{\pd{\Lambda}'\colon\pd{\Lambda}\succcurlyeq_\abb{I}\pd{\Lambda}'}\sum_{y_0,y_1}\tr\left[J_{\m{O}_{m,n}}^{B_0B_1}J_{\Lambda_{y_1|y_0}'}^{B_0B_1}\right]+cd_{B_0}\left|\idx{Y}_0\right|\right).
\end{align}
Here Eq.~\eqref{eq:c3:3} follows from Eq.~\eqref{eq:c3:2}.  It follows that
\begin{align}
	P_\abb{guess}'(\pd{\Gamma};\opt{\ens{\varsigma}},\povm{L})&=\frac{1}{c'+cd_{B_0}\left|\idx{Y}_0\right|\left|\idx{Y}_1\right|\left|\idx{L}\right|}\left(\max_{\pd{\Gamma}'\colon\pd{\Gamma}\succcurlyeq_\abb{I}\pd{\Gamma}'}\sum_{y_0,y_1}\tr\left[J_{\m{O}_{y_0,y_1}}^{B_0B_1}J_{\Gamma_{y_1|y_0}'}^{B_0B_1}\right]+cd_{B_0}\left|\idx{Y}_0\right|\right) \\
	&\geq\frac{1}{c'+cd_{B_0}\left|\idx{Y}_0\right|\left|\idx{Y}_1\right|\left|\idx{L}\right|}\left(\sum_{y_0,y_1}\tr\left[J_{\m{O}_{y_0,y_1}}^{B_0B_1}J_{\Gamma_{y_1|y_0}}^{B_0B_1}\right]+cd_{B_0}\left|\idx{Y}_0\right|\right) \\
	&>\frac{1}{c'+cd_{B_0}\left|\idx{Y}_0\right|\left|\idx{Y}_1\right|\left|\idx{L}\right|}\left(\max_{\pd{\Lambda}'\colon\pd{\Lambda}\succcurlyeq_\abb{I}\pd{\Lambda}'}\sum_{y_0,y_1}\tr\left[J_{\m{O}_{y_0,y_1}}^{B_0B_1}J_{\Lambda_{y_1|y_0}'}^{B_0B_1}\right]+cd_{B_0}\left|\idx{Y}_0\right|\right) \\
	&=P_\abb{guess}'(\pd{\Lambda};\opt{\ens{\varsigma}},\povm{L}).
\end{align}
This contradicts the assumption that $P_\abb{guess}'(\pd{\Lambda};\opt{\ens{\varsigma}},\povm{L})\geq P_\abb{guess}'(\pd{\Gamma};\opt{\ens{\varsigma}},\povm{L})$ for every $\opt{\ens{\varsigma}}$.  Therefore, we have $\pd{\Lambda}\succcurlyeq_\abb{I}\pd{\Gamma}$.  This concludes the proof of Proposition~\ref{prop:convertibility}.

\twocolumngrid


\bibliography{references}

\end{document}